\documentclass[prl,aps,twocolumn,amsmath,amsmath,amssymb,superscriptaddress]{revtex4-2}

\usepackage{graphicx}%
\usepackage{siunitx}
\usepackage{bm}
\usepackage[caption=false]{subfig}
\usepackage{mathtools}
\usepackage{multirow}
\usepackage{amssymb}
\usepackage{braket}
\usepackage{amsmath}
\usepackage{placeins}
\usepackage{cancel}
\usepackage{xcolor}
\usepackage{spreadtab}
\usepackage{dsfont}
\usepackage{tikz}
\usepackage[none]{hyphenat}
\usepackage[shortlabels]{enumitem}

\usepackage[colorlinks=true,
            linkcolor=blue,
	        citecolor=blue,
	         urlcolor=blue]{hyperref}

\newcommand\bq{{\mathbf{q}}}

\usetikzlibrary{decorations.markings}
\usetikzlibrary{decorations.pathmorphing}
\tikzset{%
    baseline=-0.43ex,
    el/.style={
        line cap=round,
        decoration={markings, mark=at position 1/2 with {\arrow[xshift=1.5pt, line width=1.5pt]>}},
        preaction={decorate},
         thick,
        },
    el1/.style={
        line cap=round,
         thick,
        },
    el_gray/.style={
        line cap=round,
        decoration={markings, mark=at position 1/2 with {\arrow[xshift=1.5pt]>}},
        preaction={decorate},
        ultra thick, color=gray
        },
    ph/.style={
        line cap=rect,
        decoration={snake, segment length=1.5mm, pre length=1mm, post length=1mm},
        decorate,
        thick,
        },
    elel/.style={
        line cap=rect,
        decoration={snake, aspect=0, pre length=0.3mm, post length=0.3mm},
        decorate,
        thick,
        },
    elhbubble_thick/.pic={
        \fill[white] (0, 0) to[out=-45, in=-135] (2, 0) to[out=135, in=45] cycle;
        \draw[el, ultra thick] (0, 0) to[out=-45, in=-135] (1.8, 0);
        \draw[el, ultra thick] (1.8, 0) to[out=+135, in=+45] (0, 0);   
        },
    elhbubble/.pic={
        \fill[white] (0, 0) to[out=-45, in=-135] (2, 0) to[out=135, in=45] cycle;
        \draw[el, double] (0, 0) to[out=-45, in=-135] (2, 0);
        \draw[el, double] (2, 0) to[out=+135, in=+45] (0, 0);
        },
    elhbubble2/.pic={
        \fill[white] (0, 0) to[out=-45, in=-135] (2, 0) to[out=135, in=45] cycle;
\draw[el1] (0,0.05) to[out=-45, in=-135]  (2,0.05);
\draw[el] (0,0)  to[out=-45, in=-135]  (2,0);
\draw[el1] (0,-0.05) to[out=-45, in=-135] (2,-0.05);,
\draw[el1] (2,0.05) to[out=135, in=45]  (0,0.05);
\draw[el] (2,0)  to[out=135, in=45]  (0,0);
\draw[el1] (2,-0.05) to[out=135, in=45] (0,-0.05);,
        },
    elhbubble_thin/.pic={
        \fill[white] (0, 0) to[out=-45, in=-135] (2, 0) to[out=135, in=45] cycle;
        \draw[el] (0, 0) to[out=-45, in=-135] (1.8, 0);
        \draw[el] (1.8, 0) to[out=+135, in=+45] (0, 0);
        },
        wiggle_circle_0/.pic={
        \draw[ph_,fill=white] (0.1, 0.5) circle (13pt); 
        \node[above=1mm] at (0.1, 0.9) {$\text{D}_0$};
         \fill[gray!80] (0,-0.15) rectangle (0.3,0.15);
        },
        wiggle_circle_small0/.pic={
        \draw[ph_,fill=white] (0.1, 0.3) circle (8.5pt); 
         \fill[gray!80] (0,-0.1) rectangle (0.2,0.1);
        },
    elhbubble_anh/.pic={
        \fill[white] (0, 0) to[out=-45, in=-135] (2, 0) to[out=135, in=45] cycle;
        \draw[el] (0, 0) to[out=-45, in=-135] (2, 0);
        \draw[el] (2, 0) to[out=+135, in=+45] (0, 0);
        },
    elhbubble_/.pic={
        \fill[white] (0, 0) to[out=-45, in=-135] (2, 0) to[out=135, in=45] cycle;
        \draw[el, ultra thick, color = gray] (0, 0) to[out=-45, in=-135] (2, 0);
        \draw[el, ultra thick, color = gray] (2, 0) to[out=+135, in=+45] (0, 0);
        },
    bubble/.pic={
        \draw[ph,double] (0, 0) to[out=120, in=80] (2, 0);
        },
    bubble_thick/.pic={
        \draw[ph,ultra thick] (0, 0) to[out=120, in=80,looseness=1.75] (1.5, 0);
        },
    bubble_bare/.pic={
        \draw[ph] (0, 0) to[out=120, in=80,looseness=1.75] (1.5, 0);
        \draw[el](0,0) to (1.5,0);
        },
    chi/.pic={\pic{bubble};
        },
    chi_thick/.pic={\pic{bubble_thick};
        },
    chiPH/.pic={\pic{elhbubble};
        },
    chiPH_thick/.pic={\pic{elhbubble_thick}; \node at (1, 0.05){$\Pi^{\text{exact}}$};
        },
    chiPH_gn/.pic={\pic{elhbubble_thick}; \node at (0.9, 0.05){$\Pi^{\text{scGscG}}$};
        },
    chig1/.pic={\pic{elhbubble}; \node at (1, 0.05) {$\Pi^{\text{G}_1\text{G}_1}$};
        },
    chig1_/.pic={\pic{elhbubble}; ;
        },
    chig2/.pic={\pic{elhbubble2}; \node at (1, 0.05) {$\Pi^{\text{G}_2\text{G}_2}$};
        },
  chi_bare/.pic={\pic{bubble_bare}; 
        },
    chi_bare_2/.pic={
        \draw[ph] (0, 0) to[out=120, in=80,looseness = 1.8] (0.8, 0);
        \draw[el](0,0) to (2,0);
        \draw[ph] (1.2, 0) to[out=120, in=80,looseness = 1.8] (2, 0);
        },
    chi_bare_wbub/.pic={        \draw[ph] (0, 0) to[out=120, in=80] (2, 0);
    \draw[el](0,0) to (2,0);
        \draw[el](0.5,0.55) to (1.4,0.55);
        \fill[white] (0.5,0.55) to[out=-45, in=-135] (1.4,0.55) to[out=135, in=45] cycle;
        \draw[el] (.5,0.55) to[out=-45, in=-135] (1.4,0.55);
        \draw[el] (1.4,0.55) to[out=+135, in=+45] (0.5,0.55);
        },
    chiPH_bare/.pic={\pic{elhbubble_thin}; \node at (1, 0.05) {$\Pi^{\text{G}_0\text{G}_0}$};
        },
    chiPH_bare_/.pic={\pic{elhbubble_thin};
        },
    chiPH_bare_sm/.pic={
        \fill[white] (0, 0) to[out=-45, in=-135] (1.5, 0) to[out=135, in=45] cycle;
        \draw[el] (0, 0) to[out=-45, in=-135] (1.5, 0);
        \draw[el] (1.5, 0) to[out=+135, in=+45] (0, 0);
        },
    chi0/.pic={\pic{bubble}; \node at (1, 0.05) {$\chi^b (\sigma, 0)$};
        },   
    chis/.pic={
        \draw[el] (0, 0) to[out=-70, in=-110] (1, 0);
        \draw[el] (1, 0) to[out=+110, in=+70] (0, 0);
        },
    Db/.pic={
        \draw[ph] (0, 0) -- (1, 0);
        \useasboundingbox (-0.15, 0) -- (1.15, 0);
        \node[below=1mm] at (0.5, 0) {D$^0$};
        },
            ph_/.style={
        line cap=rect,
        decoration={snake, segment length=1.5mm, pre length=0mm, post length=0mm},
        decorate,
        thick,
        },
    D1/.pic={
        \draw[ph, double] (0, 0) -- (1, 0);
        \useasboundingbox (-0.15, 0) -- (1.15, 0);
        \node[below=1mm] at (0.5, 0) {$\text{D}_1$};
        },
    D/.pic={
        \draw[ph, double] (0, 0) -- (1, 0);
        \useasboundingbox (-0.15, 0) -- (1.15, 0);
        \node[below=1mm] at (0.5, 0) {D};
        },
    D_thick/.pic={
        \draw[ph, ultra thick] (0, 0) -- (1, 0);
        \useasboundingbox (-0.15, 0) -- (1.15, 0);
        \node[below=1mm] at (0.5, 0) {scD};
        },
    D2/.pic={
        \draw[ph, double] (0, 0) -- (1, 0);
        \useasboundingbox (-0.15, 0) -- (1.15, 0);
        \node[below=1mm] at (0.5, 0) {$D^{\text{anh}}$};
        },
    Gb/.pic={
        \draw[el] (0, 0) -- (1, 0);
        \useasboundingbox (-0.15, 0) -- (1.15, 0);
        \node[below=1mm] at (0.5, 0) {$\text{G}_0$};
        },
    Gp/.pic={
        \draw[el, double] (0, 0) -- (1, 0);
        \useasboundingbox (-0.15, 0) -- (1.15, 0);
        \node[below=1mm] at (0.5, 0) {$G^p (T, \omega)$};
        },
    G/.pic={
        \draw[el,double] (0, 0) -- (1, 0);
        \useasboundingbox (-0.15, 0) -- (1.15, 0);
        \node[below=1mm] at (0.5, 0) {$G$};
        },
    G_thick/.pic={
        \draw[el,ultra thick] (0, 0) -- (1, 0);
        \useasboundingbox (-0.15, 0) -- (1.15, 0);
        \node[below=1mm] at (0.5, 0) {scG};
        },
    G2/.pic={
\draw[el1] (0,0.05) -- (1,0.05);
\draw[el] (0,0)    -- (1,0);
\draw[el1] (0,-0.05)-- (1,-0.05);,
        decoration={markings, mark=at position 1/2 with {\arrow[xshift=1.5pt, line width=1.5pt]>}},
        preaction={decorate},
        \useasboundingbox (-0.15, 0) -- (1.15, 0);
        \node[below=1mm] at (0.5, 0) {$\text{G}_2$};
        },    
    G1/.pic={
        \draw[el,double] (0, 0) -- (1, 0);
        \useasboundingbox (-0.15, 0) -- (1.15, 0);
        \node[below=1mm] at (0.5, 0) {$\text{G}_1$};
        },
    G_arrow/.pic={
        \draw[el, double] (0, 0) 
        -- (2, 0);
        \useasboundingbox (-0.15, 0) -- (1.15, 0);
        },
           wiggle_circle/.pic={
        \draw[ph_ ,ultra thick,fill=white] (0.1, 0.5) circle (13pt); 
        \node[above=1mm] at (0.1, 0.9) {$\text{scD}$};
         \fill[gray!80] (0,-0.15) rectangle (0.3,0.15);
        },
    G_arrow_thick/.pic={
        \draw[el, ultra thick] (0, 0) 
        -- (1.5, 0);
        \useasboundingbox (-0.15, 0) -- (1.15, 0);
        },
    op2/.pic={
        \draw[thick, fill=gray!30] (0, 0) circle (17pt);
        },
    op3/.pic={\draw[ ultra
    thick,
    fill=gray!30
] (0,0) circle (18pt);},
    op4/.pic={\draw[
    ultra thick,
    fill=gray!30
] (0,0) circle (8pt);},
    op1/.pic={
        \draw[thick, fill=gray!30] (0, 0) circle (8pt);
        },
    gb/.pic={
        \pic{ob};
        },
    grp/.pic={
        \pic{op}; 
        },
    sigma/.pic={
        \pic{op2}; 
        },
    sigma1/.pic={
        \pic{op3}; 
        },
    sigma_small/.pic={
        \pic{op1}; 
        },
    sigma1_small/.pic={
        \pic{op4}; 
        },
    ob/.pic={
        \fill (0, 0) circle (1.7pt);
        },
    o/.pic={\fill (0, 0) circle (2.6pt);},
    op/.pic={
        \draw[thick, fill=white] (0, 0) circle (7pt);
        },
    gb/.pic={
        \pic{ob};
        },
    grp/.pic={
        \pic{op}; 
        },
    small_wiggle_ud/.pic={
        \draw[ph] (0.35,0.3) to[out=120, in=80,looseness=1.8] (1.5, 0.35);
        \pic at (0.35,0.3) {gr};
        \pic at (1.5,0.35) {gr};
        \draw[ph] (0.35,-0.3) to[out=-120, in=-80,looseness=1.8] (1.5, -0.35);
        \pic at (0.35,-0.3) {gr};
        \pic at (1.5,-0.35) {gr};
        },
    small_wiggle_ud_sm/.pic={
        \draw[ph] (0.35,0.25) to[out=120, in=80,looseness=2] (1.15, 0.25);
        \pic at (0.35,0.25) {gr};
        \pic at (1.2,0.25) {gr};
        \draw[ph] (0.35,-0.25) to[out=-120, in=-80,looseness=2] (1.15, -0.25);
        \pic at (0.35,-0.25) {gr};
        \pic at (1.15,-0.25) {gr};
        },
    small_wiggle_below/.pic={
        \draw[ph] (0.35,-0.3) to[out=-120, in=-80,,looseness=1.8] (1.5, -0.35);
        \pic at (0.35,-0.3) {gr};
        \pic at (1.5,-0.35) {gr};
        },
    small_wiggle_ud1/.pic={
        \draw[ph,ultra thick] (0.35,0.3) to[out=120, in=80,looseness=1.8] (1.5, 0.35);
         \draw[el1,ultra thick] (1.5, 0.35) to[out=160, in=20] (0.35, 0.3);
        \pic at (0.35,0.3) {gr};
        \pic at (1.5,0.35) {gr};
        \draw[ph,ultra thick] (0.35,-0.3) to[out=-120, in=-80,looseness=1.8] (1.5, -0.35);
        \draw[el1,ultra thick] (0.35, -0.3) to[out=-20, in=-160] (1.5, -0.35);
        \pic at (0.35,-0.3) {gr};
        \pic at (1.5,-0.35) {gr};
        },
    small_wiggle_below1/.pic={

        \draw[ph,ultra thick ] (0.35,-0.3) to[out=-120, in=-80,,looseness=1.8] (1.5, -0.35);
        \pic at (0.35,-0.3) {gr};
        \pic at (1.5,-0.35) {gr};
        \draw[el1,ultra thick] (0.35, -0.3) to[out=-20, in=-160] (1.5, -0.35);},
    rainbow/.pic={
        \draw[ph] (0.35,0.3) to[out=120, in=80,looseness=1.8] (1.5, 0.35);
        \pic at (0.35,0.3) {gr};
        \pic at (1.5,0.35) {gr};
        \pic at (0.5,0.35) {gr};
        \pic at (1.35,0.35) {gr};
        \draw[ph] (0.5,0.35) to[out=120, in=80,looseness=1.4] (1.35, 0.35);
        },
    rainbow_sm/.pic={
        \draw[ph] (0.2,0.2) to[out=120, in=80,looseness=2] (1.35, 0.2);
        \pic at (0.2,0.2) {gr};
        \pic at (1.15,0.25) {gr};
        \pic at (0.4,0.25) {gr};
        \pic at (1.35,0.2) {gr};
        \draw[ph] (0.4,0.25) to[out=120, in=80,looseness=1.4] (1.15, 0.25);
        },
    rainbow2/.pic={
        \draw[ph] (0.,0.0) to[out=120, in=80,looseness=1.8] (1., 0.0);
        \pic at (1.0,0.0) {gr};
        \pic at (0.75,0.0) {gr};
        \pic at (0.25,0.0) {gr};
        \pic at (0.0,0.0) {gr};
        \draw[ph] (0.25,0.0) to[out=120, in=80,looseness=1.6] (0.75, 0.0);
        },
    small_wiggle/.pic={
        \draw[ph] (0.35,0.3) to[out=120, in=80,looseness=1.8] (1.5, 0.35);
        \pic at (0.35,0.3) {gr};
        \pic at (1.5,0.35) {gr};
        },
    small_wiggle_sm/.pic={
        \draw[ph] (0.33,0.25) to[out=120, in=80,looseness=2] (1.15, 0.25);
        \pic at (0.35,0.25) {gr};
        \pic at (1.15,0.25) {gr};
        },
    small_wiggle1/.pic={
        \draw[ph,ultra thick] (0.35,0.3) to[out=120, in=80,looseness=1.8] (1.5, 0.35);
        \pic at (0.35,0.3) {gr};
        \pic at (1.5,0.35) {gr};
        \draw[el1,ultra thick] (1.5, 0.35) to[out=160, in=20] (0.35, 0.3);},
    vertex/.pic={
        \draw[ph] (0.35,0.3) to(0.35, -0.3);
        \pic at (0.35,0.25) {gr};
        \pic at (0.35,-0.25) {gr};
        },
    twosmall_wiggle/.pic={
        \draw[ph] (0.18,0.15) to[out=120, in=80,looseness=1.8] (0.8, 0.4);
        \pic at (0.18,0.15) {gr};
        \pic at (0.8,0.4) {gr};
        \draw[ph] (1.3,0.4) to[out=120, in=80,looseness=1.8] (1.8, 0.1);
        \pic at (1.3,0.4) {gr};
        \pic at (1.8,0.1) {gr};
        },
    twosmall_wiggle_sm/.pic={
        \draw[ph] (0.18,0.15) to[out=120, in=80,looseness=3] (0.55, 0.3);
        \pic at (0.18,0.15) {gr};
        \pic at (0.55,0.3) {gr};
        \draw[ph] (0.95,0.3) to[out=90, in=60,looseness=2] (1.4, 0.15);
        \pic at (0.95,0.3) {gr};
        \pic at (1.35,0.15) {gr};
        },
    twosmall_wiggle1/.pic={
        \draw[ph,ultra thick] (0.18,0.15) to[out=120, in=80,looseness=1.8] (0.8, 0.4);
        \pic at (0.18,0.15) {gr};
        \pic at (0.8,0.4) {gr};
        \draw[ph,ultra thick] (1.3,0.4) to[out=120, in=80,looseness=1.8] (1.85, 0.15);
        \pic at (1.3,0.4) {gr};
        \pic at (1.85,0.15) {gr};
         \draw[el1,ultra thick] (0.18,0.15) to[out=40, in=10] (0.8, 0.40);
        \draw[el1,ultra thick] (1.3,0.4) to[out=-5, in=140] (1.85, 0.15);},
    twosmall_wigglec/.pic={
        \draw[ph] (0.35,0.25) to[out=120, in=120,looseness=1.8] (1.3, 0.3);
        \pic at (0.35,0.25) {gr};
        \pic at (1.3,0.3) {gr};

        \draw[ph] (0.75,0.35) to[out=80, in=80,looseness=1.75] (1.6, 0.16);
        \pic at (0.75,0.35) {gr};
        \pic at (1.6,0.16) {gr};
        },
    gr/.pic={
        \pic {o};
        }}
\begin{document}

\title{Higher-order nonadiabaticity governs the temperature dependence of the phonon spectrum}

\author{Nina Girotto Erhardt}
\email{nina.girotto@uclouvain.be}
\affiliation{%
European Theoretical Spectroscopy Facility, Institute of Condensed Matter and Nanosciences, Université catholique de Louvain, Chemin des Étoiles 8, B-1348 Louvain-la-Neuve, Belgium. 	
}%

\author{Samuel Ponc\'e}
\email{samuel.ponce@uclouvain.be}
\affiliation{%
European Theoretical Spectroscopy Facility, Institute of Condensed Matter and Nanosciences, Université catholique de Louvain, Chemin des Étoiles 8, B-1348 Louvain-la-Neuve, Belgium. 	
}%
\affiliation{%
WEL Research Institute, avenue Pasteur, 6, 1300 Wavre, Belgique.		
}%

\date{\today}

\begin{abstract}
Nonadiabatic effects determine the frequency and linewidth of coupled phonon modes, shape of Kohn anomalies
and have important implications on many material properties. 
State-of-the-art \textit{ab-initio} nonadiabatic phonon self-energy relies on an infinite electron lifetime
approximation, which cannot capture the temperature dependence of the phonon spectrum, neglects long-wavelength intraband phonon decay, and exhibits exaggerated phonon splitting.
In MgB$_2$, we show how the higher-order nonadiabatic phonon corrections mitigate these deficiencies, yielding linewidths in a closer experimental agreement and a dome-like coupling strength temperature dependence.
\end{abstract}

\maketitle
The electron-phonon coupling (EPC) renormalizes and broadens the electron and phonon band structures~\cite{giustino2017,mahan}.
Electron and phonon spectral functions can be computed from first-principles~\cite{giustino07,Ponce2016,giustino2017}, but almost always rely on a first-order and one-shot approximation. 
Only a few studies self-consistently include dynamical EPC renormalizations and linewidths for the electron and phonon degrees of freedom~\cite{Dee2020,bombin2023,bombin2023b,novko2019,park2025,Erhardt2025,lihm2026,xia2026}.
Additionally, phonons are frequently obtained from density-functional perturbation theory (DFPT)~\cite{Gonze1997,Baroni2001} within the adiabatic approximation, which is invalid if electronic transitions occur on the energy scale comparable to that of the phonon perturbations~\cite{Engelsberg1963,MAKSIMOV1996}.
%
%
Nonadiabatic (NA) effects can arise for coupled optical phonons around the zone center~\cite{cerdeira72,ponosov98,ponosov16,cappelluti2006,ponosov17,novko2018,novko20b} and at finite wavevectors~\cite{lazzeri06,pisana07,piscanec07,saitta08,malard2008,yan2008,das2009,calandra10,caruso17,sohier19,novko2020a,garcia2020,girotto2023,li2025}.
Many aspects of NA, including the renormalization of phonons via the phonon-electron self-energy~\cite{calandra10,berges2023,girotto2023,novko2018,garcia2020,allen74,marsiglio90,nosarzewski21,setty20,setty22,novko2020a}, vertex corrections~\cite{grimaldi95b,cappelluti00,gorkov16}, dynamical screening~\cite{krsnik2024,bauer2009,marini2025,krsnik2022}, dynamical modifications of the electron distribution~\cite{meng22a}, potential energy surface, and anharmonicity~\cite{meng22b} have been analyzed in model systems and from first-principles.
Indeed, NA effects have profound repercussions on the superconducting state~\cite{marsiglio90,nosarzewski21,gorkov2016, meng22a,girotto2023}, phase diagrams~\cite{setty22,Setty_2024,erhardt2025b}, transport properties~\cite{girotto2023}, and the total EPC strength~\cite{meng22a,girotto2023}.
However, first-order NA phonon renormalizations often predict excessive NA phonon hardening and pronounced NA splitting~\cite{garcia2020,krsnik2024,erhardt2025b}, cannot correctly capture long-wavelength phonon linewidths and temperature-dependent features~\cite{novko2018,Erhardt2025,novko20b,park2025,Dalladay2025}.
In these cases, higher-order NA effects, also called mode-mode coupling or electron-mediated anharmonicities~\cite{varma1983,Yoshiyama_1986} are needed.
So far only investigated for zone-center modes, they were found to induce two temperature regimes for the longwavelength phonon linewidth~\cite{cappelluti2006,novko2018,novko2020a,novko20b,Erhardt2025}.
With Raman measurements, this non-monotonic temperature dependence was observed in  MgB$_2$~\cite{ponosov17}, graphene~\cite{zan2024,chae10}, and a few other materials~\cite{osterhoudt2021,coulter2019,li2024,Yang2021}.

To emphasize the importance of these effects, we study MgB$_2$, a prominent superconductor~\cite{Nagamatsu2001,bohnen2001,kortus01,liu2001,yildirim2001,choi2002,margine2013,eiguren08,calandra10} with an anomalously large optical phonon linewidth, strong EPC~\cite{shukla2003,calandra2005,bohnen2001,Osborn2001,Tarenkov2009,Karapetrov2001,choi2002,Pickett2003,liu2001,Quilty2002,mou2015}, and prominent NA effects~\cite{cappelluti2006,novko2018,cappelluti2002,novko20b,girotto2023}. 
The large linewidth of the E$_{2g}$ symmetry phonon at $\bq=\Gamma$ derives precisely from the higher-order NA effects~\cite{cappelluti2006,novko2018}, which improve the agreement with temperature-dependent Raman experiments~\cite{Saitta2008,MAKSIMOV1996}.
Despite their importance, these effects and their temperature dependence have not yet been investigated across the full phonon spectrum.
The approach of Ref.~\cite{park2025} provides a framework applicable to arbitrary $\bq$, but was applied only to $\bq=\Gamma$ and to the imaginary part of the higher-order phonon self-energy.
We fill this gap by deriving the real part of the higher-order phonon self-energy within the electronic quasiparticle (QP) approximation, providing access to a full temperature-dependent analysis of phonon frequencies, linewidths, and electron-phonon coupling, which has so far remained unexplored throughout the Brillouin zone.
The concept of two temperature regimes for phonon linewidths is extended to finite wavevectors and is also observed in the temperature dependence of the phonon frequency.
In addition, we find that the temperature-induced enhancement of the electronic broadening blurs the sharp electron-hole damping continuum limit, reducing the NA splitting and hardening.
These strong temperature effects consequently renormalize the EPC strength $\lambda$, which is often computed within the adiabatic approximation~\cite{choi2002,bohnen2001,liu2001,margine2013}, leading to an intriguing dome-like temperature dependence of $\lambda$.

The exact electron and phonon propagators are connected by the self-consistent Dyson equations, see Sec.~S1 of the Supplementary Information (SI)~\cite{supplement}.
We assume the Migdal approximation and neglect vertex corrections~\cite{migdal58}.
Following the notation of Ref.~\cite{lihm2025}, the self-consistent electron and phonon propagators are denoted by scG and scD, respectively.
The fully self-consistent electron self-energy can be written as a sum of the Fan-Migdal (FM) and the static Debye-Waller (DW) terms~\cite{ponce2015,marini2015,allen1981, allen1983, Allen_1976}
\begin{equation}
\Sigma^{\text{scGscD}}_{n\mathbf{k}}(\varepsilon) = \Sigma^{\text{scGscD}-\text{FM}}_{n\mathbf{k}}(\varepsilon)+ \Sigma^{\text{DW}}_{n\mathbf{k}},
\end{equation}
where the retarded FM and DW self-energies can be written as
\begin{align}\label{eq:scSigma}
\Sigma^{\text{scGscD}-\text{FM}}_{n\mathbf{k}}(\varepsilon)=&\sum_{m\nu\mathbf{q}}|g_{mn\nu}(\mathbf{k},\mathbf{q})|^2 \iint_{-\infty}^{\infty} \!\!\!\!\! \mathrm{d}\omega\mathrm{d}\varepsilon' 
 A^{\text{scGscD}}_{m\mathbf{k}+\mathbf{q}}(\varepsilon') \nonumber\\
&\times B^{\text{scGscG}}_{\nu\mathbf{q}}(\omega)\frac{n(\omega)-f(\varepsilon')}{\varepsilon'+i0^+-\omega-\varepsilon}, \\
\Sigma^{\text{DW}}_{n\mathbf{k}} =& \sum_{\nu\mathbf{q} } \frac{\big(n(\omega_{\nu\mathbf{q}})+ \frac{1}{2}\big)}{2\omega_{\nu\mathbf{q}}}\mathfrak{g}^{2}_{n\nu}(\mathbf{k},\mathbf{q}),
\end{align}
and the retarded phonon self-energy, in the screened-screened approximation~\cite{berges2023}, reads
\begin{align}
    \Pi^{\text{scGscG}}_{\nu \mathbf{q}}(\omega)=&\sum_{nm\mathbf{k}}|{g}_{mn\nu}(\mathbf{k},\mathbf{q})|^2 \iint_{-\infty}^{\infty} \mathrm{d}\varepsilon \mathrm{d}\varepsilon'A^{\text{scGscD}}_{n\mathbf{k}} ( \varepsilon )\nonumber\\ &\times   A^{\text{scGscD}}_{m\mathbf{k}+\mathbf{q}} ( \varepsilon' ) \Big(\frac{f(\varepsilon)-f(\varepsilon')}{\varepsilon+\omega+i0^+-\varepsilon'} \Big) , \label{eq:scPi}
\end{align}
where ${g}_{mn\nu}(\mathbf{k},\mathbf{q})$ are the statically screened EPC matrix elements, $\mathfrak{g}_{n\nu}(\mathbf{k},\mathbf{q})$ the Debye-Waller matrix elements~\cite{giustino2017}, 
$f(\varepsilon)$ and $n(\omega)$ the Fermi-Dirac and the Bose-Einstein occupation functions.
$A_{n\mathbf{k}}^{\text{scGscD}}$ and $B^{\text{scGscG}}_{\nu \mathbf{q}}$ are the self-consistent electron and phonon spectral functions expressed respectively as
\begin{align}
     A^{ \text{scGscD}}_{n\mathbf{k}} ( \varepsilon ) &= -\frac{1}{\pi} \Im \frac{1}{\varepsilon - \varepsilon_{n\mathbf{k}} - \Sigma^{\text{scGscD}}_{n\mathbf{k}}(\varepsilon)}, \label{eq:scA} \\ 
    B^{\text{scGscG}}_{\nu \mathbf{q}}(\omega) &= -\frac{2\omega}{\pi} \Im \frac{1}{\omega^2 - \omega_{\nu\mathbf{q}} ^2 -2\omega_{\nu\mathbf{q}}\Pi^{\text{scGscG}}_{\nu\mathbf{q}}(\omega)}. \label{eq:scB}
\end{align}
Computing the fully self-consistent phonon self-energy requires computationally expensive iterative evaluations of the integrals in Eqs.~(\ref{eq:scSigma},~\ref{eq:scPi}).
These integrals can be solved analytically under the QP approximation~\cite{park2025}.
To go beyond the state-of-the-art first-order phonon self-energy (G$_0$G$_0$) approximation, we adopt a one-shot solution to this set of equations within a QP approximation for electrons and denote it by $\tilde{\text{G}}_1\tilde{\text{G}}_1$, where a tilde denotes the quasiparticle approximation in the electronic self-energy.
Using Eq.~\eqref{eq:scSigma} we compute the on-shell $\Sigma^{\text{G}_0\text{D}_0 }_{n\mathbf{k}}$ self-energies, yielding a QP spectral function $\tilde{A}^{\text{G}_0\text{D}_0}_{n\mathbf{k}} ( \varepsilon )$.
The QP approximation allows to conveniently rewrite the imaginary part of the phonon self-energy in terms of digamma functions $\psi$ as~\cite{park2025}: 
\begin{multline}\label{eq:img1g1}
\Im\Pi^{\rm \tilde{\text{G}}_1\tilde{\text{G}}_1}_{ \nu\mathbf{q}}(\omega)=-\frac{1}{2 \pi} \sum_{m n \mathbf{k}}\left|g_{m n \nu}(\mathbf{k}, \mathbf{q})\right|^2 \\
\times \Re\bigg[
\frac{\Psi(\mathfrak{E}^*_{m \mathbf{k}+\mathbf{q}}-\omega) \! - \! \Psi(\mathfrak{E}^*_{n \mathbf{k}})}{\mathfrak{E}^*_{n \mathbf{k}}+\omega-\mathfrak{E}^*_{m \mathbf{k}+\mathbf{q}}}  + \frac{\Psi(\mathfrak{E}^*_{n \mathbf{k}}) \!-\! \Psi(\omega-\mathfrak{E}_{m \mathbf{k}+\mathbf{q}})}{\mathfrak{E}^*_{n \mathbf{k}}+\omega-\mathfrak{E}_{m \mathbf{k}+\mathbf{q}}} \\
 +\frac{\Psi(\mathfrak{E}^*_{n \mathbf{k}} \!+\! \omega)\!-\!\Psi(\mathfrak{E}^*_{m \mathbf{k}+\mathbf{q}})}{\mathfrak{E}^*_{n \mathbf{k}}+\omega - \mathfrak{E}^*_{m \mathbf{k}+\mathbf{q}}}   +  \frac{\Psi(\mathfrak{E}^*_{m \mathbf{k}+\mathbf{q}}) \!-\! \Psi(-\omega\!-\!\mathfrak{E}_{n \mathbf{k}})}{\mathfrak{E}_{n \mathbf{k}}+\omega-\mathfrak{E}^*_{m \mathbf{k}+\mathbf{q}}}\bigg], 
\end{multline}
where 
\begin{align}
    \Psi(\omega + \mathfrak{E}_{n \mathbf{k}}) \equiv& \psi\left[\frac{1}{2}-\frac{1}{2 \pi i \text{k$_\text{B}$T}}\left(\omega + \mathfrak{E}_{n \mathbf{k}}\right)\right], \\
    \mathfrak{E}_{n \mathbf{k}} \equiv& E_{n\mathbf{k}}-\mu+i \gamma_{n \mathbf{k}}\label{eq:digamma},
\end{align}
and $\mu$ is the chemical potential.
A novel result of this work is a derivation of the real part shown in Sec.~S2 of the SI~\cite{supplement} that enables to investigate higher-order NA effects in the full phonon spectrum. 
The renormalized eigenvalues and linewidths in Eq.~\eqref{eq:digamma} are given by
\begin{align}\label{eq:Ereal}
    E_{n\mathbf{k}} =& \varepsilon_{n\mathbf{k}} + \Re \Sigma_{n\mathbf{k}}^{\rm \text{G}_0\text{D}_0}(E_{n\mathbf{k}}),\\
    \gamma_{n\mathbf{k}} =& -\Im \Sigma_{n\mathbf{k}}^{\rm \text{G}_0\text{D}_0}(E_{n\mathbf{k}}), \label{eq:Eimag}
\end{align}
where the real part of the self-energy $\Sigma_{n\mathbf{k}}^{\rm \text{G}_0\text{D}_0}(E_{n\mathbf{k}})$ includes the FM and the DW contributions
\begin{align}
\Sigma^{\text{G}_0\text{D}_0}_{n\mathbf{k}} (E_{n\mathbf{k}}) =& \Sigma^{\text{DW}}_{n\mathbf{k}} + \sum_{m\nu \textbf{q}} |g_{mn \nu} (\textbf{k},\textbf{q})|^2 \nonumber \\ &\times\sum_{\pm}
\frac{
f^{\pm}(\varepsilon_{m\mathbf{k}+\mathbf{q}}) + n(\omega_{\mathbf{q}\nu})
}{
E_{n\mathbf{k}} - \varepsilon_{m\mathbf{k}+\mathbf{q}} \pm \omega_{\mathbf{q}\nu} + i\delta
}, \label{eq:g0d0}
\end{align} 
where we use a small fixed $\delta$ = 5~meV broadening.
The DW term is computed through Wannier function perturbation theory~\cite{lihm2021} within the rigid-ion approximation~\cite{Allen_1976,ponce2014}.
The Eqs.~(\ref{eq:Ereal} -  \ref{eq:g0d0}) are solved iteratively for each temperature, under the constraint that the number of electrons ($N$) is conserved~\cite{tomczak2010,pickem2022}
\begin{equation}
    N = 2\sum_{n\mathbf{k}} \frac{1}{2} - \frac{1}{\pi}\Im \psi \Big( \frac{1}{2} - \frac{1}{2\pi i\text{k$_{\text{B}}\text{T}$}}\mathfrak{E}^*_{n \mathbf{k}}\Big)
\end{equation}
by updating the chemical potential $\mu$ inside $\mathfrak{E}^*_{n \mathbf{k}}$. 
The phonon spectral functions in the new one-shot ($B_{\nu \mathbf{q}}^{\rm \tilde{\text{G}}_1\tilde{\text{G}}_1}(\omega)$) and state-of-the-art first-order ($B_{\nu \mathbf{q}}^{\rm \text{G}_0 \text{G}_0}(\omega)$) approximation are then obtained by substituting the respective dynamical phonon self-energy ($\Pi^{\rm \tilde{\text{G}}_1\tilde{\text{G}}_1}_{\nu\mathbf{q} }(\omega)$ or $\Pi^{\rm \text{G}_0\text{G}_0}_{\nu\mathbf{q}}(\omega)$) in Eq.~\ref{eq:scB}, where $\Pi^{\rm \text{G}_0 \text{G}_0}_{\nu\mathbf{q}}(\omega)$ is given by~\cite{giustino2017,berges2023}:
\begin{align}\label{eq:pi0}
\Pi^{\rm \text{G}_0 \text{G}_0}_{\nu\mathbf{q}}(\omega) \! = \! \sum_{nm\mathbf{k}} |g_{mn\nu}(\mathbf{k},\mathbf{q})|^2\frac{f(\varepsilon_{m\mathbf{k}+\mathbf{q}}) \! - \! f(\varepsilon_{n\mathbf{k}})}{\varepsilon_{m\mathbf{k}+\mathbf{q}} \! - \!\varepsilon_{n\mathbf{k}} \! - \! \omega \! - \! i\eta},
\end{align}
where we used $\eta$ = 10~meV.

\begin{figure}[t!]
    \centering
\includegraphics[width=0.49\textwidth]{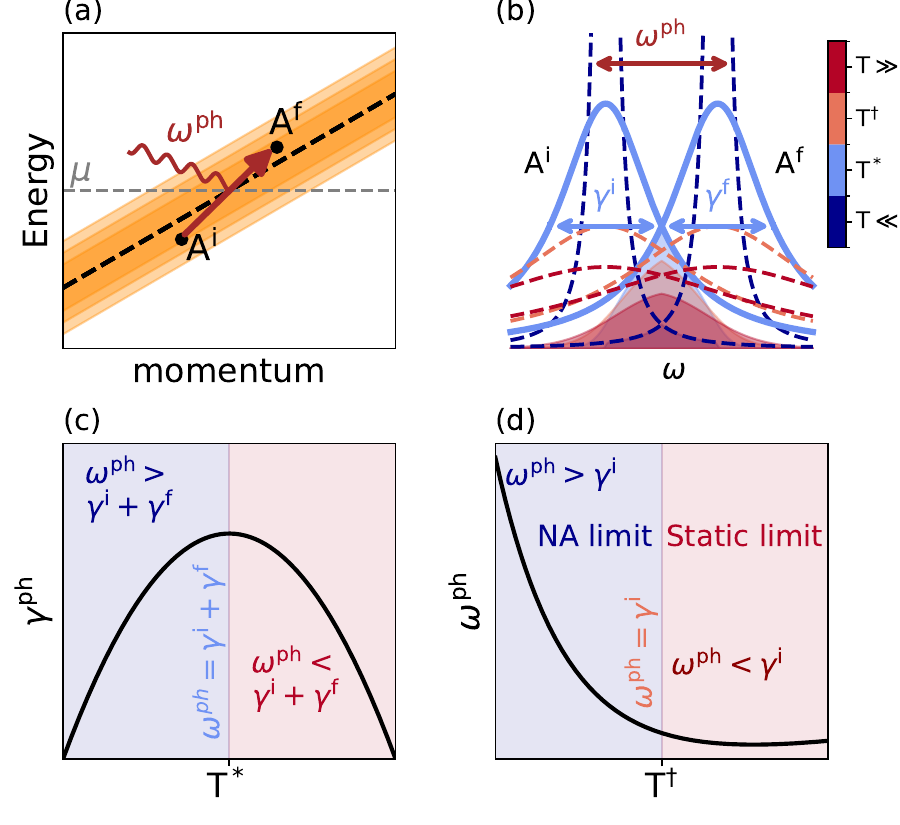}
    \caption{Schematics of the higher-order NA phonon energy and linewidth temperature dependence. (a) Phonon-assisted electron transition from the initial (A$^{\text{i}}$) to the final state (A$^{\text{f}}$). (b) A$^{\text{i}}$ and A$^{\text{f}}$ are shown for different temperatures, inducing weak (dark blue), moderate (light blue) and large (light and dark red) broadening. Their overlap multiplied by the difference of the Fermi-Dirac factors is shaded and is the largest at $\text{T}=\text{T}^*$ when $\gamma^{\text{i}}+\gamma^{\text{f}}=\omega^{\text{ph}}$. (c) Phonon linewidth with a peak at $\text{T}=\text{T}^*$.  (d) Phonon frequency, with a saturation of the softening for $\text{T}> \text{T}^{\dagger}$, when the phonon renormalization enters the adiabatic regime ($\gamma^{\text{i}} = \omega^{\text{ph}}$).}
    \label{fig:doodle}
\end{figure}

To have a deeper understanding of the temperature dependence of higher-order NA effects, we first consider the Fröhlich and Holstein models and present the results in the End Matter.
We find that finite broadening of the electronic bands naturally distinguishes two regimes: the electronic linewidth being larger or smaller than the phonon frequency, see Fig.~\ref{fig:doodle}. 
Higher-order NA effects increase the phonon linewidth with T for a specific mode, until the sum of electron broadening of the initial and final states becomes larger than the phonon frequency at a temperature that we denote by T$^*$.
This broadening increase with T corresponds to the increase of the overlap between the electron spectral functions of the initial and final states (see Sec.~S1 of the SI~\cite{supplement}), shown as the shaded region in Fig.~\ref{fig:doodle}(b). 
The phonon linewidth starts to decrease for $\text{T}>\text{T}^*$ as the phase space for electronic scattering is reduced by QP broadening and Pauli blocking.
Even though in Fig.~\ref{fig:doodle} we show the case of phonon-assisted intraband electron transition with a finite wavevector $\bq$, the generalization to the interband transition is straightforward; the A$^{\text{i}}$ and A$^{\text{f}}$ just belong to different bands.
Note that in the limit $\bq\rightarrow0$ (vertical transitions), the intraband terms still contribute, while this contribution is neglected in the G$_0$G$_0$ approximation.

The first-order NA frequency renormalization hardens the adiabatic phonons, and the addition of higher-orders diminishes this effect~\cite{novko2018}, more so as temperature increases.
However, when the electron lifetime becomes becomes shorter than the phonon period, the system enters the adiabatic limit and there is a saturation and even a weak reversal of this softening trend.
We denote this temperature with T$^{\dagger}$.
In the adiabatic limit, the static component of the $\tilde{\text{G}}_1\tilde{\text{G}}_1$ phonon self-energy determines the frequency renormalization, see Sec.~S3 of the SI~\cite{supplement}.

\begin{figure}
    \centering
\includegraphics[width=0.48\textwidth]{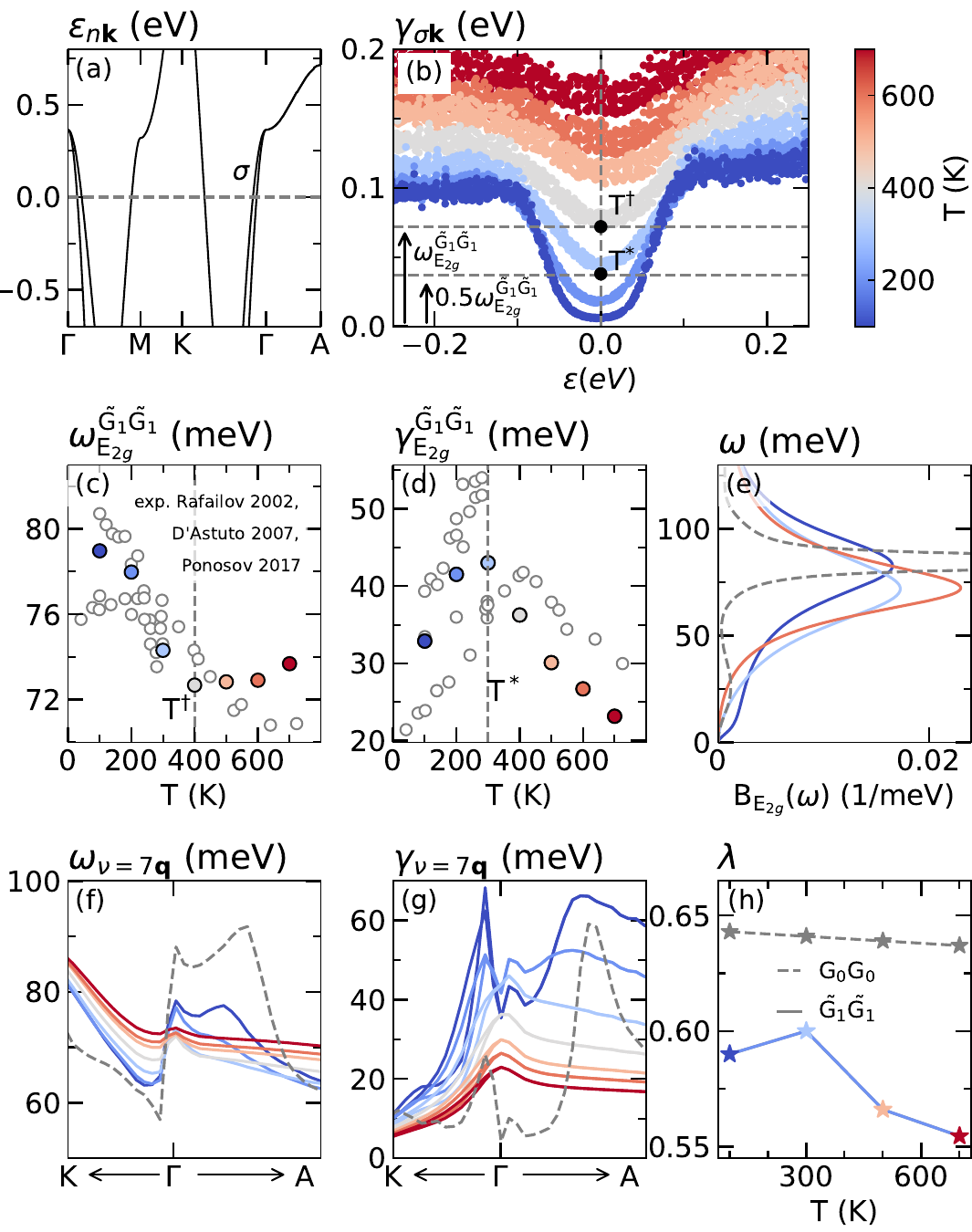}
    \caption{(a) MgB$_2$ band structure. (b) $\sigma$-band linewidth for various temperatures. Horizontal lines mark $\gamma_{\sigma\mathbf{k}}(\text{T}) = \omega^{\tilde{\text{G}}_1\tilde{\text{G}}_1}_{E_{2g}}$ and $\gamma_{\sigma\mathbf{k}}(\text{T})=0.5\omega^{\tilde{\text{G}}_1\tilde{\text{G}}_1}_{E_{2g}}$, reached at respective temperatures T$^{\dagger}$ and T$^*$.
    (c, d) Temperature-dependent Raman spectrum of MgB$_2$ compared with experiments (white discs) from Refs.~\cite{rafailov2002,astuto2007,ponosov17}. Vertical lines denote the two characteristic temperatures: T$^{\dagger}$ and T$^*$. 
    (e) Phonon spectral function of the E$_{2g}$ mode at $\bq = \Gamma$ for three different temperatures (100~K, 300~K, 500~K), compared with the G$_0$G$_0$ approximation (dashed grey lines).
    (f) Spectral function maxima of the $\nu=7$ mode close to the $\bq = \Gamma$ point for different temperatures within the $\tilde{\text{G}}_1\tilde{\text{G}}_1$ approach, compared with the G$_0$G$_0$ approach (dashed grey lines). The latter shows no temperature dependence and is shown only for 300~K.
    (g) Temperature-dependent linewidth of the $\nu=7$ mode. 
    (h) Temperature dependence of the coupling constant $\lambda$ within the $\tilde{\text{G}}_1$$\tilde{\text{G}}_1$ approach and the G$_0$G$_0$ approximation (grey symbols).}
    \label{fig:fig3}
\end{figure}
\begin{figure}[ht]
    \centering
    \includegraphics[width=\linewidth]{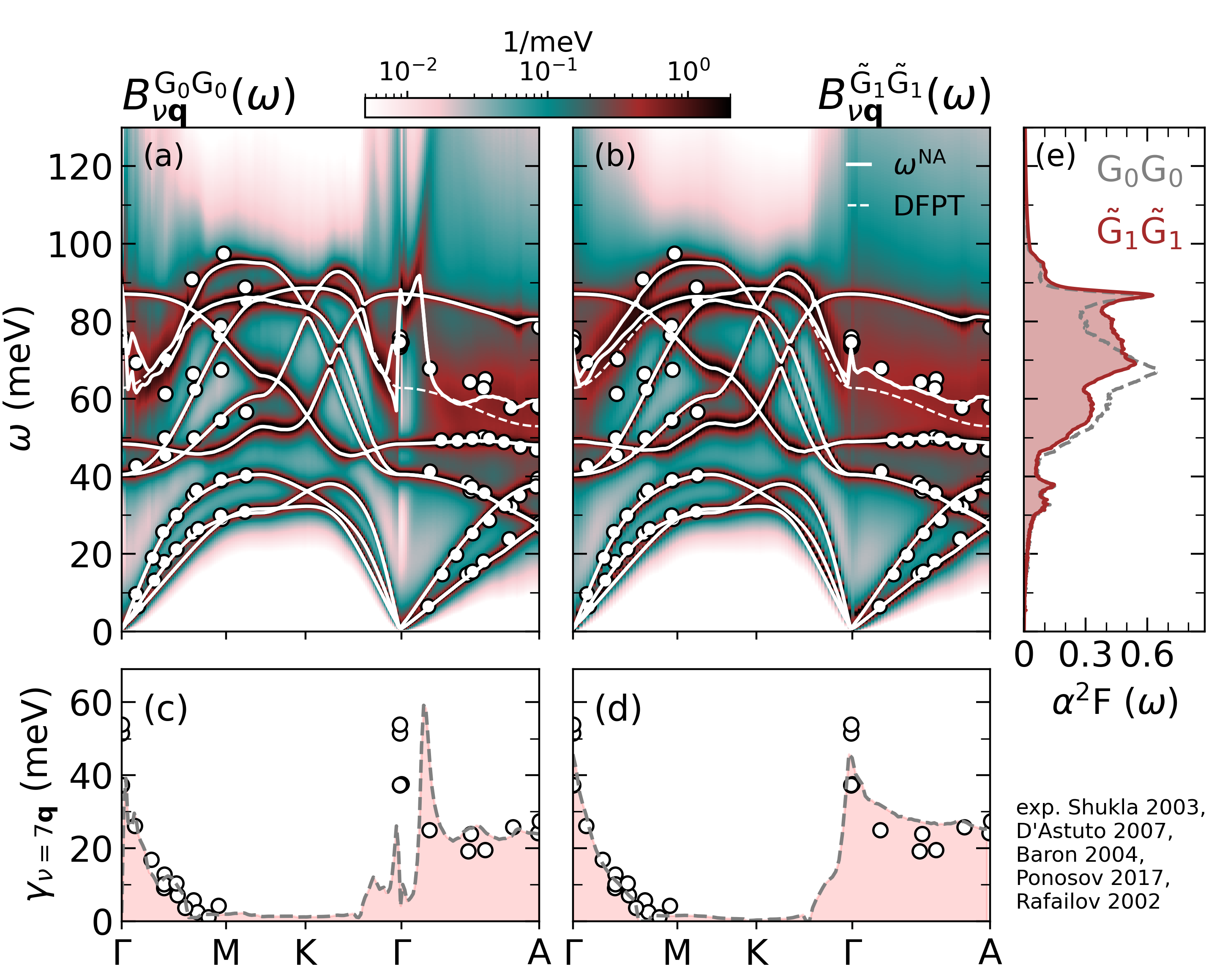}
    \caption{\label{fig:gg}  MgB$_2$ phonon spectral function compared with experiments from Refs.~\cite{shukla2003,astuto2007,baron2004,rafailov2002,ponosov17}. (a) First-order spectral function $B_{\nu\mathbf{q}}^{\text{G}_0\text{G}_0}(\omega)$ and (c) linewidth $\gamma_{\nu\mathbf{q}}^{\text{G}_0\text{G}_0}$ for the $E_{2g}$ branch, at 300~K, along high-symmetry lines. 
    (b, d) Same but for the higher-order spectral function $B_{\nu\mathbf{q}}^{\tilde{\text{G}}_1\tilde{\text{G}}_1}(\omega)$ and linewidth $\gamma_{\nu\mathbf{q}}^{\rm \tilde{\text{G}}_1\tilde{\text{G}}_1}$.
    The white line, denoted by $\omega^{\rm NA}$, follows the spectral functional maxima and the white dashed lines are the adiabatic DFPT result. 
    (e) Eliashberg spectral function obtained using the two approaches.}
\end{figure}
With this intuition, we investigate the temperature dependence of the phonon spectrum in MgB$_2$, see Fig.~\ref{fig:fig3}.
The strongly-coupled $\sigma$ band crosses the Fermi level and is broadened by the EPC.
It is responsible for the majority of phonon self-energy contributions.
In Fig.~\ref{fig:fig3}(b) with horizontal lines we show when the conditions for $\frac{\omega^{\text{ph}}}{2}=\gamma^{\text{i}}$ at $\text{T} = \text{T}^{*}$ and $\omega^{\text{ph}}=\gamma^{\text{i}}$ at $\text{T} = \text{T}^{\dagger}$ are achieved, assuming that the largest contribution to the phonon self-energy arises from the electrons at the Fermi surface and that $\gamma^{\text{i}}=\gamma^{\text{f}}$.
Temperature dependence of the Raman spectrum of the strongly-coupled optical E$_{2g}$ phonon branch is compared with experiments from Refs.~\cite{rafailov2002,astuto2007,ponosov17}.  
For the phonon frequency we observe a strong reduction of the large first-order NA correction until T$^{\dagger}=400$~K.
After 400~K the phonon frequency shows a saturation and even a weak hardening, that could be absent if anharmonic effects were included.
The E$_{2g}$ phonon linewidth follows a non-monotonic temperature trend predicted in Fig.~\ref{fig:doodle}(c) and reaches its peak value at $\text{T}^*=300$~K.
Experimentally, the peak occurs near 400~K, suggesting our calculations overestimate the electron linewidth.
Note that the G$_0$D$_0$ electron linewidth depends on the choice of smearing $\delta$, and the peak linewidth value could thereby be tuned. 
Nevertheless, both the frequency and linewidth follow the experimental temperature trends, which cannot be achieved with the temperature independent first-order NA self-energy corrections.
When the phonon enters the electron-hole damping continuum, the G$_0$G$_0$ approximation predicts a large NA splitting unaffected by temperature, visible as a two peak structure in the spectral function of the E$_{2g}$ mode in Fig.~\ref{fig:fig3}(e).
QP broadening in the $\tilde{\text{G}}_1\tilde{\text{G}}_1$ approach smears the boundaries of the damping continuum and reduces the NA splitting.
It is completely absent at 300~K, but is weakly recovered at 100~K, due to the reduced electron broadening.
NA phonon splitting could therefore be experimentally observed only in systems with a small electron linewidth or at low temperatures.
By studying the temperature dependence of the phonon frequency and linewidth at finite wavevectors in Fig.~\ref{fig:fig3}(f,g), we generalize the conclusions from the schematics in Fig.~\ref{fig:doodle} to $\mathbf{q}>0$.
The E$_{2g}$ notation is used for the zone-center $\nu=7$ phonon mode, while at finite wavevectors we refer to the corresponding $\nu=7$ phonon branch.
The peak linewidth value and the frequency hardening or saturation onset occurs for different temperatures depending on the wavevector. 
This means that the sum of electron linewidths of the initial and final electronic states becomes larger than the phonon frequency at different temperatures, depending on the wavevector $\mathbf{q}$ that connects them.
Since the interband contributions start occurring at energies larger than the phonon energy~\cite{dicastro2006}, most of the temperature dependencies we observe arise due to the intraband channel.
Furthermore, in MgB$_2$ the electron QP energy renormalization has a weak quantitative effect on the phonon spectrum, and our conclusions stem from the temperature dependent QP broadening arguments.

Vertex corrections addressed previously in MgB$_2$~\cite{cappelluti2002} might partially cancel the self-energy corrections ~\cite{grimaldi95b,pietronero96,itai92}.

From the $\tilde{\text{G}}_1\tilde{\text{G}}_1$ spectrum, we can examine the temperature-dependence of the EPC properties. 
The Eliashberg spectral function is computed as~\cite{allen1974}
\begin{equation} \label{eq:a2f}
    \alpha^2F(\omega) = \frac{1}{\pi n_0}\sum_{\nu\mathbf{q}}\frac{\gamma_{\nu\mathbf{q}}}{\omega_{\nu\mathbf{q}}}B_{\nu\mathbf{q}}(\omega),
\end{equation}
where $B_{\nu\mathbf{q}}(\omega)$ is given by Eq.~\eqref{eq:scB}, and  the phonon linewidths from the prefactor $\gamma_{\nu\mathbf{q}}$ are computed within the double delta approximation~\cite{allen1972}. $n_0$ is the density of states at the Fermi level.
Its integral gives the EPC constant $\lambda = 2\int_0^{\infty} \frac{\alpha^2F(\omega)}{\omega} \mathrm{d}\omega$.
The $\lambda^{\tilde{\text{G}}_1\tilde{\text{G}}_1}$ has a dome-like temperature dependence with the largest value achieved at 300~K (see Fig.~\ref{fig:fig3}(h)).
Temperature dependence is lacking in the G$_0$G$_0$ approximation. 
As shown in  Fig.~\ref{fig:fig3}(f), at 300~K the $\nu=7$ phonon along the $\Gamma-\text{A}$ direction, from where the dominant EPC contributions arise, has a minimal frequency for many wavevectors, producing the largest EPC for this particular temperature. 
EPC strength temperature dependence insights might be important for high-temperature superconductors~\cite{pickett80,appel68}.
For example, in superconducting hydrides, where the nonadiabatic effects are crucial and EPC is strong~\cite{mishra2026,Mishra2025}, we expect higher-order NA effects to renormalize the phonon spectrum in the superconducting temperature range.

In  Fig.~\ref{fig:gg}, we compare the spectral function of MgB$_2$ within the $\text{G}_0\text{G}_0$ and $\tilde{\text{G}}_1\tilde{\text{G}}_1$ approximations.
Strong NA effects occur in the adiabatically softened optical $\nu=7$ phonon branch when approaching the BZ center from the M, K, or A point. 
The flatness of the strongly-coupled $\sigma$ band along the $\Gamma-A$ direction gives the strongest renormalization effects along the $\Gamma-\text{A}$ direction~\cite{zhang2005}.
The failure of the G$_0$G$_0$ approximation is most apparent at the $\Gamma$ point, where it overestimates the phonon frequency by almost 20~meV and underestimates its broadening by 33~meV.
For finite wavevectors, a large NA renormalization is replaced by strong damping once the phonon is inside the electron-hole continuum~\cite{MAKSIMOV1996,calandra2007}.
For this reason, the strongly-coupled G$_0$G$_0$ phonon frequency and linewidth show nonanalytic wavevector dependence, see Fig.~\ref{fig:gg}(a,c), and NA phonon splitting~\cite{krsnik2024,garcia2020,berges2023,Engelsberg1963}.

In contrast, the $\tilde{\text{G}}_1\tilde{\text{G}}_1$ approximation provides a smoother spectrum and a better agreement with the experiments from Refs.~\cite{shukla2003,astuto2007,baron2004,rafailov2002,ponosov17}.
The overall broadening of the $\nu=7$ mode along the $\Gamma-\text{A}$ direction remains large, but the sharp variations with the wavevector are substantially reduced at 300~K.
The largest $\tilde{\text{G}}_1\tilde{\text{G}}_1$ phonon broadening contribution arises at the $\Gamma$ point where it surpasses the G$_0$G$_0$ approach by 33~meV.
The striking NA phonon splitting predicted in the G$_0$G$_0$ approximation is strongly suppressed in the $\tilde{\text{G}}_1\tilde{\text{G}}_1$ approach.
Unlike the G$_0$G$_0$, the $\tilde{\text{G}}_1\tilde{\text{G}}_1$ spectrum is robust against changes in the smearing parameter and is even improved if additional energy-independent electron broadening is included, see Sec.~S4 of the SI~\cite{supplement}.

Finally, the first-order NA Eliashberg spectral function in Fig.~\ref{fig:gg}(e) indicates that the strongly coupled phonons reside at 65~meV.
Including higher-order effects shifts the main peak to 70-80~meV, in better experimental agreement~\cite{mou2015,Osborn2001,mor2025,Adriano2025}, showing that the $\tilde{\text{G}}_1\tilde{\text{G}}_1$ approach provides spectral renormalizations across the whole Brillouin zone.

In conclusion, we study higher-order NA effects across the full phonon spectrum of MgB$_2$.
The temperature evolution of the phonon frequencies and linewidths follows the temperature scale set by the electron linewidth.
The phonon linewidth exhibits a peak at the temperature for which the phonon frequency becomes equal to the sum of the QP linewidths for the initial and final states of the relevant transitions. 
The phonon frequency, after initial temperature-induced softening, saturates in the adiabatic regime when the electronic linewidth near the chemical potential matches the phonon frequency. 
Higher-order NA effects could be crucial in interpreting the experimentally obtained temperature-dependent Raman spectra in many novel materials such as Weyl semimetals, ABC trilayer graphene, 2M-WS$_2$ and NbGe$_2$~\cite{osterhoudt2021,coulter2019,li2024,Yang2021,zan2024}. 
Furthermore, for systems undergoing soft-mode-driven phase transitions, their inclusion could also yield more accurate transition temperatures~\cite{Yoshiyama_1986}, and mitigate the overestimated $\lambda$, compared with a G$_0$G$_0$ approximation~\cite{erhardt2025b}.
The temperature dependence of $\lambda$ could be particularly relevant for high-temperature superconductors~\cite{pickett80,appel68}.
Especially promising are the superconducting hydrides, with strong EPC and NA effects~\cite{mishra2026,Mishra2025}.

\begin{acknowledgments}
\textit{Acknowledgments—} We thank Dr. Dino Novko for useful discussions.
N. G. E. is a Postdoctoral Researcher and S. P. is a Research Associate of the Fonds de la Recherche Scientifique-FNRS. 
This work was supported by the Fonds de la Recherche Scientifique-FNRS under Grants No. T.0183.23 (PDR) and No. T.W011.23 (PDR-WEAVE).
Computational resources have been provided by the Consortium des Équipements de Calcul Intensif (CÉCI), funded by the Fonds de la Recherche Scientifique de Belgique (F.R.S.-FNRS) under Grant No. 2.5020.11 and computational resources on Lucia, the Tier-1 supercomputer of the Walloon Region with infrastructure funded by the Walloon Region under the Grant Agreement No. 1910247.

\end{acknowledgments}
\textit{Data availability}—The modified \texttt{EPW} software and the input files used in this Letter are openly available~\cite{dataset_}. Additional computational details are provided in Sec.~S5 of the SI~\cite{supplement}.
\bibliography{ref}

\providecommand{\noopsort}[1]{}\providecommand{\singleletter}[1]{#1}%
\begin{thebibliography}{111}%
\makeatletter
\providecommand \@ifxundefined [1]{%
 \@ifx{#1\undefined}
}%
\providecommand \@ifnum [1]{%
 \ifnum #1\expandafter \@firstoftwo
 \else \expandafter \@secondoftwo
 \fi
}%
\providecommand \@ifx [1]{%
 \ifx #1\expandafter \@firstoftwo
 \else \expandafter \@secondoftwo
 \fi
}%
\providecommand \natexlab [1]{#1}%
\providecommand \enquote  [1]{``#1''}%
\providecommand \bibnamefont  [1]{#1}%
\providecommand \bibfnamefont [1]{#1}%
\providecommand \citenamefont [1]{#1}%
\providecommand \href@noop [0]{\@secondoftwo}%
\providecommand \href [0]{\begingroup \@sanitize@url \@href}%
\providecommand \@href[1]{\@@startlink{#1}\@@href}%
\providecommand \@@href[1]{\endgroup#1\@@endlink}%
\providecommand \@sanitize@url [0]{\catcode `\\12\catcode `\$12\catcode
  `\&12\catcode `\#12\catcode `\^12\catcode `\_12\catcode `\%12\relax}%
\providecommand \@@startlink[1]{}%
\providecommand \@@endlink[0]{}%
\providecommand \url  [0]{\begingroup\@sanitize@url \@url }%
\providecommand \@url [1]{\endgroup\@href {#1}{\urlprefix }}%
\providecommand \urlprefix  [0]{URL }%
\providecommand \Eprint [0]{\href }%
\providecommand \doibase [0]{https://doi.org/}%
\providecommand \selectlanguage [0]{\@gobble}%
\providecommand \bibinfo  [0]{\@secondoftwo}%
\providecommand \bibfield  [0]{\@secondoftwo}%
\providecommand \translation [1]{[#1]}%
\providecommand \BibitemOpen [0]{}%
\providecommand \bibitemStop [0]{}%
\providecommand \bibitemNoStop [0]{.\EOS\space}%
\providecommand \EOS [0]{\spacefactor3000\relax}%
\providecommand \BibitemShut  [1]{\csname bibitem#1\endcsname}%
\let\auto@bib@innerbib\@empty
\bibitem [{\citenamefont {Giustino}(2017)}]{giustino2017}%
  \BibitemOpen
  \bibfield  {author} {\bibinfo {author} {\bibfnamefont {F.}~\bibnamefont
  {Giustino}},\ }\bibfield  {title} {\bibinfo {title} {Electron-phonon
  interactions from first principles},\ }\href
  {https://doi.org/10.1103/RevModPhys.89.015003} {\bibfield  {journal}
  {\bibinfo  {journal} {Rev. Mod. Phys.}\ }\textbf {\bibinfo {volume} {89}},\
  \bibinfo {pages} {015003} (\bibinfo {year} {2017})}\BibitemShut {NoStop}%
\bibitem [{\citenamefont {Mahan}(2013)}]{mahan}%
  \BibitemOpen
  \bibfield  {author} {\bibinfo {author} {\bibfnamefont {G.~D.}\ \bibnamefont
  {Mahan}},\ }\bibinfo {title} {Physics of solids and liquids},\ in\ \href
  {https://doi.org/https://doi.org/10.1007/978-1-4757-5714-9} {\emph {\bibinfo
  {booktitle} {Many-Particle Physics}}}\ (\bibinfo  {publisher} {Springer New
  York, NY},\ \bibinfo {year} {2013})\ p.\ \bibinfo {pages} {785}\BibitemShut
  {NoStop}%
\bibitem [{\citenamefont {Giustino}\ \emph {et~al.}(2007)\citenamefont
  {Giustino}, \citenamefont {Cohen},\ and\ \citenamefont {Louie}}]{giustino07}%
  \BibitemOpen
  \bibfield  {author} {\bibinfo {author} {\bibfnamefont {F.}~\bibnamefont
  {Giustino}}, \bibinfo {author} {\bibfnamefont {M.~L.}\ \bibnamefont
  {Cohen}},\ and\ \bibinfo {author} {\bibfnamefont {S.~G.}\ \bibnamefont
  {Louie}},\ }\bibfield  {title} {\bibinfo {title} {Electron-phonon interaction
  using {W}annier functions},\ }\href
  {https://doi.org/10.1103/PhysRevB.76.165108} {\bibfield  {journal} {\bibinfo
  {journal} {Phys. Rev. B}\ }\textbf {\bibinfo {volume} {76}},\ \bibinfo
  {pages} {165108} (\bibinfo {year} {2007})}\BibitemShut {NoStop}%
\bibitem [{\citenamefont {Ponc\'e}\ \emph {et~al.}(2016)\citenamefont
  {Ponc\'e}, \citenamefont {Margine}, \citenamefont {Verdi},\ and\
  \citenamefont {Giustino}}]{Ponce2016}%
  \BibitemOpen
  \bibfield  {author} {\bibinfo {author} {\bibfnamefont {S.}~\bibnamefont
  {Ponc\'e}}, \bibinfo {author} {\bibfnamefont {E.}~\bibnamefont {Margine}},
  \bibinfo {author} {\bibfnamefont {C.}~\bibnamefont {Verdi}},\ and\ \bibinfo
  {author} {\bibfnamefont {F.}~\bibnamefont {Giustino}},\ }\bibfield  {title}
  {\bibinfo {title} {{EPW}: Electron--phonon coupling, transport and
  superconducting properties using maximally localized {Wannier} functions},\
  }\href {https://doi.org/10.1016/j.cpc.2016.07.028} {\bibfield  {journal}
  {\bibinfo  {journal} {Comput. Phys. Commun.}\ }\textbf {\bibinfo {volume}
  {209}},\ \bibinfo {pages} {116} (\bibinfo {year} {2016})}\BibitemShut
  {NoStop}%
\bibitem [{\citenamefont {Dee}\ \emph {et~al.}(2020)\citenamefont {Dee},
  \citenamefont {Coulter}, \citenamefont {Kleiner},\ and\ \citenamefont
  {Johnston}}]{Dee2020}%
  \BibitemOpen
  \bibfield  {author} {\bibinfo {author} {\bibfnamefont {P.~M.}\ \bibnamefont
  {Dee}}, \bibinfo {author} {\bibfnamefont {J.}~\bibnamefont {Coulter}},
  \bibinfo {author} {\bibfnamefont {K.~G.}\ \bibnamefont {Kleiner}},\ and\
  \bibinfo {author} {\bibfnamefont {S.}~\bibnamefont {Johnston}},\ }\bibfield
  {title} {\bibinfo {title} {Relative importance of nonlinear electron-phonon
  coupling and vertex corrections in the {H}olstein model},\ }\href
  {https://doi.org/10.1038/s42005-020-00413-2} {\bibfield  {journal} {\bibinfo
  {journal} {Communications Physics}\ }\textbf {\bibinfo {volume} {3}},\
  \bibinfo {pages} {145} (\bibinfo {year} {2020})}\BibitemShut {NoStop}%
\bibitem [{\citenamefont {Bomb\'{\i}n}\ \emph
  {et~al.}(2023{\natexlab{a}})\citenamefont {Bomb\'{\i}n}, \citenamefont
  {Muzas}, \citenamefont {Novko}, \citenamefont {Juaristi},\ and\ \citenamefont
  {Alducin}}]{bombin2023}%
  \BibitemOpen
  \bibfield  {author} {\bibinfo {author} {\bibfnamefont {R.}~\bibnamefont
  {Bomb\'{\i}n}}, \bibinfo {author} {\bibfnamefont {A.~S.}\ \bibnamefont
  {Muzas}}, \bibinfo {author} {\bibfnamefont {D.}~\bibnamefont {Novko}},
  \bibinfo {author} {\bibfnamefont {J.~I.~n.}\ \bibnamefont {Juaristi}},\ and\
  \bibinfo {author} {\bibfnamefont {M.}~\bibnamefont {Alducin}},\ }\bibfield
  {title} {\bibinfo {title} {Anomalous transient blueshift in the internal
  stretch mode of {CO/Pd}(111)},\ }\href
  {https://doi.org/10.1103/PhysRevB.107.L121404} {\bibfield  {journal}
  {\bibinfo  {journal} {Phys. Rev. B}\ }\textbf {\bibinfo {volume} {107}},\
  \bibinfo {pages} {L121404} (\bibinfo {year}
  {2023}{\natexlab{a}})}\BibitemShut {NoStop}%
\bibitem [{\citenamefont {Bomb\'{\i}n}\ \emph
  {et~al.}(2023{\natexlab{b}})\citenamefont {Bomb\'{\i}n}, \citenamefont
  {Muzas}, \citenamefont {Novko}, \citenamefont {Juaristi},\ and\ \citenamefont
  {Alducin}}]{bombin2023b}%
  \BibitemOpen
  \bibfield  {author} {\bibinfo {author} {\bibfnamefont {R.}~\bibnamefont
  {Bomb\'{\i}n}}, \bibinfo {author} {\bibfnamefont {A.~S.}\ \bibnamefont
  {Muzas}}, \bibinfo {author} {\bibfnamefont {D.}~\bibnamefont {Novko}},
  \bibinfo {author} {\bibfnamefont {J.~I.~n.}\ \bibnamefont {Juaristi}},\ and\
  \bibinfo {author} {\bibfnamefont {M.}~\bibnamefont {Alducin}},\ }\bibfield
  {title} {\bibinfo {title} {Vibrational dynamics of {CO} on {Pd}(111) in and
  out of thermal equilibrium},\ }\href
  {https://doi.org/10.1103/PhysRevB.108.045409} {\bibfield  {journal} {\bibinfo
   {journal} {Phys. Rev. B}\ }\textbf {\bibinfo {volume} {108}},\ \bibinfo
  {pages} {045409} (\bibinfo {year} {2023}{\natexlab{b}})}\BibitemShut
  {NoStop}%
\bibitem [{\citenamefont {Novko}\ \emph {et~al.}(2019)\citenamefont {Novko},
  \citenamefont {Tremblay}, \citenamefont {Alducin},\ and\ \citenamefont
  {Juaristi}}]{novko2019}%
  \BibitemOpen
  \bibfield  {author} {\bibinfo {author} {\bibfnamefont {D.}~\bibnamefont
  {Novko}}, \bibinfo {author} {\bibfnamefont {J.~C.}\ \bibnamefont {Tremblay}},
  \bibinfo {author} {\bibfnamefont {M.}~\bibnamefont {Alducin}},\ and\ \bibinfo
  {author} {\bibfnamefont {J.~I.}\ \bibnamefont {Juaristi}},\ }\bibfield
  {title} {\bibinfo {title} {Ultrafast transient dynamics of adsorbates on
  surfaces deciphered: The case of {CO} on {Cu}(100)},\ }\href
  {https://doi.org/10.1103/PhysRevLett.122.016806} {\bibfield  {journal}
  {\bibinfo  {journal} {Phys. Rev. Lett.}\ }\textbf {\bibinfo {volume} {122}},\
  \bibinfo {pages} {016806} (\bibinfo {year} {2019})}\BibitemShut {NoStop}%
\bibitem [{\citenamefont {Park}(2025)}]{park2025}%
  \BibitemOpen
  \bibfield  {author} {\bibinfo {author} {\bibfnamefont {C.-H.}\ \bibnamefont
  {Park}},\ }\bibfield  {title} {\bibinfo {title} {Nonadiabatic phonon
  self-energy due to electrons with finite linewidths},\ }\href
  {https://doi.org/10.1103/6jq1-cbwq} {\bibfield  {journal} {\bibinfo
  {journal} {Phys. Rev. B}\ }\textbf {\bibinfo {volume} {112}},\ \bibinfo
  {pages} {104314} (\bibinfo {year} {2025})}\BibitemShut {NoStop}%
\bibitem [{\citenamefont {Erhardt}\ \emph {et~al.}(2025)\citenamefont
  {Erhardt}, \citenamefont {Castellano}, \citenamefont {Batista}, \citenamefont
  {Bianco}, \citenamefont {Lončarić}, \citenamefont {Verstraete},\ and\
  \citenamefont {Novko}}]{Erhardt2025}%
  \BibitemOpen
  \bibfield  {author} {\bibinfo {author} {\bibfnamefont {N.~G.}\ \bibnamefont
  {Erhardt}}, \bibinfo {author} {\bibfnamefont {A.}~\bibnamefont {Castellano}},
  \bibinfo {author} {\bibfnamefont {J.~P.~A.}\ \bibnamefont {Batista}},
  \bibinfo {author} {\bibfnamefont {R.}~\bibnamefont {Bianco}}, \bibinfo
  {author} {\bibfnamefont {I.}~\bibnamefont {Lončarić}}, \bibinfo {author}
  {\bibfnamefont {M.~J.}\ \bibnamefont {Verstraete}},\ and\ \bibinfo {author}
  {\bibfnamefont {D.}~\bibnamefont {Novko}},\ }\bibfield  {title} {\bibinfo
  {title} {Electron-mediated anharmonicity and its role in the {R}aman spectrum
  of graphene},\ }\href {https://doi.org/10.1038/s41524-025-01610-9} {\bibfield
   {journal} {\bibinfo  {journal} {npj Computational Materials}\ }\textbf
  {\bibinfo {volume} {11}},\ \bibinfo {pages} {114} (\bibinfo {year}
  {2025})}\BibitemShut {NoStop}%
\bibitem [{\citenamefont {Lihm}\ and\ \citenamefont
  {Ponc\'e}(2026)}]{lihm2026}%
  \BibitemOpen
  \bibfield  {author} {\bibinfo {author} {\bibfnamefont {J.-M.}\ \bibnamefont
  {Lihm}}\ and\ \bibinfo {author} {\bibfnamefont {S.}~\bibnamefont {Ponc\'e}},\
  }\bibfield  {title} {\bibinfo {title} {Beyond-quasiparticle transport with
  vertex correction: Self-consistent ladder formalism for electron-phonon
  interactions},\ }\href {https://doi.org/10.1103/mtjf-ylf7} {\bibfield
  {journal} {\bibinfo  {journal} {Phys. Rev. X}\ }\textbf {\bibinfo {volume}
  {16}},\ \bibinfo {pages} {011008} (\bibinfo {year} {2026})}\BibitemShut
  {NoStop}%
\bibitem [{\citenamefont {Xia}(2026)}]{xia2026}%
  \BibitemOpen
  \bibfield  {author} {\bibinfo {author} {\bibfnamefont {Y.}~\bibnamefont
  {Xia}},\ }\href {https://arxiv.org/abs/2607.10211} {\bibinfo {title}
  {Self-consistent phonon spectral functions and thermal transport beyond the
  quasiparticle approximation}} (\bibinfo {year} {2026}),\ \Eprint
  {https://arxiv.org/abs/2607.10211} {arXiv:2607.10211} \BibitemShut {NoStop}%
\bibitem [{\citenamefont {Gonze}\ and\ \citenamefont {Lee}(1997)}]{Gonze1997}%
  \BibitemOpen
  \bibfield  {author} {\bibinfo {author} {\bibfnamefont {X.}~\bibnamefont
  {Gonze}}\ and\ \bibinfo {author} {\bibfnamefont {C.}~\bibnamefont {Lee}},\
  }\bibfield  {title} {\bibinfo {title} {Dynamical matrices, born effective
  charges, dielectric permittivity tensors, and interatomic force constants
  from density-functional perturbation theory},\ }\href
  {https://doi.org/10.1103/PhysRevB.55.10355} {\bibfield  {journal} {\bibinfo
  {journal} {Phys. Rev. B}\ }\textbf {\bibinfo {volume} {55}},\ \bibinfo
  {pages} {10355} (\bibinfo {year} {1997})}\BibitemShut {NoStop}%
\bibitem [{\citenamefont {Baroni}\ \emph {et~al.}(2001)\citenamefont {Baroni},
  \citenamefont {de~Gironcoli}, \citenamefont {Dal~Corso},\ and\ \citenamefont
  {Giannozzi}}]{Baroni2001}%
  \BibitemOpen
  \bibfield  {author} {\bibinfo {author} {\bibfnamefont {S.}~\bibnamefont
  {Baroni}}, \bibinfo {author} {\bibfnamefont {S.}~\bibnamefont
  {de~Gironcoli}}, \bibinfo {author} {\bibfnamefont {A.}~\bibnamefont
  {Dal~Corso}},\ and\ \bibinfo {author} {\bibfnamefont {P.}~\bibnamefont
  {Giannozzi}},\ }\bibfield  {title} {\bibinfo {title} {Phonons and related
  crystal properties from density-functional perturbation theory},\ }\href
  {https://doi.org/10.1103/RevModPhys.73.515} {\bibfield  {journal} {\bibinfo
  {journal} {Rev. Mod. Phys.}\ }\textbf {\bibinfo {volume} {73}},\ \bibinfo
  {pages} {515} (\bibinfo {year} {2001})}\BibitemShut {NoStop}%
\bibitem [{\citenamefont {Engelsberg}\ and\ \citenamefont
  {Schrieffer}(1963)}]{Engelsberg1963}%
  \BibitemOpen
  \bibfield  {author} {\bibinfo {author} {\bibfnamefont {S.}~\bibnamefont
  {Engelsberg}}\ and\ \bibinfo {author} {\bibfnamefont {J.~R.}\ \bibnamefont
  {Schrieffer}},\ }\bibfield  {title} {\bibinfo {title} {Coupled
  electron-phonon system},\ }\href {https://doi.org/10.1103/PhysRev.131.993}
  {\bibfield  {journal} {\bibinfo  {journal} {Phys. Rev.}\ }\textbf {\bibinfo
  {volume} {131}},\ \bibinfo {pages} {993} (\bibinfo {year}
  {1963})}\BibitemShut {NoStop}%
\bibitem [{\citenamefont {Maksimov}\ and\ \citenamefont
  {Shulga}(1996)}]{MAKSIMOV1996}%
  \BibitemOpen
  \bibfield  {author} {\bibinfo {author} {\bibfnamefont {E.}~\bibnamefont
  {Maksimov}}\ and\ \bibinfo {author} {\bibfnamefont {S.}~\bibnamefont
  {Shulga}},\ }\bibfield  {title} {\bibinfo {title} {Nonadiabatic effects in
  optical phonon self-energy},\ }\href
  {https://doi.org/https://doi.org/10.1016/0038-1098(95)00745-8} {\bibfield
  {journal} {\bibinfo  {journal} {Solid State Communications}\ }\textbf
  {\bibinfo {volume} {97}},\ \bibinfo {pages} {553} (\bibinfo {year}
  {1996})}\BibitemShut {NoStop}%
\bibitem [{\citenamefont {Cerdeira}\ and\ \citenamefont
  {Cardona}(1972)}]{cerdeira72}%
  \BibitemOpen
  \bibfield  {author} {\bibinfo {author} {\bibfnamefont {F.}~\bibnamefont
  {Cerdeira}}\ and\ \bibinfo {author} {\bibfnamefont {M.}~\bibnamefont
  {Cardona}},\ }\bibfield  {title} {\bibinfo {title} {Effect of carrier
  concentration on the {R}aman frequencies of {Si} and {Ge}},\ }\href
  {https://doi.org/10.1103/PhysRevB.5.1440} {\bibfield  {journal} {\bibinfo
  {journal} {Phys. Rev. B}\ }\textbf {\bibinfo {volume} {5}},\ \bibinfo {pages}
  {1440} (\bibinfo {year} {1972})}\BibitemShut {NoStop}%
\bibitem [{\citenamefont {Ponosov}\ \emph {et~al.}(1998)\citenamefont
  {Ponosov}, \citenamefont {Bolotin}, \citenamefont {Thomsen},\ and\
  \citenamefont {Cardona}}]{ponosov98}%
  \BibitemOpen
  \bibfield  {author} {\bibinfo {author} {\bibfnamefont {Y.~S.}\ \bibnamefont
  {Ponosov}}, \bibinfo {author} {\bibfnamefont {G.~A.}\ \bibnamefont
  {Bolotin}}, \bibinfo {author} {\bibfnamefont {C.}~\bibnamefont {Thomsen}},\
  and\ \bibinfo {author} {\bibfnamefont {M.}~\bibnamefont {Cardona}},\
  }\bibfield  {title} {\bibinfo {title} {Raman scattering in {Os}: Nonadiabatic
  renormalization of the optical phonon self-energies},\ }\href
  {https://doi.org/10.1002/(SICI)1521-3951(199807)208:1<257::AID-PSSB257>3.0.CO;2-F}
  {\bibfield  {journal} {\bibinfo  {journal} {physica status solidi (b)}\
  }\textbf {\bibinfo {volume} {208}},\ \bibinfo {pages} {257} (\bibinfo {year}
  {1998})}\BibitemShut {NoStop}%
\bibitem [{\citenamefont {Ponosov}\ and\ \citenamefont
  {Streltsov}(2016)}]{ponosov16}%
  \BibitemOpen
  \bibfield  {author} {\bibinfo {author} {\bibfnamefont {Y.~S.}\ \bibnamefont
  {Ponosov}}\ and\ \bibinfo {author} {\bibfnamefont {S.~V.}\ \bibnamefont
  {Streltsov}},\ }\bibfield  {title} {\bibinfo {title} {Raman evidence for
  nonadiabatic effects in optical phonon self-energies of transition metals},\
  }\href {https://doi.org/10.1103/PhysRevB.94.214302} {\bibfield  {journal}
  {\bibinfo  {journal} {Phys. Rev. B}\ }\textbf {\bibinfo {volume} {94}},\
  \bibinfo {pages} {214302} (\bibinfo {year} {2016})}\BibitemShut {NoStop}%
\bibitem [{\citenamefont {Cappelluti}(2006)}]{cappelluti2006}%
  \BibitemOpen
  \bibfield  {author} {\bibinfo {author} {\bibfnamefont {E.}~\bibnamefont
  {Cappelluti}},\ }\bibfield  {title} {\bibinfo {title} {Electron-phonon
  effects on the {R}aman spectrum in {$\mathrm{Mg}{\mathrm{B}}_{2}$}},\ }\href
  {https://doi.org/10.1103/PhysRevB.73.140505} {\bibfield  {journal} {\bibinfo
  {journal} {Phys. Rev. B}\ }\textbf {\bibinfo {volume} {73}},\ \bibinfo
  {pages} {140505} (\bibinfo {year} {2006})}\BibitemShut {NoStop}%
\bibitem [{\citenamefont {Ponosov}\ and\ \citenamefont
  {Streltsov}(2017)}]{ponosov17}%
  \BibitemOpen
  \bibfield  {author} {\bibinfo {author} {\bibfnamefont {Y.~S.}\ \bibnamefont
  {Ponosov}}\ and\ \bibinfo {author} {\bibfnamefont {S.~V.}\ \bibnamefont
  {Streltsov}},\ }\bibfield  {title} {\bibinfo {title} {Raman-active ${E}_{2g}$
  phonon in {${\mathrm{MgB}}_{2}$}: Electron-phonon interaction and
  anharmonicity},\ }\href {https://doi.org/10.1103/PhysRevB.96.214503}
  {\bibfield  {journal} {\bibinfo  {journal} {Phys. Rev. B}\ }\textbf {\bibinfo
  {volume} {96}},\ \bibinfo {pages} {214503} (\bibinfo {year}
  {2017})}\BibitemShut {NoStop}%
\bibitem [{\citenamefont {Novko}(2018)}]{novko2018}%
  \BibitemOpen
  \bibfield  {author} {\bibinfo {author} {\bibfnamefont {D.}~\bibnamefont
  {Novko}},\ }\bibfield  {title} {\bibinfo {title} {Nonadiabatic coupling
  effects in {${\mathrm{MgB}}_{2}$} reexamined},\ }\href
  {https://doi.org/10.1103/PhysRevB.98.041112} {\bibfield  {journal} {\bibinfo
  {journal} {Phys. Rev. B}\ }\textbf {\bibinfo {volume} {98}},\ \bibinfo
  {pages} {041112} (\bibinfo {year} {2018})}\BibitemShut {NoStop}%
\bibitem [{\citenamefont {Novko}\ \emph {et~al.}(2020)\citenamefont {Novko},
  \citenamefont {Caruso}, \citenamefont {Draxl},\ and\ \citenamefont
  {Cappelluti}}]{novko20b}%
  \BibitemOpen
  \bibfield  {author} {\bibinfo {author} {\bibfnamefont {D.}~\bibnamefont
  {Novko}}, \bibinfo {author} {\bibfnamefont {F.}~\bibnamefont {Caruso}},
  \bibinfo {author} {\bibfnamefont {C.}~\bibnamefont {Draxl}},\ and\ \bibinfo
  {author} {\bibfnamefont {E.}~\bibnamefont {Cappelluti}},\ }\bibfield  {title}
  {\bibinfo {title} {Ultrafast hot phonon dynamics in {${\mathrm{MgB}}_{2}$}
  driven by anisotropic electron-phonon coupling},\ }\href
  {https://doi.org/10.1103/PhysRevLett.124.077001} {\bibfield  {journal}
  {\bibinfo  {journal} {Phys. Rev. Lett.}\ }\textbf {\bibinfo {volume} {124}},\
  \bibinfo {pages} {077001} (\bibinfo {year} {2020})}\BibitemShut {NoStop}%
\bibitem [{\citenamefont {Lazzeri}\ and\ \citenamefont
  {Mauri}(2006)}]{lazzeri06}%
  \BibitemOpen
  \bibfield  {author} {\bibinfo {author} {\bibfnamefont {M.}~\bibnamefont
  {Lazzeri}}\ and\ \bibinfo {author} {\bibfnamefont {F.}~\bibnamefont
  {Mauri}},\ }\bibfield  {title} {\bibinfo {title} {Nonadiabatic {K}ohn anomaly
  in a doped graphene monolayer},\ }\href
  {https://doi.org/10.1103/PhysRevLett.97.266407} {\bibfield  {journal}
  {\bibinfo  {journal} {Phys. Rev. Lett.}\ }\textbf {\bibinfo {volume} {97}},\
  \bibinfo {pages} {266407} (\bibinfo {year} {2006})}\BibitemShut {NoStop}%
\bibitem [{\citenamefont {Pisana}\ \emph {et~al.}(2007)\citenamefont {Pisana},
  \citenamefont {Lazzeri}, \citenamefont {Casiraghi}, \citenamefont
  {Novoselov}, \citenamefont {Geim}, \citenamefont {Ferrari},\ and\
  \citenamefont {Mauri}}]{pisana07}%
  \BibitemOpen
  \bibfield  {author} {\bibinfo {author} {\bibfnamefont {S.}~\bibnamefont
  {Pisana}}, \bibinfo {author} {\bibfnamefont {M.}~\bibnamefont {Lazzeri}},
  \bibinfo {author} {\bibfnamefont {C.}~\bibnamefont {Casiraghi}}, \bibinfo
  {author} {\bibfnamefont {K.~S.}\ \bibnamefont {Novoselov}}, \bibinfo {author}
  {\bibfnamefont {A.~K.}\ \bibnamefont {Geim}}, \bibinfo {author}
  {\bibfnamefont {A.~C.}\ \bibnamefont {Ferrari}},\ and\ \bibinfo {author}
  {\bibfnamefont {F.}~\bibnamefont {Mauri}},\ }\bibfield  {title} {\bibinfo
  {title} {Breakdown of the adiabatic {B}orn--{O}ppenheimer approximation in
  graphene},\ }\href {https://doi.org/10.1038/nmat1846} {\bibfield  {journal}
  {\bibinfo  {journal} {Nature Materials}\ }\textbf {\bibinfo {volume} {6}},\
  \bibinfo {pages} {198} (\bibinfo {year} {2007})}\BibitemShut {NoStop}%
\bibitem [{\citenamefont {Piscanec}\ \emph {et~al.}(2007)\citenamefont
  {Piscanec}, \citenamefont {Lazzeri}, \citenamefont {Robertson}, \citenamefont
  {Ferrari},\ and\ \citenamefont {Mauri}}]{piscanec07}%
  \BibitemOpen
  \bibfield  {author} {\bibinfo {author} {\bibfnamefont {S.}~\bibnamefont
  {Piscanec}}, \bibinfo {author} {\bibfnamefont {M.}~\bibnamefont {Lazzeri}},
  \bibinfo {author} {\bibfnamefont {J.}~\bibnamefont {Robertson}}, \bibinfo
  {author} {\bibfnamefont {A.~C.}\ \bibnamefont {Ferrari}},\ and\ \bibinfo
  {author} {\bibfnamefont {F.}~\bibnamefont {Mauri}},\ }\bibfield  {title}
  {\bibinfo {title} {Optical phonons in carbon nanotubes: Kohn anomalies,
  peierls distortions, and dynamic effects},\ }\href
  {https://doi.org/10.1103/PhysRevB.75.035427} {\bibfield  {journal} {\bibinfo
  {journal} {Phys. Rev. B}\ }\textbf {\bibinfo {volume} {75}},\ \bibinfo
  {pages} {035427} (\bibinfo {year} {2007})}\BibitemShut {NoStop}%
\bibitem [{\citenamefont {Saitta}\ \emph
  {et~al.}(2008{\natexlab{a}})\citenamefont {Saitta}, \citenamefont {Lazzeri},
  \citenamefont {Calandra},\ and\ \citenamefont {Mauri}}]{saitta08}%
  \BibitemOpen
  \bibfield  {author} {\bibinfo {author} {\bibfnamefont {A.~M.}\ \bibnamefont
  {Saitta}}, \bibinfo {author} {\bibfnamefont {M.}~\bibnamefont {Lazzeri}},
  \bibinfo {author} {\bibfnamefont {M.}~\bibnamefont {Calandra}},\ and\
  \bibinfo {author} {\bibfnamefont {F.}~\bibnamefont {Mauri}},\ }\bibfield
  {title} {\bibinfo {title} {Giant nonadiabatic effects in layer metals:
  {R}aman spectra of intercalated graphite explained},\ }\href
  {https://doi.org/10.1103/PhysRevLett.100.226401} {\bibfield  {journal}
  {\bibinfo  {journal} {Phys. Rev. Lett.}\ }\textbf {\bibinfo {volume} {100}},\
  \bibinfo {pages} {226401} (\bibinfo {year} {2008}{\natexlab{a}})}\BibitemShut
  {NoStop}%
\bibitem [{\citenamefont {Malard}\ \emph {et~al.}(2008)\citenamefont {Malard},
  \citenamefont {Elias}, \citenamefont {Alves},\ and\ \citenamefont
  {Pimenta}}]{malard2008}%
  \BibitemOpen
  \bibfield  {author} {\bibinfo {author} {\bibfnamefont {L.~M.}\ \bibnamefont
  {Malard}}, \bibinfo {author} {\bibfnamefont {D.~C.}\ \bibnamefont {Elias}},
  \bibinfo {author} {\bibfnamefont {E.~S.}\ \bibnamefont {Alves}},\ and\
  \bibinfo {author} {\bibfnamefont {M.~A.}\ \bibnamefont {Pimenta}},\
  }\bibfield  {title} {\bibinfo {title} {Observation of distinct
  electron-phonon couplings in gated bilayer graphene},\ }\href
  {https://doi.org/10.1103/PhysRevLett.101.257401} {\bibfield  {journal}
  {\bibinfo  {journal} {Phys. Rev. Lett.}\ }\textbf {\bibinfo {volume} {101}},\
  \bibinfo {pages} {257401} (\bibinfo {year} {2008})}\BibitemShut {NoStop}%
\bibitem [{\citenamefont {Yan}\ \emph {et~al.}(2008)\citenamefont {Yan},
  \citenamefont {Henriksen}, \citenamefont {Kim},\ and\ \citenamefont
  {Pinczuk}}]{yan2008}%
  \BibitemOpen
  \bibfield  {author} {\bibinfo {author} {\bibfnamefont {J.}~\bibnamefont
  {Yan}}, \bibinfo {author} {\bibfnamefont {E.~A.}\ \bibnamefont {Henriksen}},
  \bibinfo {author} {\bibfnamefont {P.}~\bibnamefont {Kim}},\ and\ \bibinfo
  {author} {\bibfnamefont {A.}~\bibnamefont {Pinczuk}},\ }\bibfield  {title}
  {\bibinfo {title} {Observation of anomalous phonon softening in bilayer
  graphene},\ }\href {https://doi.org/10.1103/PhysRevLett.101.136804}
  {\bibfield  {journal} {\bibinfo  {journal} {Phys. Rev. Lett.}\ }\textbf
  {\bibinfo {volume} {101}},\ \bibinfo {pages} {136804} (\bibinfo {year}
  {2008})}\BibitemShut {NoStop}%
\bibitem [{\citenamefont {Das}\ \emph {et~al.}(2009)\citenamefont {Das},
  \citenamefont {Chakraborty}, \citenamefont {Piscanec}, \citenamefont
  {Pisana}, \citenamefont {Sood},\ and\ \citenamefont {Ferrari}}]{das2009}%
  \BibitemOpen
  \bibfield  {author} {\bibinfo {author} {\bibfnamefont {A.}~\bibnamefont
  {Das}}, \bibinfo {author} {\bibfnamefont {B.}~\bibnamefont {Chakraborty}},
  \bibinfo {author} {\bibfnamefont {S.}~\bibnamefont {Piscanec}}, \bibinfo
  {author} {\bibfnamefont {S.}~\bibnamefont {Pisana}}, \bibinfo {author}
  {\bibfnamefont {A.~K.}\ \bibnamefont {Sood}},\ and\ \bibinfo {author}
  {\bibfnamefont {A.~C.}\ \bibnamefont {Ferrari}},\ }\bibfield  {title}
  {\bibinfo {title} {Phonon renormalization in doped bilayer graphene},\ }\href
  {https://doi.org/10.1103/PhysRevB.79.155417} {\bibfield  {journal} {\bibinfo
  {journal} {Phys. Rev. B}\ }\textbf {\bibinfo {volume} {79}},\ \bibinfo
  {pages} {155417} (\bibinfo {year} {2009})}\BibitemShut {NoStop}%
\bibitem [{\citenamefont {Calandra}\ \emph {et~al.}(2010)\citenamefont
  {Calandra}, \citenamefont {Profeta},\ and\ \citenamefont
  {Mauri}}]{calandra10}%
  \BibitemOpen
  \bibfield  {author} {\bibinfo {author} {\bibfnamefont {M.}~\bibnamefont
  {Calandra}}, \bibinfo {author} {\bibfnamefont {G.}~\bibnamefont {Profeta}},\
  and\ \bibinfo {author} {\bibfnamefont {F.}~\bibnamefont {Mauri}},\ }\bibfield
   {title} {\bibinfo {title} {Adiabatic and nonadiabatic phonon dispersion in a
  {W}annier function approach},\ }\href
  {https://doi.org/10.1103/PhysRevB.82.165111} {\bibfield  {journal} {\bibinfo
  {journal} {Phys. Rev. B}\ }\textbf {\bibinfo {volume} {82}},\ \bibinfo
  {pages} {165111} (\bibinfo {year} {2010})}\BibitemShut {NoStop}%
\bibitem [{\citenamefont {Caruso}\ \emph {et~al.}(2017)\citenamefont {Caruso},
  \citenamefont {Hoesch}, \citenamefont {Achatz}, \citenamefont {Serrano},
  \citenamefont {Krisch}, \citenamefont {Bustarret},\ and\ \citenamefont
  {Giustino}}]{caruso17}%
  \BibitemOpen
  \bibfield  {author} {\bibinfo {author} {\bibfnamefont {F.}~\bibnamefont
  {Caruso}}, \bibinfo {author} {\bibfnamefont {M.}~\bibnamefont {Hoesch}},
  \bibinfo {author} {\bibfnamefont {P.}~\bibnamefont {Achatz}}, \bibinfo
  {author} {\bibfnamefont {J.}~\bibnamefont {Serrano}}, \bibinfo {author}
  {\bibfnamefont {M.}~\bibnamefont {Krisch}}, \bibinfo {author} {\bibfnamefont
  {E.}~\bibnamefont {Bustarret}},\ and\ \bibinfo {author} {\bibfnamefont
  {F.}~\bibnamefont {Giustino}},\ }\bibfield  {title} {\bibinfo {title}
  {Nonadiabatic {K}ohn anomaly in heavily boron-doped diamond},\ }\href
  {https://doi.org/10.1103/PhysRevLett.119.017001} {\bibfield  {journal}
  {\bibinfo  {journal} {Phys. Rev. Lett.}\ }\textbf {\bibinfo {volume} {119}},\
  \bibinfo {pages} {017001} (\bibinfo {year} {2017})}\BibitemShut {NoStop}%
\bibitem [{\citenamefont {Sohier}\ \emph {et~al.}(2019)\citenamefont {Sohier},
  \citenamefont {Ponomarev}, \citenamefont {Gibertini}, \citenamefont {Berger},
  \citenamefont {Marzari}, \citenamefont {Ubrig},\ and\ \citenamefont
  {Morpurgo}}]{sohier19}%
  \BibitemOpen
  \bibfield  {author} {\bibinfo {author} {\bibfnamefont {T.}~\bibnamefont
  {Sohier}}, \bibinfo {author} {\bibfnamefont {E.}~\bibnamefont {Ponomarev}},
  \bibinfo {author} {\bibfnamefont {M.}~\bibnamefont {Gibertini}}, \bibinfo
  {author} {\bibfnamefont {H.}~\bibnamefont {Berger}}, \bibinfo {author}
  {\bibfnamefont {N.}~\bibnamefont {Marzari}}, \bibinfo {author} {\bibfnamefont
  {N.}~\bibnamefont {Ubrig}},\ and\ \bibinfo {author} {\bibfnamefont {A.~F.}\
  \bibnamefont {Morpurgo}},\ }\bibfield  {title} {\bibinfo {title} {Enhanced
  electron-phonon interaction in multivalley materials},\ }\href
  {https://doi.org/10.1103/PhysRevX.9.031019} {\bibfield  {journal} {\bibinfo
  {journal} {Phys. Rev. X}\ }\textbf {\bibinfo {volume} {9}},\ \bibinfo {pages}
  {031019} (\bibinfo {year} {2019})}\BibitemShut {NoStop}%
\bibitem [{\citenamefont {Novko}(2020)}]{novko2020a}%
  \BibitemOpen
  \bibfield  {author} {\bibinfo {author} {\bibfnamefont {D.}~\bibnamefont
  {Novko}},\ }\bibfield  {title} {\bibinfo {title} {Broken adiabaticity induced
  by {L}ifshitz transition in {MoS$_2$} and {WS$_2$} single layers},\ }\href
  {https://doi.org/10.1038/s42005-020-0299-1} {\bibfield  {journal} {\bibinfo
  {journal} {Communications Physics}\ }\textbf {\bibinfo {volume} {3}},\
  \bibinfo {pages} {1} (\bibinfo {year} {2020})}\BibitemShut {NoStop}%
\bibitem [{\citenamefont {Garcia-Goiricelaya}\ \emph
  {et~al.}(2020)\citenamefont {Garcia-Goiricelaya}, \citenamefont
  {Lafuente-Bartolome}, \citenamefont {Gurtubay},\ and\ \citenamefont
  {Eiguren}}]{garcia2020}%
  \BibitemOpen
  \bibfield  {author} {\bibinfo {author} {\bibfnamefont {P.}~\bibnamefont
  {Garcia-Goiricelaya}}, \bibinfo {author} {\bibfnamefont {J.}~\bibnamefont
  {Lafuente-Bartolome}}, \bibinfo {author} {\bibfnamefont {I.~G.}\ \bibnamefont
  {Gurtubay}},\ and\ \bibinfo {author} {\bibfnamefont {A.}~\bibnamefont
  {Eiguren}},\ }\bibfield  {title} {\bibinfo {title} {Emergence of large
  nonadiabatic effects induced by the electron-phonon interaction on the
  complex vibrational quasiparticle spectrum of doped monolayer
  ${\mathrm{mos}}_{2}$},\ }\href {https://doi.org/10.1103/PhysRevB.101.054304}
  {\bibfield  {journal} {\bibinfo  {journal} {Phys. Rev. B}\ }\textbf {\bibinfo
  {volume} {101}},\ \bibinfo {pages} {054304} (\bibinfo {year}
  {2020})}\BibitemShut {NoStop}%
\bibitem [{\citenamefont {Girotto}\ and\ \citenamefont
  {Novko}(2023)}]{girotto2023}%
  \BibitemOpen
  \bibfield  {author} {\bibinfo {author} {\bibfnamefont {N.}~\bibnamefont
  {Girotto}}\ and\ \bibinfo {author} {\bibfnamefont {D.}~\bibnamefont
  {Novko}},\ }\bibfield  {title} {\bibinfo {title} {Dynamical renormalization
  of electron-phonon coupling in conventional superconductors},\ }\href
  {https://doi.org/10.1103/PhysRevB.107.064310} {\bibfield  {journal} {\bibinfo
   {journal} {Phys. Rev. B}\ }\textbf {\bibinfo {volume} {107}},\ \bibinfo
  {pages} {064310} (\bibinfo {year} {2023})}\BibitemShut {NoStop}%
\bibitem [{\citenamefont {Li}\ \emph {et~al.}(2025)\citenamefont {Li},
  \citenamefont {Tang}, \citenamefont {Tao}, \citenamefont {Shi}, \citenamefont
  {Gao}, \citenamefont {Xue}, \citenamefont {Gao}, \citenamefont {Sun},
  \citenamefont {Miao}, \citenamefont {Su}, \citenamefont {Shi}, \citenamefont
  {Hu}, \citenamefont {Mao}, \citenamefont {Zhang}, \citenamefont {He},
  \citenamefont {Wang}, \citenamefont {Gao}, \citenamefont {Peng},
  \citenamefont {Lian}, \citenamefont {Guo},\ and\ \citenamefont
  {Zhu}}]{li2025}%
  \BibitemOpen
  \bibfield  {author} {\bibinfo {author} {\bibfnamefont {J.}~\bibnamefont
  {Li}}, \bibinfo {author} {\bibfnamefont {J.}~\bibnamefont {Tang}}, \bibinfo
  {author} {\bibfnamefont {Z.}~\bibnamefont {Tao}}, \bibinfo {author}
  {\bibfnamefont {R.}~\bibnamefont {Shi}}, \bibinfo {author} {\bibfnamefont
  {X.}~\bibnamefont {Gao}}, \bibinfo {author} {\bibfnamefont {S.}~\bibnamefont
  {Xue}}, \bibinfo {author} {\bibfnamefont {X.}~\bibnamefont {Gao}}, \bibinfo
  {author} {\bibfnamefont {W.}~\bibnamefont {Sun}}, \bibinfo {author}
  {\bibfnamefont {G.}~\bibnamefont {Miao}}, \bibinfo {author} {\bibfnamefont
  {Z.}~\bibnamefont {Su}}, \bibinfo {author} {\bibfnamefont {X.}~\bibnamefont
  {Shi}}, \bibinfo {author} {\bibfnamefont {S.}~\bibnamefont {Hu}}, \bibinfo
  {author} {\bibfnamefont {R.}~\bibnamefont {Mao}}, \bibinfo {author}
  {\bibfnamefont {X.}~\bibnamefont {Zhang}}, \bibinfo {author} {\bibfnamefont
  {P.}~\bibnamefont {He}}, \bibinfo {author} {\bibfnamefont {W.}~\bibnamefont
  {Wang}}, \bibinfo {author} {\bibfnamefont {P.}~\bibnamefont {Gao}}, \bibinfo
  {author} {\bibfnamefont {H.}~\bibnamefont {Peng}}, \bibinfo {author}
  {\bibfnamefont {C.}~\bibnamefont {Lian}}, \bibinfo {author} {\bibfnamefont
  {J.}~\bibnamefont {Guo}},\ and\ \bibinfo {author} {\bibfnamefont
  {X.}~\bibnamefont {Zhu}},\ }\bibfield  {title} {\bibinfo {title}
  {Nonadiabatic renormalization of the phonon dispersion in monolayer and
  bilayer graphene},\ }\href {https://doi.org/10.1103/4x9y-txyy} {\bibfield
  {journal} {\bibinfo  {journal} {Phys. Rev. Lett.}\ }\textbf {\bibinfo
  {volume} {135}},\ \bibinfo {pages} {026204} (\bibinfo {year}
  {2025})}\BibitemShut {NoStop}%
\bibitem [{\citenamefont {Berges}\ \emph {et~al.}(2023)\citenamefont {Berges},
  \citenamefont {Girotto}, \citenamefont {Wehling}, \citenamefont {Marzari},\
  and\ \citenamefont {Ponc\'e}}]{berges2023}%
  \BibitemOpen
  \bibfield  {author} {\bibinfo {author} {\bibfnamefont {J.}~\bibnamefont
  {Berges}}, \bibinfo {author} {\bibfnamefont {N.}~\bibnamefont {Girotto}},
  \bibinfo {author} {\bibfnamefont {T.}~\bibnamefont {Wehling}}, \bibinfo
  {author} {\bibfnamefont {N.}~\bibnamefont {Marzari}},\ and\ \bibinfo {author}
  {\bibfnamefont {S.}~\bibnamefont {Ponc\'e}},\ }\bibfield  {title} {\bibinfo
  {title} {Phonon self-energy corrections: To screen, or not to screen},\
  }\href {https://doi.org/10.1103/PhysRevX.13.041009} {\bibfield  {journal}
  {\bibinfo  {journal} {Phys. Rev. X}\ }\textbf {\bibinfo {volume} {13}},\
  \bibinfo {pages} {041009} (\bibinfo {year} {2023})}\BibitemShut {NoStop}%
\bibitem [{\citenamefont {Allen}\ and\ \citenamefont
  {Silberglitt}(1974{\natexlab{a}})}]{allen74}%
  \BibitemOpen
  \bibfield  {author} {\bibinfo {author} {\bibfnamefont {P.~B.}\ \bibnamefont
  {Allen}}\ and\ \bibinfo {author} {\bibfnamefont {R.}~\bibnamefont
  {Silberglitt}},\ }\bibfield  {title} {\bibinfo {title} {Some effects of
  phonon dynamics on electron lifetime, mass renormalization, and
  superconducting transition temperature},\ }\href
  {https://doi.org/10.1103/PhysRevB.9.4733} {\bibfield  {journal} {\bibinfo
  {journal} {Phys. Rev. B}\ }\textbf {\bibinfo {volume} {9}},\ \bibinfo {pages}
  {4733} (\bibinfo {year} {1974}{\natexlab{a}})}\BibitemShut {NoStop}%
\bibitem [{\citenamefont {Marsiglio}(1990)}]{marsiglio90}%
  \BibitemOpen
  \bibfield  {author} {\bibinfo {author} {\bibfnamefont {F.}~\bibnamefont
  {Marsiglio}},\ }\bibfield  {title} {\bibinfo {title} {Pairing and
  charge-density-wave correlations in the {H}olstein model at half-filling},\
  }\href {https://doi.org/10.1103/PhysRevB.42.2416} {\bibfield  {journal}
  {\bibinfo  {journal} {Phys. Rev. B}\ }\textbf {\bibinfo {volume} {42}},\
  \bibinfo {pages} {2416} (\bibinfo {year} {1990})}\BibitemShut {NoStop}%
\bibitem [{\citenamefont {Nosarzewski}\ \emph {et~al.}(2021)\citenamefont
  {Nosarzewski}, \citenamefont {Sch\"uler},\ and\ \citenamefont
  {Devereaux}}]{nosarzewski21}%
  \BibitemOpen
  \bibfield  {author} {\bibinfo {author} {\bibfnamefont {B.}~\bibnamefont
  {Nosarzewski}}, \bibinfo {author} {\bibfnamefont {M.}~\bibnamefont
  {Sch\"uler}},\ and\ \bibinfo {author} {\bibfnamefont {T.~P.}\ \bibnamefont
  {Devereaux}},\ }\bibfield  {title} {\bibinfo {title} {Spectral properties and
  enhanced superconductivity in renormalized {M}igdal-{E}liashberg theory},\
  }\href {https://doi.org/10.1103/PhysRevB.103.024520} {\bibfield  {journal}
  {\bibinfo  {journal} {Phys. Rev. B}\ }\textbf {\bibinfo {volume} {103}},\
  \bibinfo {pages} {024520} (\bibinfo {year} {2021})}\BibitemShut {NoStop}%
\bibitem [{\citenamefont {Setty}\ \emph {et~al.}(2020)\citenamefont {Setty},
  \citenamefont {Baggioli},\ and\ \citenamefont {Zaccone}}]{setty20}%
  \BibitemOpen
  \bibfield  {author} {\bibinfo {author} {\bibfnamefont {C.}~\bibnamefont
  {Setty}}, \bibinfo {author} {\bibfnamefont {M.}~\bibnamefont {Baggioli}},\
  and\ \bibinfo {author} {\bibfnamefont {A.}~\bibnamefont {Zaccone}},\
  }\bibfield  {title} {\bibinfo {title} {Anharmonic phonon damping enhances the
  ${T}_{c}$ of {BCS}-type superconductors},\ }\href
  {https://doi.org/10.1103/PhysRevB.102.174506} {\bibfield  {journal} {\bibinfo
   {journal} {Phys. Rev. B}\ }\textbf {\bibinfo {volume} {102}},\ \bibinfo
  {pages} {174506} (\bibinfo {year} {2020})}\BibitemShut {NoStop}%
\bibitem [{\citenamefont {Setty}\ \emph {et~al.}(2022)\citenamefont {Setty},
  \citenamefont {Baggioli},\ and\ \citenamefont {Zaccone}}]{setty22}%
  \BibitemOpen
  \bibfield  {author} {\bibinfo {author} {\bibfnamefont {C.}~\bibnamefont
  {Setty}}, \bibinfo {author} {\bibfnamefont {M.}~\bibnamefont {Baggioli}},\
  and\ \bibinfo {author} {\bibfnamefont {A.}~\bibnamefont {Zaccone}},\
  }\bibfield  {title} {\bibinfo {title} {Superconducting dome in
  ferroelectric-type materials from soft mode instability},\ }\href
  {https://doi.org/10.1103/PhysRevB.105.L020506} {\bibfield  {journal}
  {\bibinfo  {journal} {Phys. Rev. B}\ }\textbf {\bibinfo {volume} {105}},\
  \bibinfo {pages} {L020506} (\bibinfo {year} {2022})}\BibitemShut {NoStop}%
\bibitem [{\citenamefont {Pietronero}\ \emph {et~al.}(1995)\citenamefont
  {Pietronero}, \citenamefont {Str\"assler},\ and\ \citenamefont
  {Grimaldi}}]{grimaldi95b}%
  \BibitemOpen
  \bibfield  {author} {\bibinfo {author} {\bibfnamefont {L.}~\bibnamefont
  {Pietronero}}, \bibinfo {author} {\bibfnamefont {S.}~\bibnamefont
  {Str\"assler}},\ and\ \bibinfo {author} {\bibfnamefont {C.}~\bibnamefont
  {Grimaldi}},\ }\bibfield  {title} {\bibinfo {title} {Nonadiabatic
  superconductivity. i. vertex corrections for the electron-phonon
  interactions},\ }\href {https://doi.org/10.1103/PhysRevB.52.10516} {\bibfield
   {journal} {\bibinfo  {journal} {Phys. Rev. B}\ }\textbf {\bibinfo {volume}
  {52}},\ \bibinfo {pages} {10516} (\bibinfo {year} {1995})}\BibitemShut
  {NoStop}%
\bibitem [{\citenamefont {Cappelluti}\ \emph {et~al.}(2000)\citenamefont
  {Cappelluti}, \citenamefont {Grimaldi}, \citenamefont {Pietronero},\ and\
  \citenamefont {Str\"assler}}]{cappelluti00}%
  \BibitemOpen
  \bibfield  {author} {\bibinfo {author} {\bibfnamefont {E.}~\bibnamefont
  {Cappelluti}}, \bibinfo {author} {\bibfnamefont {C.}~\bibnamefont
  {Grimaldi}}, \bibinfo {author} {\bibfnamefont {L.}~\bibnamefont
  {Pietronero}},\ and\ \bibinfo {author} {\bibfnamefont {S.}~\bibnamefont
  {Str\"assler}},\ }\bibfield  {title} {\bibinfo {title} {Nonadiabatic channels
  in the superconducting pairing of fullerides},\ }\href
  {https://doi.org/10.1103/PhysRevLett.85.4771} {\bibfield  {journal} {\bibinfo
   {journal} {Phys. Rev. Lett.}\ }\textbf {\bibinfo {volume} {85}},\ \bibinfo
  {pages} {4771} (\bibinfo {year} {2000})}\BibitemShut {NoStop}%
\bibitem [{\citenamefont {Gor'kov}(2016)}]{gorkov16}%
  \BibitemOpen
  \bibfield  {author} {\bibinfo {author} {\bibfnamefont {L.~P.}\ \bibnamefont
  {Gor'kov}},\ }\bibfield  {title} {\bibinfo {title} {Superconducting
  transition temperature: Interacting fermi gas and phonon mechanisms in the
  nonadiabatic regime},\ }\href {https://doi.org/10.1103/PhysRevB.93.054517}
  {\bibfield  {journal} {\bibinfo  {journal} {Phys. Rev. B}\ }\textbf {\bibinfo
  {volume} {93}},\ \bibinfo {pages} {054517} (\bibinfo {year}
  {2016})}\BibitemShut {NoStop}%
\bibitem [{\citenamefont {Krsnik}\ \emph {et~al.}(2024)\citenamefont {Krsnik},
  \citenamefont {Novko},\ and\ \citenamefont {Bari\ifmmode \check{s}\else
  \v{s}\fi{}i\ifmmode~\acute{c}\else \'{c}\fi{}}}]{krsnik2024}%
  \BibitemOpen
  \bibfield  {author} {\bibinfo {author} {\bibfnamefont {J.}~\bibnamefont
  {Krsnik}}, \bibinfo {author} {\bibfnamefont {D.}~\bibnamefont {Novko}},\ and\
  \bibinfo {author} {\bibfnamefont {O.~S.}\ \bibnamefont {Bari\ifmmode
  \check{s}\else \v{s}\fi{}i\ifmmode~\acute{c}\else \'{c}\fi{}}},\ }\bibfield
  {title} {\bibinfo {title} {Superconductivity in two-dimensional systems
  enhanced by nonadiabatic phonon-production effects},\ }\href
  {https://doi.org/10.1103/PhysRevB.110.L180505} {\bibfield  {journal}
  {\bibinfo  {journal} {Phys. Rev. B}\ }\textbf {\bibinfo {volume} {110}},\
  \bibinfo {pages} {L180505} (\bibinfo {year} {2024})}\BibitemShut {NoStop}%
\bibitem [{\citenamefont {Bauer}\ and\ \citenamefont
  {Falter}(2009)}]{bauer2009}%
  \BibitemOpen
  \bibfield  {author} {\bibinfo {author} {\bibfnamefont {T.}~\bibnamefont
  {Bauer}}\ and\ \bibinfo {author} {\bibfnamefont {C.}~\bibnamefont {Falter}},\
  }\bibfield  {title} {\bibinfo {title} {Impact of dynamical screening on the
  phonon dynamics of metallic {${\text{La}}_{2}{\text{CuO}}_{4}$}},\ }\href
  {https://doi.org/10.1103/PhysRevB.80.094525} {\bibfield  {journal} {\bibinfo
  {journal} {Phys. Rev. B}\ }\textbf {\bibinfo {volume} {80}},\ \bibinfo
  {pages} {094525} (\bibinfo {year} {2009})}\BibitemShut {NoStop}%
\bibitem [{\citenamefont {Marini}(2025)}]{marini2025}%
  \BibitemOpen
  \bibfield  {author} {\bibinfo {author} {\bibfnamefont {A.}~\bibnamefont
  {Marini}},\ }\href {https://arxiv.org/abs/2501.01866} {\bibinfo {title}
  {Dynamical electron-phonon vertex correction}} (\bibinfo {year} {2025}),\
  \Eprint {https://arxiv.org/abs/2501.01866} {arXiv:2501.01866} \BibitemShut
  {NoStop}%
\bibitem [{\citenamefont {Krsnik}\ and\ \citenamefont {Bari\ifmmode
  \check{s}\else \v{s}\fi{}i\ifmmode~\acute{c}\else
  \'{c}\fi{}}(2022)}]{krsnik2022}%
  \BibitemOpen
  \bibfield  {author} {\bibinfo {author} {\bibfnamefont {J.}~\bibnamefont
  {Krsnik}}\ and\ \bibinfo {author} {\bibfnamefont {O.~S.}\ \bibnamefont
  {Bari\ifmmode \check{s}\else \v{s}\fi{}i\ifmmode~\acute{c}\else
  \'{c}\fi{}}},\ }\bibfield  {title} {\bibinfo {title} {Importance of coupling
  strength in shaping electron energy loss and phonon spectra of phonon-plasmon
  systems},\ }\href {https://doi.org/10.1103/PhysRevB.106.075207} {\bibfield
  {journal} {\bibinfo  {journal} {Phys. Rev. B}\ }\textbf {\bibinfo {volume}
  {106}},\ \bibinfo {pages} {075207} (\bibinfo {year} {2022})}\BibitemShut
  {NoStop}%
\bibitem [{\citenamefont {Hu}\ \emph {et~al.}(2022{\natexlab{a}})\citenamefont
  {Hu}, \citenamefont {Liu}, \citenamefont {Chen}, \citenamefont {Lian},
  \citenamefont {Wang},\ and\ \citenamefont {Meng}}]{meng22a}%
  \BibitemOpen
  \bibfield  {author} {\bibinfo {author} {\bibfnamefont {S.-Q.}\ \bibnamefont
  {Hu}}, \bibinfo {author} {\bibfnamefont {X.-B.}\ \bibnamefont {Liu}},
  \bibinfo {author} {\bibfnamefont {D.-Q.}\ \bibnamefont {Chen}}, \bibinfo
  {author} {\bibfnamefont {C.}~\bibnamefont {Lian}}, \bibinfo {author}
  {\bibfnamefont {E.-G.}\ \bibnamefont {Wang}},\ and\ \bibinfo {author}
  {\bibfnamefont {S.}~\bibnamefont {Meng}},\ }\bibfield  {title} {\bibinfo
  {title} {Nonadiabatic electron-phonon coupling and its effects on
  superconductivity},\ }\href {https://doi.org/10.1103/PhysRevB.105.224311}
  {\bibfield  {journal} {\bibinfo  {journal} {Phys. Rev. B}\ }\textbf {\bibinfo
  {volume} {105}},\ \bibinfo {pages} {224311} (\bibinfo {year}
  {2022}{\natexlab{a}})}\BibitemShut {NoStop}%
\bibitem [{\citenamefont {Hu}\ \emph {et~al.}(2022{\natexlab{b}})\citenamefont
  {Hu}, \citenamefont {Chen}, \citenamefont {Zhang}, \citenamefont {Liu},\ and\
  \citenamefont {Meng}}]{meng22b}%
  \BibitemOpen
  \bibfield  {author} {\bibinfo {author} {\bibfnamefont {S.-Q.}\ \bibnamefont
  {Hu}}, \bibinfo {author} {\bibfnamefont {D.-Q.}\ \bibnamefont {Chen}},
  \bibinfo {author} {\bibfnamefont {S.-J.}\ \bibnamefont {Zhang}}, \bibinfo
  {author} {\bibfnamefont {X.-B.}\ \bibnamefont {Liu}},\ and\ \bibinfo {author}
  {\bibfnamefont {S.}~\bibnamefont {Meng}},\ }\bibfield  {title} {\bibinfo
  {title} {Probing precise interatomic potentials by nonadiabatic nonlinear
  phonons},\ }\href
  {https://doi.org/https://doi.org/10.1016/j.mtphys.2022.100790} {\bibfield
  {journal} {\bibinfo  {journal} {Materials Today Physics}\ }\textbf {\bibinfo
  {volume} {27}},\ \bibinfo {pages} {100790} (\bibinfo {year}
  {2022}{\natexlab{b}})}\BibitemShut {NoStop}%
\bibitem [{\citenamefont {Gor’kov}(2016)}]{gorkov2016}%
  \BibitemOpen
  \bibfield  {author} {\bibinfo {author} {\bibfnamefont {L.~P.}\ \bibnamefont
  {Gor’kov}},\ }\bibfield  {title} {\bibinfo {title} {Phonon mechanism in the
  most dilute superconductor n-type {SrTiO$_3$}},\ }\href
  {https://doi.org/10.1073/pnas.1604145113} {\bibfield  {journal} {\bibinfo
  {journal} {Proceedings of the National Academy of Sciences}\ }\textbf
  {\bibinfo {volume} {113}},\ \bibinfo {pages} {4646} (\bibinfo {year}
  {2016})}\BibitemShut {NoStop}%
\bibitem [{\citenamefont {Setty}\ \emph {et~al.}(2024)\citenamefont {Setty},
  \citenamefont {Baggioli},\ and\ \citenamefont {Zaccone}}]{Setty_2024}%
  \BibitemOpen
  \bibfield  {author} {\bibinfo {author} {\bibfnamefont {C.}~\bibnamefont
  {Setty}}, \bibinfo {author} {\bibfnamefont {M.}~\bibnamefont {Baggioli}},\
  and\ \bibinfo {author} {\bibfnamefont {A.}~\bibnamefont {Zaccone}},\
  }\bibfield  {title} {\bibinfo {title} {Anharmonic theory of superconductivity
  and its applications to emerging quantum materials},\ }\href
  {https://doi.org/10.1088/1361-648X/ad2159} {\bibfield  {journal} {\bibinfo
  {journal} {Journal of Physics: Condensed Matter}\ }\textbf {\bibinfo {volume}
  {36}},\ \bibinfo {pages} {173002} (\bibinfo {year} {2024})}\BibitemShut
  {NoStop}%
\bibitem [{\citenamefont {Girotto~Erhardt}\ \emph {et~al.}(2025)\citenamefont
  {Girotto~Erhardt}, \citenamefont {Berges}, \citenamefont {Poncé},\ and\
  \citenamefont {Novko}}]{erhardt2025b}%
  \BibitemOpen
  \bibfield  {author} {\bibinfo {author} {\bibfnamefont {N.}~\bibnamefont
  {Girotto~Erhardt}}, \bibinfo {author} {\bibfnamefont {J.}~\bibnamefont
  {Berges}}, \bibinfo {author} {\bibfnamefont {S.}~\bibnamefont {Poncé}},\
  and\ \bibinfo {author} {\bibfnamefont {D.}~\bibnamefont {Novko}},\ }\bibfield
   {title} {\bibinfo {title} {Understanding the origin of superconducting dome
  in electron-doped {MoS$_2$} monolayer},\ }\href
  {https://doi.org/https://doi.org/10.1038/s41699-025-00563-3} {\bibfield
  {journal} {\bibinfo  {journal} {npj 2D Materials and Applications}\ }\textbf
  {\bibinfo {volume} {9}},\ \bibinfo {pages} {44} (\bibinfo {year}
  {2025})}\BibitemShut {NoStop}%
\bibitem [{\citenamefont {Dalladay-Simpson}\ \emph {et~al.}(2025)\citenamefont
  {Dalladay-Simpson}, \citenamefont {Marchese}, \citenamefont {Cao},
  \citenamefont {Barone}, \citenamefont {Benfatto}, \citenamefont {Garbarino},
  \citenamefont {Mauri},\ and\ \citenamefont {Gorelli}}]{Dalladay2025}%
  \BibitemOpen
  \bibfield  {author} {\bibinfo {author} {\bibfnamefont {P.}~\bibnamefont
  {Dalladay-Simpson}}, \bibinfo {author} {\bibfnamefont {G.}~\bibnamefont
  {Marchese}}, \bibinfo {author} {\bibfnamefont {Z.-Y.}\ \bibnamefont {Cao}},
  \bibinfo {author} {\bibfnamefont {P.}~\bibnamefont {Barone}}, \bibinfo
  {author} {\bibfnamefont {L.}~\bibnamefont {Benfatto}}, \bibinfo {author}
  {\bibfnamefont {G.}~\bibnamefont {Garbarino}}, \bibinfo {author}
  {\bibfnamefont {F.}~\bibnamefont {Mauri}},\ and\ \bibinfo {author}
  {\bibfnamefont {F.~A.}\ \bibnamefont {Gorelli}},\ }\href
  {https://arxiv.org/abs/2511.10784} {\bibinfo {title} {Raman fingerprint of
  high-temperature superconductivity in compressed hydrides}} (\bibinfo {year}
  {2025}),\ \Eprint {https://arxiv.org/abs/2511.10784} {arXiv:2511.10784}
  \BibitemShut {NoStop}%
\bibitem [{\citenamefont {Varma}\ and\ \citenamefont
  {Simons}(1983)}]{varma1983}%
  \BibitemOpen
  \bibfield  {author} {\bibinfo {author} {\bibfnamefont {C.~M.}\ \bibnamefont
  {Varma}}\ and\ \bibinfo {author} {\bibfnamefont {A.~L.}\ \bibnamefont
  {Simons}},\ }\bibfield  {title} {\bibinfo {title} {Strong-coupling theory of
  charge-density-wave transitions},\ }\href
  {https://doi.org/10.1103/PhysRevLett.51.138} {\bibfield  {journal} {\bibinfo
  {journal} {Phys. Rev. Lett.}\ }\textbf {\bibinfo {volume} {51}},\ \bibinfo
  {pages} {138} (\bibinfo {year} {1983})}\BibitemShut {NoStop}%
\bibitem [{\citenamefont {Yoshiyama}\ \emph {et~al.}(1986)\citenamefont
  {Yoshiyama}, \citenamefont {Takaoka}, \citenamefont {Suzuki},\ and\
  \citenamefont {Motizuki}}]{Yoshiyama_1986}%
  \BibitemOpen
  \bibfield  {author} {\bibinfo {author} {\bibfnamefont {H.}~\bibnamefont
  {Yoshiyama}}, \bibinfo {author} {\bibfnamefont {Y.}~\bibnamefont {Takaoka}},
  \bibinfo {author} {\bibfnamefont {N.}~\bibnamefont {Suzuki}},\ and\ \bibinfo
  {author} {\bibfnamefont {K.}~\bibnamefont {Motizuki}},\ }\bibfield  {title}
  {\bibinfo {title} {Effects on lattice fluctuations on the charge-density-wave
  transition in transition-metal dichalcogenides},\ }\href
  {https://doi.org/10.1088/0022-3719/19/28/011} {\bibfield  {journal} {\bibinfo
   {journal} {Journal of Physics C: Solid State Physics}\ }\textbf {\bibinfo
  {volume} {19}},\ \bibinfo {pages} {5591} (\bibinfo {year}
  {1986})}\BibitemShut {NoStop}%
\bibitem [{\citenamefont {Zan}\ \emph {et~al.}(2024)\citenamefont {Zan},
  \citenamefont {Guo}, \citenamefont {Deng}, \citenamefont {Huang},
  \citenamefont {Liu}, \citenamefont {Wu}, \citenamefont {Yuan}, \citenamefont
  {Jiaojiao}, \citenamefont {Peng}, \citenamefont {Li}, \citenamefont {Zhang},
  \citenamefont {Li}, \citenamefont {Zhu}, \citenamefont {Dong}, \citenamefont
  {Shi}, \citenamefont {Yang}, \citenamefont {Yang}, \citenamefont {Shi},
  \citenamefont {Du},\ and\ \citenamefont {Zhang}}]{zan2024}%
  \BibitemOpen
  \bibfield  {author} {\bibinfo {author} {\bibfnamefont {X.}~\bibnamefont
  {Zan}}, \bibinfo {author} {\bibfnamefont {X.}~\bibnamefont {Guo}}, \bibinfo
  {author} {\bibfnamefont {A.}~\bibnamefont {Deng}}, \bibinfo {author}
  {\bibfnamefont {Z.}~\bibnamefont {Huang}}, \bibinfo {author} {\bibfnamefont
  {L.}~\bibnamefont {Liu}}, \bibinfo {author} {\bibfnamefont {F.}~\bibnamefont
  {Wu}}, \bibinfo {author} {\bibfnamefont {Y.}~\bibnamefont {Yuan}}, \bibinfo
  {author} {\bibfnamefont {Z.}~\bibnamefont {Jiaojiao}}, \bibinfo {author}
  {\bibfnamefont {Y.}~\bibnamefont {Peng}}, \bibinfo {author} {\bibfnamefont
  {L.}~\bibnamefont {Li}}, \bibinfo {author} {\bibfnamefont {Y.}~\bibnamefont
  {Zhang}}, \bibinfo {author} {\bibfnamefont {X.}~\bibnamefont {Li}}, \bibinfo
  {author} {\bibfnamefont {J.}~\bibnamefont {Zhu}}, \bibinfo {author}
  {\bibfnamefont {J.}~\bibnamefont {Dong}}, \bibinfo {author} {\bibfnamefont
  {D.-X.}\ \bibnamefont {Shi}}, \bibinfo {author} {\bibfnamefont
  {W.}~\bibnamefont {Yang}}, \bibinfo {author} {\bibfnamefont {X.}~\bibnamefont
  {Yang}}, \bibinfo {author} {\bibfnamefont {Z.}~\bibnamefont {Shi}}, \bibinfo
  {author} {\bibfnamefont {L.}~\bibnamefont {Du}},\ and\ \bibinfo {author}
  {\bibfnamefont {G.}~\bibnamefont {Zhang}},\ }\bibfield  {title} {\bibinfo
  {title} {Electron/infrared-phonon coupling in {ABC} trilayer graphene},\
  }\href {https://doi.org/10.1038/s41467-024-46129-7} {\bibfield  {journal}
  {\bibinfo  {journal} {Nature Communications}\ }\textbf {\bibinfo {volume}
  {15}},\ \bibinfo {pages} {1888} (\bibinfo {year} {2024})}\BibitemShut
  {NoStop}%
\bibitem [{\citenamefont {Chae}\ \emph {et~al.}(2010)\citenamefont {Chae},
  \citenamefont {Krauss}, \citenamefont {Klitzing},\ and\ \citenamefont
  {Smet}}]{chae10}%
  \BibitemOpen
  \bibfield  {author} {\bibinfo {author} {\bibfnamefont {D.-H.}\ \bibnamefont
  {Chae}}, \bibinfo {author} {\bibfnamefont {B.}~\bibnamefont {Krauss}},
  \bibinfo {author} {\bibfnamefont {K.}~\bibnamefont {Klitzing}},\ and\
  \bibinfo {author} {\bibfnamefont {J.}~\bibnamefont {Smet}},\ }\bibfield
  {title} {\bibinfo {title} {Hot phonons in an electrically biased graphene
  constriction},\ }\href {https://doi.org/10.1021/nl903167f} {\bibfield
  {journal} {\bibinfo  {journal} {Nano letters}\ }\textbf {\bibinfo {volume}
  {10}},\ \bibinfo {pages} {466} (\bibinfo {year} {2010})}\BibitemShut
  {NoStop}%
\bibitem [{\citenamefont {Osterhoudt}\ \emph {et~al.}(2021)\citenamefont
  {Osterhoudt}, \citenamefont {Wang}, \citenamefont {Garcia}, \citenamefont
  {Plisson}, \citenamefont {Gooth}, \citenamefont {Felser}, \citenamefont
  {Narang},\ and\ \citenamefont {Burch}}]{osterhoudt2021}%
  \BibitemOpen
  \bibfield  {author} {\bibinfo {author} {\bibfnamefont {G.~B.}\ \bibnamefont
  {Osterhoudt}}, \bibinfo {author} {\bibfnamefont {Y.}~\bibnamefont {Wang}},
  \bibinfo {author} {\bibfnamefont {C.~A.~C.}\ \bibnamefont {Garcia}}, \bibinfo
  {author} {\bibfnamefont {V.~M.}\ \bibnamefont {Plisson}}, \bibinfo {author}
  {\bibfnamefont {J.}~\bibnamefont {Gooth}}, \bibinfo {author} {\bibfnamefont
  {C.}~\bibnamefont {Felser}}, \bibinfo {author} {\bibfnamefont
  {P.}~\bibnamefont {Narang}},\ and\ \bibinfo {author} {\bibfnamefont {K.~S.}\
  \bibnamefont {Burch}},\ }\bibfield  {title} {\bibinfo {title} {Evidence for
  dominant phonon-electron scattering in {W}eyl semimetal
  {${\mathrm{WP}}_{2}$}},\ }\href {https://doi.org/10.1103/PhysRevX.11.011017}
  {\bibfield  {journal} {\bibinfo  {journal} {Phys. Rev. X}\ }\textbf {\bibinfo
  {volume} {11}},\ \bibinfo {pages} {011017} (\bibinfo {year}
  {2021})}\BibitemShut {NoStop}%
\bibitem [{\citenamefont {Coulter}\ \emph {et~al.}(2019)\citenamefont
  {Coulter}, \citenamefont {Osterhoudt}, \citenamefont {Garcia}, \citenamefont
  {Wang}, \citenamefont {Plisson}, \citenamefont {Shen}, \citenamefont {Ni},
  \citenamefont {Burch},\ and\ \citenamefont {Narang}}]{coulter2019}%
  \BibitemOpen
  \bibfield  {author} {\bibinfo {author} {\bibfnamefont {J.}~\bibnamefont
  {Coulter}}, \bibinfo {author} {\bibfnamefont {G.~B.}\ \bibnamefont
  {Osterhoudt}}, \bibinfo {author} {\bibfnamefont {C.~A.~C.}\ \bibnamefont
  {Garcia}}, \bibinfo {author} {\bibfnamefont {Y.}~\bibnamefont {Wang}},
  \bibinfo {author} {\bibfnamefont {V.~M.}\ \bibnamefont {Plisson}}, \bibinfo
  {author} {\bibfnamefont {B.}~\bibnamefont {Shen}}, \bibinfo {author}
  {\bibfnamefont {N.}~\bibnamefont {Ni}}, \bibinfo {author} {\bibfnamefont
  {K.~S.}\ \bibnamefont {Burch}},\ and\ \bibinfo {author} {\bibfnamefont
  {P.}~\bibnamefont {Narang}},\ }\bibfield  {title} {\bibinfo {title}
  {Uncovering electron-phonon scattering and phonon dynamics in type-{I} {W}eyl
  semimetals},\ }\href {https://doi.org/10.1103/PhysRevB.100.220301} {\bibfield
   {journal} {\bibinfo  {journal} {Phys. Rev. B}\ }\textbf {\bibinfo {volume}
  {100}},\ \bibinfo {pages} {220301(R)} (\bibinfo {year} {2019})}\BibitemShut
  {NoStop}%
\bibitem [{\citenamefont {Li}\ \emph {et~al.}(2024)\citenamefont {Li},
  \citenamefont {Xu}, \citenamefont {Liu}, \citenamefont {Fang}, \citenamefont
  {Zheng}, \citenamefont {Dai}, \citenamefont {Li}, \citenamefont {Zhu},
  \citenamefont {Zhang}, \citenamefont {Liang}, \citenamefont {Yang},
  \citenamefont {Huang}, \citenamefont {Xi}, \citenamefont {Liu}, \citenamefont
  {Xu},\ and\ \citenamefont {Chen}}]{li2024}%
  \BibitemOpen
  \bibfield  {author} {\bibinfo {author} {\bibfnamefont {Y.}~\bibnamefont
  {Li}}, \bibinfo {author} {\bibfnamefont {L.}~\bibnamefont {Xu}}, \bibinfo
  {author} {\bibfnamefont {G.}~\bibnamefont {Liu}}, \bibinfo {author}
  {\bibfnamefont {Y.}~\bibnamefont {Fang}}, \bibinfo {author} {\bibfnamefont
  {H.}~\bibnamefont {Zheng}}, \bibinfo {author} {\bibfnamefont
  {S.}~\bibnamefont {Dai}}, \bibinfo {author} {\bibfnamefont {E.}~\bibnamefont
  {Li}}, \bibinfo {author} {\bibfnamefont {G.}~\bibnamefont {Zhu}}, \bibinfo
  {author} {\bibfnamefont {S.}~\bibnamefont {Zhang}}, \bibinfo {author}
  {\bibfnamefont {S.}~\bibnamefont {Liang}}, \bibinfo {author} {\bibfnamefont
  {L.}~\bibnamefont {Yang}}, \bibinfo {author} {\bibfnamefont {F.}~\bibnamefont
  {Huang}}, \bibinfo {author} {\bibfnamefont {X.}~\bibnamefont {Xi}}, \bibinfo
  {author} {\bibfnamefont {Z.}~\bibnamefont {Liu}}, \bibinfo {author}
  {\bibfnamefont {N.}~\bibnamefont {Xu}},\ and\ \bibinfo {author}
  {\bibfnamefont {Y.}~\bibnamefont {Chen}},\ }\bibfield  {title} {\bibinfo
  {title} {Evidence of strong and mode-selective electron–phonon coupling in
  the topological superconductor candidate {2M-WS$_2$}},\ }\href
  {https://doi.org/10.1038/s41467-024-50590-9} {\bibfield  {journal} {\bibinfo
  {journal} {Nature Communications}\ }\textbf {\bibinfo {volume} {15}}
  (\bibinfo {year} {2024})}\BibitemShut {NoStop}%
\bibitem [{\citenamefont {Yang}\ \emph {et~al.}(2021)\citenamefont {Yang},
  \citenamefont {Yao}, \citenamefont {Plisson}, \citenamefont {Mozaffari},
  \citenamefont {Scheifers}, \citenamefont {Savvidou}, \citenamefont {Choi},
  \citenamefont {McCandless}, \citenamefont {Padlewski}, \citenamefont
  {Putzke}, \citenamefont {Moll}, \citenamefont {Chan}, \citenamefont
  {Balicas}, \citenamefont {Burch},\ and\ \citenamefont {Tafti}}]{Yang2021}%
  \BibitemOpen
  \bibfield  {author} {\bibinfo {author} {\bibfnamefont {H.-Y.}\ \bibnamefont
  {Yang}}, \bibinfo {author} {\bibfnamefont {X.}~\bibnamefont {Yao}}, \bibinfo
  {author} {\bibfnamefont {V.}~\bibnamefont {Plisson}}, \bibinfo {author}
  {\bibfnamefont {S.}~\bibnamefont {Mozaffari}}, \bibinfo {author}
  {\bibfnamefont {J.~P.}\ \bibnamefont {Scheifers}}, \bibinfo {author}
  {\bibfnamefont {A.~F.}\ \bibnamefont {Savvidou}}, \bibinfo {author}
  {\bibfnamefont {E.~S.}\ \bibnamefont {Choi}}, \bibinfo {author}
  {\bibfnamefont {G.~T.}\ \bibnamefont {McCandless}}, \bibinfo {author}
  {\bibfnamefont {M.~F.}\ \bibnamefont {Padlewski}}, \bibinfo {author}
  {\bibfnamefont {C.}~\bibnamefont {Putzke}}, \bibinfo {author} {\bibfnamefont
  {P.~J.~W.}\ \bibnamefont {Moll}}, \bibinfo {author} {\bibfnamefont {J.~Y.}\
  \bibnamefont {Chan}}, \bibinfo {author} {\bibfnamefont {L.}~\bibnamefont
  {Balicas}}, \bibinfo {author} {\bibfnamefont {K.~S.}\ \bibnamefont {Burch}},\
  and\ \bibinfo {author} {\bibfnamefont {F.}~\bibnamefont {Tafti}},\ }\bibfield
   {title} {\bibinfo {title} {Evidence of a coupled electron-phonon liquid in
  {NbGe2}},\ }\href {https://doi.org/10.1038/s41467-021-25547-x} {\bibfield
  {journal} {\bibinfo  {journal} {Nature Communications}\ }\textbf {\bibinfo
  {volume} {12}},\ \bibinfo {pages} {5292} (\bibinfo {year}
  {2021})}\BibitemShut {NoStop}%
\bibitem [{\citenamefont {Nagamatsu}\ \emph {et~al.}(2001)\citenamefont
  {Nagamatsu}, \citenamefont {Nakagawa}, \citenamefont {Muranaka},
  \citenamefont {Zenitani},\ and\ \citenamefont {Akimitsu}}]{Nagamatsu2001}%
  \BibitemOpen
  \bibfield  {author} {\bibinfo {author} {\bibfnamefont {J.}~\bibnamefont
  {Nagamatsu}}, \bibinfo {author} {\bibfnamefont {N.}~\bibnamefont {Nakagawa}},
  \bibinfo {author} {\bibfnamefont {T.}~\bibnamefont {Muranaka}}, \bibinfo
  {author} {\bibfnamefont {Y.}~\bibnamefont {Zenitani}},\ and\ \bibinfo
  {author} {\bibfnamefont {J.}~\bibnamefont {Akimitsu}},\ }\bibfield  {title}
  {\bibinfo {title} {Superconductivity at 39 {K} in magnesium diboride},\
  }\href {https://doi.org/10.1038/35065039} {\bibfield  {journal} {\bibinfo
  {journal} {Nature}\ }\textbf {\bibinfo {volume} {410}},\ \bibinfo {pages}
  {63} (\bibinfo {year} {2001})}\BibitemShut {NoStop}%
\bibitem [{\citenamefont {Bohnen}\ \emph {et~al.}(2001)\citenamefont {Bohnen},
  \citenamefont {Heid},\ and\ \citenamefont {Renker}}]{bohnen2001}%
  \BibitemOpen
  \bibfield  {author} {\bibinfo {author} {\bibfnamefont {K.-P.}\ \bibnamefont
  {Bohnen}}, \bibinfo {author} {\bibfnamefont {R.}~\bibnamefont {Heid}},\ and\
  \bibinfo {author} {\bibfnamefont {B.}~\bibnamefont {Renker}},\ }\bibfield
  {title} {\bibinfo {title} {Phonon dispersion and electron-phonon coupling in
  {${\mathrm{MgB}}_{2}$ and ${\mathrm{AlB}}_{2}$}},\ }\href
  {https://doi.org/10.1103/PhysRevLett.86.5771} {\bibfield  {journal} {\bibinfo
   {journal} {Phys. Rev. Lett.}\ }\textbf {\bibinfo {volume} {86}},\ \bibinfo
  {pages} {5771} (\bibinfo {year} {2001})}\BibitemShut {NoStop}%
\bibitem [{\citenamefont {Kortus}\ \emph {et~al.}(2001)\citenamefont {Kortus},
  \citenamefont {Mazin}, \citenamefont {Belashchenko}, \citenamefont
  {Antropov},\ and\ \citenamefont {Boyer}}]{kortus01}%
  \BibitemOpen
  \bibfield  {author} {\bibinfo {author} {\bibfnamefont {J.}~\bibnamefont
  {Kortus}}, \bibinfo {author} {\bibfnamefont {I.~I.}\ \bibnamefont {Mazin}},
  \bibinfo {author} {\bibfnamefont {K.~D.}\ \bibnamefont {Belashchenko}},
  \bibinfo {author} {\bibfnamefont {V.~P.}\ \bibnamefont {Antropov}},\ and\
  \bibinfo {author} {\bibfnamefont {L.~L.}\ \bibnamefont {Boyer}},\ }\bibfield
  {title} {\bibinfo {title} {Superconductivity of metallic boron in
  {${\mathrm{MgB}}_{2}$}},\ }\href
  {https://doi.org/10.1103/PhysRevLett.86.4656} {\bibfield  {journal} {\bibinfo
   {journal} {Phys. Rev. Lett.}\ }\textbf {\bibinfo {volume} {86}},\ \bibinfo
  {pages} {4656} (\bibinfo {year} {2001})}\BibitemShut {NoStop}%
\bibitem [{\citenamefont {Liu}\ \emph {et~al.}(2001)\citenamefont {Liu},
  \citenamefont {Mazin},\ and\ \citenamefont {Kortus}}]{liu2001}%
  \BibitemOpen
  \bibfield  {author} {\bibinfo {author} {\bibfnamefont {A.~Y.}\ \bibnamefont
  {Liu}}, \bibinfo {author} {\bibfnamefont {I.~I.}\ \bibnamefont {Mazin}},\
  and\ \bibinfo {author} {\bibfnamefont {J.}~\bibnamefont {Kortus}},\
  }\bibfield  {title} {\bibinfo {title} {Beyond {E}liashberg superconductivity
  in {MgB}$_{2}$: Anharmonicity, two-phonon scattering, and multiple gaps},\
  }\href {https://doi.org/10.1103/PhysRevLett.87.087005} {\bibfield  {journal}
  {\bibinfo  {journal} {Phys. Rev. Lett.}\ }\textbf {\bibinfo {volume} {87}},\
  \bibinfo {pages} {087005} (\bibinfo {year} {2001})}\BibitemShut {NoStop}%
\bibitem [{\citenamefont {Yildirim}\ \emph {et~al.}(2001)\citenamefont
  {Yildirim}, \citenamefont {G\"ulseren}, \citenamefont {Lynn}, \citenamefont
  {Brown}, \citenamefont {Udovic}, \citenamefont {Huang}, \citenamefont
  {Rogado}, \citenamefont {Regan}, \citenamefont {Hayward}, \citenamefont
  {Slusky}, \citenamefont {He}, \citenamefont {Haas}, \citenamefont {Khalifah},
  \citenamefont {Inumaru},\ and\ \citenamefont {Cava}}]{yildirim2001}%
  \BibitemOpen
  \bibfield  {author} {\bibinfo {author} {\bibfnamefont {T.}~\bibnamefont
  {Yildirim}}, \bibinfo {author} {\bibfnamefont {O.}~\bibnamefont
  {G\"ulseren}}, \bibinfo {author} {\bibfnamefont {J.~W.}\ \bibnamefont
  {Lynn}}, \bibinfo {author} {\bibfnamefont {C.~M.}\ \bibnamefont {Brown}},
  \bibinfo {author} {\bibfnamefont {T.~J.}\ \bibnamefont {Udovic}}, \bibinfo
  {author} {\bibfnamefont {Q.}~\bibnamefont {Huang}}, \bibinfo {author}
  {\bibfnamefont {N.}~\bibnamefont {Rogado}}, \bibinfo {author} {\bibfnamefont
  {K.~A.}\ \bibnamefont {Regan}}, \bibinfo {author} {\bibfnamefont {M.~A.}\
  \bibnamefont {Hayward}}, \bibinfo {author} {\bibfnamefont {J.~S.}\
  \bibnamefont {Slusky}}, \bibinfo {author} {\bibfnamefont {T.}~\bibnamefont
  {He}}, \bibinfo {author} {\bibfnamefont {M.~K.}\ \bibnamefont {Haas}},
  \bibinfo {author} {\bibfnamefont {P.}~\bibnamefont {Khalifah}}, \bibinfo
  {author} {\bibfnamefont {K.}~\bibnamefont {Inumaru}},\ and\ \bibinfo {author}
  {\bibfnamefont {R.~J.}\ \bibnamefont {Cava}},\ }\bibfield  {title} {\bibinfo
  {title} {Giant anharmonicity and nonlinear electron-phonon coupling in
  {${\mathrm{MgB}}_{2}$}: A combined first-principles calculation and neutron
  scattering study},\ }\href {https://doi.org/10.1103/PhysRevLett.87.037001}
  {\bibfield  {journal} {\bibinfo  {journal} {Phys. Rev. Lett.}\ }\textbf
  {\bibinfo {volume} {87}},\ \bibinfo {pages} {037001} (\bibinfo {year}
  {2001})}\BibitemShut {NoStop}%
\bibitem [{\citenamefont {Choi}\ \emph {et~al.}(2002)\citenamefont {Choi},
  \citenamefont {Roundy}, \citenamefont {Sun}, \citenamefont {Cohen},\ and\
  \citenamefont {Louie}}]{choi2002}%
  \BibitemOpen
  \bibfield  {author} {\bibinfo {author} {\bibfnamefont {H.~J.}\ \bibnamefont
  {Choi}}, \bibinfo {author} {\bibfnamefont {D.}~\bibnamefont {Roundy}},
  \bibinfo {author} {\bibfnamefont {H.}~\bibnamefont {Sun}}, \bibinfo {author}
  {\bibfnamefont {M.~L.}\ \bibnamefont {Cohen}},\ and\ \bibinfo {author}
  {\bibfnamefont {S.~G.}\ \bibnamefont {Louie}},\ }\bibfield  {title} {\bibinfo
  {title} {First-principles calculation of the superconducting transition in
  {${\mathrm{MgB}}_{2}$} within the anisotropic eliashberg formalism},\ }\href
  {https://doi.org/10.1103/PhysRevB.66.020513} {\bibfield  {journal} {\bibinfo
  {journal} {Phys. Rev. B}\ }\textbf {\bibinfo {volume} {66}},\ \bibinfo
  {pages} {020513} (\bibinfo {year} {2002})}\BibitemShut {NoStop}%
\bibitem [{\citenamefont {Margine}\ and\ \citenamefont
  {Giustino}(2013)}]{margine2013}%
  \BibitemOpen
  \bibfield  {author} {\bibinfo {author} {\bibfnamefont {E.~R.}\ \bibnamefont
  {Margine}}\ and\ \bibinfo {author} {\bibfnamefont {F.}~\bibnamefont
  {Giustino}},\ }\bibfield  {title} {\bibinfo {title} {Anisotropic
  {M}igdal-{E}liashberg theory using {W}annier functions},\ }\href
  {https://doi.org/10.1103/PhysRevB.87.024505} {\bibfield  {journal} {\bibinfo
  {journal} {Phys. Rev. B}\ }\textbf {\bibinfo {volume} {87}},\ \bibinfo
  {pages} {024505} (\bibinfo {year} {2013})}\BibitemShut {NoStop}%
\bibitem [{\citenamefont {Eiguren}\ and\ \citenamefont
  {Ambrosch-Draxl}(2008)}]{eiguren08}%
  \BibitemOpen
  \bibfield  {author} {\bibinfo {author} {\bibfnamefont {A.}~\bibnamefont
  {Eiguren}}\ and\ \bibinfo {author} {\bibfnamefont {C.}~\bibnamefont
  {Ambrosch-Draxl}},\ }\bibfield  {title} {\bibinfo {title} {Wannier
  interpolation scheme for phonon-induced potentials: Application to bulk
  {${\text{MgB}}_{2}$}, {W}, and the {$(1\ifmmode\times\else\texttimes\fi{}1)$
  H-covered W(110)} surface},\ }\href
  {https://doi.org/10.1103/PhysRevB.78.045124} {\bibfield  {journal} {\bibinfo
  {journal} {Phys. Rev. B}\ }\textbf {\bibinfo {volume} {78}},\ \bibinfo
  {pages} {045124} (\bibinfo {year} {2008})}\BibitemShut {NoStop}%
\bibitem [{\citenamefont {Shukla}\ \emph {et~al.}(2003)\citenamefont {Shukla},
  \citenamefont {Calandra}, \citenamefont {d'Astuto}, \citenamefont {Lazzeri},
  \citenamefont {Mauri}, \citenamefont {Bellin}, \citenamefont {Krisch},
  \citenamefont {Karpinski}, \citenamefont {Kazakov}, \citenamefont {Jun},
  \citenamefont {Daghero},\ and\ \citenamefont {Parlinski}}]{shukla2003}%
  \BibitemOpen
  \bibfield  {author} {\bibinfo {author} {\bibfnamefont {A.}~\bibnamefont
  {Shukla}}, \bibinfo {author} {\bibfnamefont {M.}~\bibnamefont {Calandra}},
  \bibinfo {author} {\bibfnamefont {M.}~\bibnamefont {d'Astuto}}, \bibinfo
  {author} {\bibfnamefont {M.}~\bibnamefont {Lazzeri}}, \bibinfo {author}
  {\bibfnamefont {F.}~\bibnamefont {Mauri}}, \bibinfo {author} {\bibfnamefont
  {C.}~\bibnamefont {Bellin}}, \bibinfo {author} {\bibfnamefont
  {M.}~\bibnamefont {Krisch}}, \bibinfo {author} {\bibfnamefont
  {J.}~\bibnamefont {Karpinski}}, \bibinfo {author} {\bibfnamefont {S.~M.}\
  \bibnamefont {Kazakov}}, \bibinfo {author} {\bibfnamefont {J.}~\bibnamefont
  {Jun}}, \bibinfo {author} {\bibfnamefont {D.}~\bibnamefont {Daghero}},\ and\
  \bibinfo {author} {\bibfnamefont {K.}~\bibnamefont {Parlinski}},\ }\bibfield
  {title} {\bibinfo {title} {Phonon dispersion and lifetimes in
  {${\mathrm{M}\mathrm{g}\mathrm{B}}_{2}$}},\ }\href
  {https://doi.org/10.1103/PhysRevLett.90.095506} {\bibfield  {journal}
  {\bibinfo  {journal} {Phys. Rev. Lett.}\ }\textbf {\bibinfo {volume} {90}},\
  \bibinfo {pages} {095506} (\bibinfo {year} {2003})}\BibitemShut {NoStop}%
\bibitem [{\citenamefont {Calandra}\ and\ \citenamefont
  {Mauri}(2005)}]{calandra2005}%
  \BibitemOpen
  \bibfield  {author} {\bibinfo {author} {\bibfnamefont {M.}~\bibnamefont
  {Calandra}}\ and\ \bibinfo {author} {\bibfnamefont {F.}~\bibnamefont
  {Mauri}},\ }\bibfield  {title} {\bibinfo {title} {Electron-phonon coupling
  and phonon self-energy in {${\mathrm{MgB}}_{2}$}: Interpretation of
  {${\mathrm{MgB}}_{2}$ {R}aman spectra}},\ }\href
  {https://doi.org/10.1103/PhysRevB.71.064501} {\bibfield  {journal} {\bibinfo
  {journal} {Phys. Rev. B}\ }\textbf {\bibinfo {volume} {71}},\ \bibinfo
  {pages} {064501} (\bibinfo {year} {2005})}\BibitemShut {NoStop}%
\bibitem [{\citenamefont {Osborn}\ \emph {et~al.}(2001)\citenamefont {Osborn},
  \citenamefont {Goremychkin}, \citenamefont {Kolesnikov},\ and\ \citenamefont
  {Hinks}}]{Osborn2001}%
  \BibitemOpen
  \bibfield  {author} {\bibinfo {author} {\bibfnamefont {R.}~\bibnamefont
  {Osborn}}, \bibinfo {author} {\bibfnamefont {E.~A.}\ \bibnamefont
  {Goremychkin}}, \bibinfo {author} {\bibfnamefont {A.~I.}\ \bibnamefont
  {Kolesnikov}},\ and\ \bibinfo {author} {\bibfnamefont {D.~G.}\ \bibnamefont
  {Hinks}},\ }\bibfield  {title} {\bibinfo {title} {Phonon density of states in
  {MgB$_2$}},\ }\href {https://doi.org/10.1103/physrevlett.87.017005}
  {\bibfield  {journal} {\bibinfo  {journal} {Physical Review Letters}\
  }\textbf {\bibinfo {volume} {87}},\ \bibinfo {pages} {017005} (\bibinfo
  {year} {2001})}\BibitemShut {NoStop}%
\bibitem [{\citenamefont {Tarenkov}\ \emph {et~al.}(2009)\citenamefont
  {Tarenkov}, \citenamefont {D'yachenko}, \citenamefont {Sidorov},
  \citenamefont {Bo{\u{\i}}chenko}, \citenamefont {Bo{\u{\i}}chenko},
  \citenamefont {Chromik}, \citenamefont {{\v{S}}trb{\'i}k}, \citenamefont
  {Ga{\v{z}}i}, \citenamefont {{\v{S}}pankov{\'a}},\ and\ \citenamefont
  {Be{\v{n}}a{\v{c}}ka}}]{Tarenkov2009}%
  \BibitemOpen
  \bibfield  {author} {\bibinfo {author} {\bibfnamefont {V.~Y.}\ \bibnamefont
  {Tarenkov}}, \bibinfo {author} {\bibfnamefont {A.~I.}\ \bibnamefont
  {D'yachenko}}, \bibinfo {author} {\bibfnamefont {S.~L.}\ \bibnamefont
  {Sidorov}}, \bibinfo {author} {\bibfnamefont {V.~A.}\ \bibnamefont
  {Bo{\u{\i}}chenko}}, \bibinfo {author} {\bibfnamefont {D.~I.}\ \bibnamefont
  {Bo{\u{\i}}chenko}}, \bibinfo {author} {\bibfnamefont {{\v{S}}.}~\bibnamefont
  {Chromik}}, \bibinfo {author} {\bibfnamefont {V.}~\bibnamefont
  {{\v{S}}trb{\'i}k}}, \bibinfo {author} {\bibfnamefont {{\v{S}}.}~\bibnamefont
  {Ga{\v{z}}i}}, \bibinfo {author} {\bibfnamefont {M.}~\bibnamefont
  {{\v{S}}pankov{\'a}}},\ and\ \bibinfo {author} {\bibfnamefont
  {{\v{S}}.}~\bibnamefont {Be{\v{n}}a{\v{c}}ka}},\ }\bibfield  {title}
  {\bibinfo {title} {Electron tunneling spectroscopy of the phonon spectrum of
  {${\mathrm{Mg}}{\mathrm{B}}_{2}$}},\ }\href
  {https://doi.org/10.1134/S1063783409090030} {\bibfield  {journal} {\bibinfo
  {journal} {Physics of the Solid State}\ }\textbf {\bibinfo {volume} {51}},\
  \bibinfo {pages} {1778} (\bibinfo {year} {2009})}\BibitemShut {NoStop}%
\bibitem [{\citenamefont {Karapetrov}\ \emph {et~al.}(2001)\citenamefont
  {Karapetrov}, \citenamefont {Iavarone}, \citenamefont {Kwok}, \citenamefont
  {Crabtree},\ and\ \citenamefont {Hinks}}]{Karapetrov2001}%
  \BibitemOpen
  \bibfield  {author} {\bibinfo {author} {\bibfnamefont {G.}~\bibnamefont
  {Karapetrov}}, \bibinfo {author} {\bibfnamefont {M.}~\bibnamefont
  {Iavarone}}, \bibinfo {author} {\bibfnamefont {W.~K.}\ \bibnamefont {Kwok}},
  \bibinfo {author} {\bibfnamefont {G.~W.}\ \bibnamefont {Crabtree}},\ and\
  \bibinfo {author} {\bibfnamefont {D.~G.}\ \bibnamefont {Hinks}},\ }\bibfield
  {title} {\bibinfo {title} {Scanning tunneling spectroscopy in
  {${\mathrm{MgB}}_{2}$}},\ }\href
  {https://doi.org/10.1103/PhysRevLett.86.4374} {\bibfield  {journal} {\bibinfo
   {journal} {Phys. Rev. Lett.}\ }\textbf {\bibinfo {volume} {86}},\ \bibinfo
  {pages} {4374} (\bibinfo {year} {2001})}\BibitemShut {NoStop}%
\bibitem [{\citenamefont {Pickett}\ \emph {et~al.}(2003)\citenamefont
  {Pickett}, \citenamefont {An}, \citenamefont {Rosner},\ and\ \citenamefont
  {Savrasov}}]{Pickett2003}%
  \BibitemOpen
  \bibfield  {author} {\bibinfo {author} {\bibfnamefont {W.}~\bibnamefont
  {Pickett}}, \bibinfo {author} {\bibfnamefont {J.}~\bibnamefont {An}},
  \bibinfo {author} {\bibfnamefont {H.}~\bibnamefont {Rosner}},\ and\ \bibinfo
  {author} {\bibfnamefont {S.}~\bibnamefont {Savrasov}},\ }\bibfield  {title}
  {\bibinfo {title} {Role of two dimensionality in
  {${\mathrm{Mg}}{\mathrm{B}}_{2}$}},\ }\href
  {https://doi.org/https://doi.org/10.1016/S0921-4534(03)00656-7} {\bibfield
  {journal} {\bibinfo  {journal} {Physica C: Superconductivity}\ }\textbf
  {\bibinfo {volume} {387}},\ \bibinfo {pages} {117} (\bibinfo {year}
  {2003})},\ \bibinfo {note} {proceedings of the 3rd Polish-US Workshop on
  Superconductivity and Magnetism of Advanced Materials}\BibitemShut {NoStop}%
\bibitem [{\citenamefont {Quilty}\ \emph {et~al.}(2002)\citenamefont {Quilty},
  \citenamefont {Lee}, \citenamefont {Yamamoto},\ and\ \citenamefont
  {Tajima}}]{Quilty2002}%
  \BibitemOpen
  \bibfield  {author} {\bibinfo {author} {\bibfnamefont {J.~W.}\ \bibnamefont
  {Quilty}}, \bibinfo {author} {\bibfnamefont {S.}~\bibnamefont {Lee}},
  \bibinfo {author} {\bibfnamefont {A.}~\bibnamefont {Yamamoto}},\ and\
  \bibinfo {author} {\bibfnamefont {S.}~\bibnamefont {Tajima}},\ }\bibfield
  {title} {\bibinfo {title} {Superconducting gap in {${\mathrm{MgB}}_{2}$}:
  Electronic {R}aman scattering measurements of single crystals},\ }\href
  {https://doi.org/10.1103/PhysRevLett.88.087001} {\bibfield  {journal}
  {\bibinfo  {journal} {Phys. Rev. Lett.}\ }\textbf {\bibinfo {volume} {88}},\
  \bibinfo {pages} {087001} (\bibinfo {year} {2002})}\BibitemShut {NoStop}%
\bibitem [{\citenamefont {Mou}\ \emph {et~al.}(2015)\citenamefont {Mou},
  \citenamefont {Jiang}, \citenamefont {Taufour}, \citenamefont {Flint},
  \citenamefont {Bud'ko}, \citenamefont {Canfield}, \citenamefont {Wen},
  \citenamefont {Xu}, \citenamefont {Gu},\ and\ \citenamefont
  {Kaminski}}]{mou2015}%
  \BibitemOpen
  \bibfield  {author} {\bibinfo {author} {\bibfnamefont {D.}~\bibnamefont
  {Mou}}, \bibinfo {author} {\bibfnamefont {R.}~\bibnamefont {Jiang}}, \bibinfo
  {author} {\bibfnamefont {V.}~\bibnamefont {Taufour}}, \bibinfo {author}
  {\bibfnamefont {R.}~\bibnamefont {Flint}}, \bibinfo {author} {\bibfnamefont
  {S.~L.}\ \bibnamefont {Bud'ko}}, \bibinfo {author} {\bibfnamefont {P.~C.}\
  \bibnamefont {Canfield}}, \bibinfo {author} {\bibfnamefont {J.~S.}\
  \bibnamefont {Wen}}, \bibinfo {author} {\bibfnamefont {Z.~J.}\ \bibnamefont
  {Xu}}, \bibinfo {author} {\bibfnamefont {G.}~\bibnamefont {Gu}},\ and\
  \bibinfo {author} {\bibfnamefont {A.}~\bibnamefont {Kaminski}},\ }\bibfield
  {title} {\bibinfo {title} {Strong interaction between electrons and
  collective excitations in the multiband superconductor
  {${\mathrm{MgB}}_{2}$}},\ }\href {https://doi.org/10.1103/PhysRevB.91.140502}
  {\bibfield  {journal} {\bibinfo  {journal} {Phys. Rev. B}\ }\textbf {\bibinfo
  {volume} {91}},\ \bibinfo {pages} {140502} (\bibinfo {year}
  {2015})}\BibitemShut {NoStop}%
\bibitem [{\citenamefont {Cappelluti}\ \emph {et~al.}(2002)\citenamefont
  {Cappelluti}, \citenamefont {Ciuchi}, \citenamefont {Grimaldi}, \citenamefont
  {Pietronero},\ and\ \citenamefont {Str\"assler}}]{cappelluti2002}%
  \BibitemOpen
  \bibfield  {author} {\bibinfo {author} {\bibfnamefont {E.}~\bibnamefont
  {Cappelluti}}, \bibinfo {author} {\bibfnamefont {S.}~\bibnamefont {Ciuchi}},
  \bibinfo {author} {\bibfnamefont {C.}~\bibnamefont {Grimaldi}}, \bibinfo
  {author} {\bibfnamefont {L.}~\bibnamefont {Pietronero}},\ and\ \bibinfo
  {author} {\bibfnamefont {S.}~\bibnamefont {Str\"assler}},\ }\bibfield
  {title} {\bibinfo {title} {High ${T}_{c}$ superconductivity in
  {${\mathrm{MgB}}_{2}$} by nonadiabatic pairing},\ }\href
  {https://doi.org/10.1103/PhysRevLett.88.117003} {\bibfield  {journal}
  {\bibinfo  {journal} {Phys. Rev. Lett.}\ }\textbf {\bibinfo {volume} {88}},\
  \bibinfo {pages} {117003} (\bibinfo {year} {2002})}\BibitemShut {NoStop}%
\bibitem [{\citenamefont {Saitta}\ \emph
  {et~al.}(2008{\natexlab{b}})\citenamefont {Saitta}, \citenamefont {Lazzeri},
  \citenamefont {Calandra},\ and\ \citenamefont {Mauri}}]{Saitta2008}%
  \BibitemOpen
  \bibfield  {author} {\bibinfo {author} {\bibfnamefont {A.~M.}\ \bibnamefont
  {Saitta}}, \bibinfo {author} {\bibfnamefont {M.}~\bibnamefont {Lazzeri}},
  \bibinfo {author} {\bibfnamefont {M.}~\bibnamefont {Calandra}},\ and\
  \bibinfo {author} {\bibfnamefont {F.}~\bibnamefont {Mauri}},\ }\bibfield
  {title} {\bibinfo {title} {Giant nonadiabatic effects in layer metals:
  {R}aman spectra of intercalated graphite explained},\ }\href
  {https://doi.org/10.1103/PhysRevLett.100.226401} {\bibfield  {journal}
  {\bibinfo  {journal} {Phys. Rev. Lett.}\ }\textbf {\bibinfo {volume} {100}},\
  \bibinfo {pages} {226401} (\bibinfo {year} {2008}{\natexlab{b}})}\BibitemShut
  {NoStop}%
\bibitem [{sup()}]{supplement}%
  \BibitemOpen
  \href@noop {} {}\bibinfo {note} {See Supplemental Material for the derivation
  of the formula and additional results.}\BibitemShut {Stop}%
\bibitem [{\citenamefont {Migdal}(1958)}]{migdal58}%
  \BibitemOpen
  \bibfield  {author} {\bibinfo {author} {\bibfnamefont {A.~B.}\ \bibnamefont
  {Migdal}},\ }\bibfield  {title} {\bibinfo {title} {Interaction between
  electrons and lattice vibrations in a normal metal},\ }\href
  {http://jetp.ras.ru/cgi-bin/e/index/e/7/6/p996?a=list} {\bibfield  {journal}
  {\bibinfo  {journal} {Sov. Phys. JETP}\ }\textbf {\bibinfo {volume} {7}},\
  \bibinfo {pages} {996} (\bibinfo {year} {1958})}\BibitemShut {NoStop}%
\bibitem [{\citenamefont {Lihm}\ and\ \citenamefont
  {Ponc\'e}(2025)}]{lihm2025}%
  \BibitemOpen
  \bibfield  {author} {\bibinfo {author} {\bibfnamefont {J.-M.}\ \bibnamefont
  {Lihm}}\ and\ \bibinfo {author} {\bibfnamefont {S.}~\bibnamefont {Ponc\'e}},\
  }\bibfield  {title} {\bibinfo {title} {Nonperturbative self-consistent
  electron-phonon spectral functions and transport},\ }\href
  {https://doi.org/10.1103/PhysRevLett.134.186401} {\bibfield  {journal}
  {\bibinfo  {journal} {Phys. Rev. Lett.}\ }\textbf {\bibinfo {volume} {134}},\
  \bibinfo {pages} {186401} (\bibinfo {year} {2025})}\BibitemShut {NoStop}%
\bibitem [{\citenamefont {Poncé}\ \emph {et~al.}(2015)\citenamefont {Poncé},
  \citenamefont {Gillet}, \citenamefont {Laflamme~Janssen}, \citenamefont
  {Marini}, \citenamefont {Verstraete},\ and\ \citenamefont
  {Gonze}}]{ponce2015}%
  \BibitemOpen
  \bibfield  {author} {\bibinfo {author} {\bibfnamefont {S.}~\bibnamefont
  {Poncé}}, \bibinfo {author} {\bibfnamefont {Y.}~\bibnamefont {Gillet}},
  \bibinfo {author} {\bibfnamefont {J.}~\bibnamefont {Laflamme~Janssen}},
  \bibinfo {author} {\bibfnamefont {A.}~\bibnamefont {Marini}}, \bibinfo
  {author} {\bibfnamefont {M.}~\bibnamefont {Verstraete}},\ and\ \bibinfo
  {author} {\bibfnamefont {X.}~\bibnamefont {Gonze}},\ }\bibfield  {title}
  {\bibinfo {title} {Temperature dependence of the electronic structure of
  semiconductors and insulators},\ }\href {https://doi.org/10.1063/1.4927081}
  {\bibfield  {journal} {\bibinfo  {journal} {The Journal of Chemical Physics}\
  }\textbf {\bibinfo {volume} {143}},\ \bibinfo {pages} {102813} (\bibinfo
  {year} {2015})}\BibitemShut {NoStop}%
\bibitem [{\citenamefont {Marini}\ \emph {et~al.}(2015)\citenamefont {Marini},
  \citenamefont {Ponc\'e},\ and\ \citenamefont {Gonze}}]{marini2015}%
  \BibitemOpen
  \bibfield  {author} {\bibinfo {author} {\bibfnamefont {A.}~\bibnamefont
  {Marini}}, \bibinfo {author} {\bibfnamefont {S.}~\bibnamefont {Ponc\'e}},\
  and\ \bibinfo {author} {\bibfnamefont {X.}~\bibnamefont {Gonze}},\ }\bibfield
   {title} {\bibinfo {title} {Many-body perturbation theory approach to the
  electron-phonon interaction with density-functional theory as a starting
  point},\ }\href {https://doi.org/10.1103/PhysRevB.91.224310} {\bibfield
  {journal} {\bibinfo  {journal} {Phys. Rev. B}\ }\textbf {\bibinfo {volume}
  {91}},\ \bibinfo {pages} {224310} (\bibinfo {year} {2015})}\BibitemShut
  {NoStop}%
\bibitem [{\citenamefont {Allen}\ and\ \citenamefont
  {Cardona}(1981)}]{allen1981}%
  \BibitemOpen
  \bibfield  {author} {\bibinfo {author} {\bibfnamefont {P.~B.}\ \bibnamefont
  {Allen}}\ and\ \bibinfo {author} {\bibfnamefont {M.}~\bibnamefont
  {Cardona}},\ }\bibfield  {title} {\bibinfo {title} {Theory of the temperature
  dependence of the direct gap of germanium},\ }\href
  {https://doi.org/10.1103/PhysRevB.23.1495} {\bibfield  {journal} {\bibinfo
  {journal} {Phys. Rev. B}\ }\textbf {\bibinfo {volume} {23}},\ \bibinfo
  {pages} {1495} (\bibinfo {year} {1981})}\BibitemShut {NoStop}%
\bibitem [{\citenamefont {Allen}\ and\ \citenamefont
  {Cardona}(1983)}]{allen1983}%
  \BibitemOpen
  \bibfield  {author} {\bibinfo {author} {\bibfnamefont {P.~B.}\ \bibnamefont
  {Allen}}\ and\ \bibinfo {author} {\bibfnamefont {M.}~\bibnamefont
  {Cardona}},\ }\bibfield  {title} {\bibinfo {title} {Temperature dependence of
  the direct gap of {Si and Ge}},\ }\href
  {https://doi.org/10.1103/PhysRevB.27.4760} {\bibfield  {journal} {\bibinfo
  {journal} {Phys. Rev. B}\ }\textbf {\bibinfo {volume} {27}},\ \bibinfo
  {pages} {4760} (\bibinfo {year} {1983})}\BibitemShut {NoStop}%
\bibitem [{\citenamefont {Allen}\ and\ \citenamefont
  {Heine}(1976)}]{Allen_1976}%
  \BibitemOpen
  \bibfield  {author} {\bibinfo {author} {\bibfnamefont {P.~B.}\ \bibnamefont
  {Allen}}\ and\ \bibinfo {author} {\bibfnamefont {V.}~\bibnamefont {Heine}},\
  }\bibfield  {title} {\bibinfo {title} {Theory of the temperature dependence
  of electronic band structures},\ }\href
  {https://doi.org/10.1088/0022-3719/9/12/013} {\bibfield  {journal} {\bibinfo
  {journal} {Journal of Physics C: Solid State Physics}\ }\textbf {\bibinfo
  {volume} {9}},\ \bibinfo {pages} {2305} (\bibinfo {year} {1976})}\BibitemShut
  {NoStop}%
\bibitem [{\citenamefont {Lihm}\ and\ \citenamefont {Park}(2021)}]{lihm2021}%
  \BibitemOpen
  \bibfield  {author} {\bibinfo {author} {\bibfnamefont {J.-M.}\ \bibnamefont
  {Lihm}}\ and\ \bibinfo {author} {\bibfnamefont {C.-H.}\ \bibnamefont
  {Park}},\ }\bibfield  {title} {\bibinfo {title} {Wannier function
  perturbation theory: Localized representation and interpolation of wave
  function perturbation},\ }\href {https://doi.org/10.1103/PhysRevX.11.041053}
  {\bibfield  {journal} {\bibinfo  {journal} {Phys. Rev. X}\ }\textbf {\bibinfo
  {volume} {11}},\ \bibinfo {pages} {041053} (\bibinfo {year}
  {2021})}\BibitemShut {NoStop}%
\bibitem [{\citenamefont {Ponc\'e}\ \emph {et~al.}(2014)\citenamefont
  {Ponc\'e}, \citenamefont {Antonius}, \citenamefont {Gillet}, \citenamefont
  {Boulanger}, \citenamefont {Laflamme~Janssen}, \citenamefont {Marini},
  \citenamefont {C\^ot\'e},\ and\ \citenamefont {Gonze}}]{ponce2014}%
  \BibitemOpen
  \bibfield  {author} {\bibinfo {author} {\bibfnamefont {S.}~\bibnamefont
  {Ponc\'e}}, \bibinfo {author} {\bibfnamefont {G.}~\bibnamefont {Antonius}},
  \bibinfo {author} {\bibfnamefont {Y.}~\bibnamefont {Gillet}}, \bibinfo
  {author} {\bibfnamefont {P.}~\bibnamefont {Boulanger}}, \bibinfo {author}
  {\bibfnamefont {J.}~\bibnamefont {Laflamme~Janssen}}, \bibinfo {author}
  {\bibfnamefont {A.}~\bibnamefont {Marini}}, \bibinfo {author} {\bibfnamefont
  {M.}~\bibnamefont {C\^ot\'e}},\ and\ \bibinfo {author} {\bibfnamefont
  {X.}~\bibnamefont {Gonze}},\ }\bibfield  {title} {\bibinfo {title}
  {Temperature dependence of electronic eigenenergies in the adiabatic harmonic
  approximation},\ }\href {https://doi.org/10.1103/PhysRevB.90.214304}
  {\bibfield  {journal} {\bibinfo  {journal} {Phys. Rev. B}\ }\textbf {\bibinfo
  {volume} {90}},\ \bibinfo {pages} {214304} (\bibinfo {year}
  {2014})}\BibitemShut {NoStop}%
\bibitem [{\citenamefont {Tomczak}\ \emph {et~al.}(2010)\citenamefont
  {Tomczak}, \citenamefont {Haule}, \citenamefont {Miyake}, \citenamefont
  {Georges},\ and\ \citenamefont {Kotliar}}]{tomczak2010}%
  \BibitemOpen
  \bibfield  {author} {\bibinfo {author} {\bibfnamefont {J.~M.}\ \bibnamefont
  {Tomczak}}, \bibinfo {author} {\bibfnamefont {K.}~\bibnamefont {Haule}},
  \bibinfo {author} {\bibfnamefont {T.}~\bibnamefont {Miyake}}, \bibinfo
  {author} {\bibfnamefont {A.}~\bibnamefont {Georges}},\ and\ \bibinfo {author}
  {\bibfnamefont {G.}~\bibnamefont {Kotliar}},\ }\bibfield  {title} {\bibinfo
  {title} {Thermopower of correlated semiconductors: Application to
  {${\text{FeAs}}_{2}$ and ${\text{FeSb}}_{2}$}},\ }\href
  {https://doi.org/10.1103/PhysRevB.82.085104} {\bibfield  {journal} {\bibinfo
  {journal} {Phys. Rev. B}\ }\textbf {\bibinfo {volume} {82}},\ \bibinfo
  {pages} {085104} (\bibinfo {year} {2010})}\BibitemShut {NoStop}%
\bibitem [{\citenamefont {Pickem}\ \emph {et~al.}(2022)\citenamefont {Pickem},
  \citenamefont {Maggio},\ and\ \citenamefont {Tomczak}}]{pickem2022}%
  \BibitemOpen
  \bibfield  {author} {\bibinfo {author} {\bibfnamefont {M.}~\bibnamefont
  {Pickem}}, \bibinfo {author} {\bibfnamefont {E.}~\bibnamefont {Maggio}},\
  and\ \bibinfo {author} {\bibfnamefont {J.~M.}\ \bibnamefont {Tomczak}},\
  }\bibfield  {title} {\bibinfo {title} {Prototypical many-body signatures in
  transport properties of semiconductors},\ }\href
  {https://doi.org/10.1103/PhysRevB.105.085139} {\bibfield  {journal} {\bibinfo
   {journal} {Phys. Rev. B}\ }\textbf {\bibinfo {volume} {105}},\ \bibinfo
  {pages} {085139} (\bibinfo {year} {2022})}\BibitemShut {NoStop}%
\bibitem [{\citenamefont {Rafailov}\ \emph {et~al.}(2002)\citenamefont
  {Rafailov}, \citenamefont {Dworzak},\ and\ \citenamefont
  {Thomsen}}]{rafailov2002}%
  \BibitemOpen
  \bibfield  {author} {\bibinfo {author} {\bibfnamefont {P.}~\bibnamefont
  {Rafailov}}, \bibinfo {author} {\bibfnamefont {M.}~\bibnamefont {Dworzak}},\
  and\ \bibinfo {author} {\bibfnamefont {C.}~\bibnamefont {Thomsen}},\
  }\bibfield  {title} {\bibinfo {title} {Luminescence and {R}aman spectroscopy
  on {${\mathrm{Mg}}{\mathrm{B}}_{2}$}},\ }\href
  {https://doi.org/https://doi.org/10.1016/S0038-1098(02)00158-8} {\bibfield
  {journal} {\bibinfo  {journal} {Solid State Communications}\ }\textbf
  {\bibinfo {volume} {122}},\ \bibinfo {pages} {455} (\bibinfo {year}
  {2002})}\BibitemShut {NoStop}%
\bibitem [{\citenamefont {d'Astuto}\ \emph {et~al.}(2007)\citenamefont
  {d'Astuto}, \citenamefont {Calandra}, \citenamefont {Reich}, \citenamefont
  {Shukla}, \citenamefont {Lazzeri}, \citenamefont {Mauri}, \citenamefont
  {Karpinski}, \citenamefont {Zhigadlo}, \citenamefont {Bossak},\ and\
  \citenamefont {Krisch}}]{astuto2007}%
  \BibitemOpen
  \bibfield  {author} {\bibinfo {author} {\bibfnamefont {M.}~\bibnamefont
  {d'Astuto}}, \bibinfo {author} {\bibfnamefont {M.}~\bibnamefont {Calandra}},
  \bibinfo {author} {\bibfnamefont {S.}~\bibnamefont {Reich}}, \bibinfo
  {author} {\bibfnamefont {A.}~\bibnamefont {Shukla}}, \bibinfo {author}
  {\bibfnamefont {M.}~\bibnamefont {Lazzeri}}, \bibinfo {author} {\bibfnamefont
  {F.}~\bibnamefont {Mauri}}, \bibinfo {author} {\bibfnamefont
  {J.}~\bibnamefont {Karpinski}}, \bibinfo {author} {\bibfnamefont {N.~D.}\
  \bibnamefont {Zhigadlo}}, \bibinfo {author} {\bibfnamefont {A.}~\bibnamefont
  {Bossak}},\ and\ \bibinfo {author} {\bibfnamefont {M.}~\bibnamefont
  {Krisch}},\ }\bibfield  {title} {\bibinfo {title} {Weak anharmonic effects in
  {$\mathrm{Mg}{\mathrm{B}}_{2}$}: A comparative inelastic {X}-ray scattering
  and {R}aman study},\ }\href {https://doi.org/10.1103/PhysRevB.75.174508}
  {\bibfield  {journal} {\bibinfo  {journal} {Phys. Rev. B}\ }\textbf {\bibinfo
  {volume} {75}},\ \bibinfo {pages} {174508} (\bibinfo {year}
  {2007})}\BibitemShut {NoStop}%
\bibitem [{\citenamefont {Baron}\ \emph {et~al.}(2004)\citenamefont {Baron},
  \citenamefont {Uchiyama}, \citenamefont {Tanaka}, \citenamefont {Tsutsui},
  \citenamefont {Ishikawa}, \citenamefont {Lee}, \citenamefont {Heid},
  \citenamefont {Bohnen}, \citenamefont {Tajima},\ and\ \citenamefont
  {Ishikawa}}]{baron2004}%
  \BibitemOpen
  \bibfield  {author} {\bibinfo {author} {\bibfnamefont {A.~Q.~R.}\
  \bibnamefont {Baron}}, \bibinfo {author} {\bibfnamefont {H.}~\bibnamefont
  {Uchiyama}}, \bibinfo {author} {\bibfnamefont {Y.}~\bibnamefont {Tanaka}},
  \bibinfo {author} {\bibfnamefont {S.}~\bibnamefont {Tsutsui}}, \bibinfo
  {author} {\bibfnamefont {D.}~\bibnamefont {Ishikawa}}, \bibinfo {author}
  {\bibfnamefont {S.}~\bibnamefont {Lee}}, \bibinfo {author} {\bibfnamefont
  {R.}~\bibnamefont {Heid}}, \bibinfo {author} {\bibfnamefont {K.-P.}\
  \bibnamefont {Bohnen}}, \bibinfo {author} {\bibfnamefont {S.}~\bibnamefont
  {Tajima}},\ and\ \bibinfo {author} {\bibfnamefont {T.}~\bibnamefont
  {Ishikawa}},\ }\bibfield  {title} {\bibinfo {title} {Kohn anomaly in
  {${\mathrm{MgB}}_{2}$} by inelastic {X}-ray scattering},\ }\href
  {https://doi.org/10.1103/PhysRevLett.92.197004} {\bibfield  {journal}
  {\bibinfo  {journal} {Phys. Rev. Lett.}\ }\textbf {\bibinfo {volume} {92}},\
  \bibinfo {pages} {197004} (\bibinfo {year} {2004})}\BibitemShut {NoStop}%
\bibitem [{\citenamefont {Di~Castro}\ \emph {et~al.}(2006)\citenamefont
  {Di~Castro}, \citenamefont {Ortolani}, \citenamefont {Cappelluti},
  \citenamefont {Schade}, \citenamefont {Zhigadlo},\ and\ \citenamefont
  {Karpinski}}]{dicastro2006}%
  \BibitemOpen
  \bibfield  {author} {\bibinfo {author} {\bibfnamefont {D.}~\bibnamefont
  {Di~Castro}}, \bibinfo {author} {\bibfnamefont {M.}~\bibnamefont {Ortolani}},
  \bibinfo {author} {\bibfnamefont {E.}~\bibnamefont {Cappelluti}}, \bibinfo
  {author} {\bibfnamefont {U.}~\bibnamefont {Schade}}, \bibinfo {author}
  {\bibfnamefont {N.~D.}\ \bibnamefont {Zhigadlo}},\ and\ \bibinfo {author}
  {\bibfnamefont {J.}~\bibnamefont {Karpinski}},\ }\bibfield  {title} {\bibinfo
  {title} {Infrared properties of
  {${\mathrm{Mg}}_{1\ensuremath{-}x}{\mathrm{Al}}_{x}{({\mathrm{B}}_{1\ensuremath{-}y}{\mathrm{C}}_{y})}_{2}$}
  single crystals in the normal and superconducting state},\ }\href
  {https://doi.org/10.1103/PhysRevB.73.174509} {\bibfield  {journal} {\bibinfo
  {journal} {Phys. Rev. B}\ }\textbf {\bibinfo {volume} {73}},\ \bibinfo
  {pages} {174509} (\bibinfo {year} {2006})}\BibitemShut {NoStop}%
\bibitem [{\citenamefont {Cappelluti}\ and\ \citenamefont
  {Pietronero}(1996)}]{pietronero96}%
  \BibitemOpen
  \bibfield  {author} {\bibinfo {author} {\bibfnamefont {E.}~\bibnamefont
  {Cappelluti}}\ and\ \bibinfo {author} {\bibfnamefont {L.}~\bibnamefont
  {Pietronero}},\ }\bibfield  {title} {\bibinfo {title} {Nonadiabatic
  superconductivity: The role of van {H}ove singularities},\ }\href
  {https://doi.org/10.1103/PhysRevB.53.932} {\bibfield  {journal} {\bibinfo
  {journal} {Phys. Rev. B}\ }\textbf {\bibinfo {volume} {53}},\ \bibinfo
  {pages} {932} (\bibinfo {year} {1996})}\BibitemShut {NoStop}%
\bibitem [{\citenamefont {Itai}(1992)}]{itai92}%
  \BibitemOpen
  \bibfield  {author} {\bibinfo {author} {\bibfnamefont {K.}~\bibnamefont
  {Itai}},\ }\bibfield  {title} {\bibinfo {title} {Theory of {R}aman scattering
  in coupled electron-phonon systems},\ }\href
  {https://doi.org/10.1103/PhysRevB.45.707} {\bibfield  {journal} {\bibinfo
  {journal} {Phys. Rev. B}\ }\textbf {\bibinfo {volume} {45}},\ \bibinfo
  {pages} {707} (\bibinfo {year} {1992})}\BibitemShut {NoStop}%
\bibitem [{\citenamefont {Allen}\ and\ \citenamefont
  {Silberglitt}(1974{\natexlab{b}})}]{allen1974}%
  \BibitemOpen
  \bibfield  {author} {\bibinfo {author} {\bibfnamefont {P.~B.}\ \bibnamefont
  {Allen}}\ and\ \bibinfo {author} {\bibfnamefont {R.}~\bibnamefont
  {Silberglitt}},\ }\bibfield  {title} {\bibinfo {title} {Some effects of
  phonon dynamics on electron lifetime, mass renormalization, and
  superconducting transition temperature},\ }\href
  {https://doi.org/10.1103/PhysRevB.9.4733} {\bibfield  {journal} {\bibinfo
  {journal} {Phys. Rev. B}\ }\textbf {\bibinfo {volume} {9}},\ \bibinfo {pages}
  {4733} (\bibinfo {year} {1974}{\natexlab{b}})}\BibitemShut {NoStop}%
\bibitem [{\citenamefont {Allen}(1972)}]{allen1972}%
  \BibitemOpen
  \bibfield  {author} {\bibinfo {author} {\bibfnamefont {P.~B.}\ \bibnamefont
  {Allen}},\ }\bibfield  {title} {\bibinfo {title} {Neutron spectroscopy of
  superconductors},\ }\href {https://doi.org/10.1103/PhysRevB.6.2577}
  {\bibfield  {journal} {\bibinfo  {journal} {Phys. Rev. B}\ }\textbf {\bibinfo
  {volume} {6}},\ \bibinfo {pages} {2577} (\bibinfo {year} {1972})}\BibitemShut
  {NoStop}%
\bibitem [{\citenamefont {Pickett}(1980)}]{pickett80}%
  \BibitemOpen
  \bibfield  {author} {\bibinfo {author} {\bibfnamefont {W.~E.}\ \bibnamefont
  {Pickett}},\ }\bibfield  {title} {\bibinfo {title} {Effect of a varying
  density of states on superconductivity},\ }\href
  {https://doi.org/10.1103/PhysRevB.21.3897} {\bibfield  {journal} {\bibinfo
  {journal} {Phys. Rev. B}\ }\textbf {\bibinfo {volume} {21}},\ \bibinfo
  {pages} {3897} (\bibinfo {year} {1980})}\BibitemShut {NoStop}%
\bibitem [{\citenamefont {Appel}(1968)}]{appel68}%
  \BibitemOpen
  \bibfield  {author} {\bibinfo {author} {\bibfnamefont {J.}~\bibnamefont
  {Appel}},\ }\bibfield  {title} {\bibinfo {title} {Role of thermal phonons in
  high-temperature superconductivity},\ }\href
  {https://doi.org/10.1103/PhysRevLett.21.1164} {\bibfield  {journal} {\bibinfo
   {journal} {Phys. Rev. Lett.}\ }\textbf {\bibinfo {volume} {21}},\ \bibinfo
  {pages} {1164} (\bibinfo {year} {1968})}\BibitemShut {NoStop}%
\bibitem [{\citenamefont {Mishra}\ and\ \citenamefont
  {Margine}(2026)}]{mishra2026}%
  \BibitemOpen
  \bibfield  {author} {\bibinfo {author} {\bibfnamefont {S.~B.}\ \bibnamefont
  {Mishra}}\ and\ \bibinfo {author} {\bibfnamefont {E.~R.}\ \bibnamefont
  {Margine}},\ }\bibfield  {title} {\bibinfo {title} {Nonadiabatic and
  anharmonic effects in high-pressure {H$_3$S} and {D$_3$S} superconductors},\
  }\href {https://doi.org/https://doi.org/10.1002/andp.202500553} {\bibfield
  {journal} {\bibinfo  {journal} {Annalen der Physik}\ }\textbf {\bibinfo
  {volume} {538}},\ \bibinfo {pages} {e00553} (\bibinfo {year}
  {2026})}\BibitemShut {NoStop}%
\bibitem [{\citenamefont {Mishra}\ \emph {et~al.}(2025)\citenamefont {Mishra},
  \citenamefont {Mori},\ and\ \citenamefont {Margine}}]{Mishra2025}%
  \BibitemOpen
  \bibfield  {author} {\bibinfo {author} {\bibfnamefont {S.~B.}\ \bibnamefont
  {Mishra}}, \bibinfo {author} {\bibfnamefont {H.}~\bibnamefont {Mori}},\ and\
  \bibinfo {author} {\bibfnamefont {E.~R.}\ \bibnamefont {Margine}},\
  }\bibfield  {title} {\bibinfo {title} {Electron–phonon vertex correction
  effect in superconducting {H$_3$S}},\ }\bibfield  {journal} {\bibinfo
  {journal} {npj Computational Materials}\ }\textbf {\bibinfo {volume} {11}},\
  \href {https://doi.org/10.1038/s41524-025-01818-9}
  {10.1038/s41524-025-01818-9} (\bibinfo {year} {2025})\BibitemShut {NoStop}%
\bibitem [{\citenamefont {Zhang}\ \emph {et~al.}(2005)\citenamefont {Zhang},
  \citenamefont {Louie},\ and\ \citenamefont {Cohen}}]{zhang2005}%
  \BibitemOpen
  \bibfield  {author} {\bibinfo {author} {\bibfnamefont {P.}~\bibnamefont
  {Zhang}}, \bibinfo {author} {\bibfnamefont {S.~G.}\ \bibnamefont {Louie}},\
  and\ \bibinfo {author} {\bibfnamefont {M.~L.}\ \bibnamefont {Cohen}},\
  }\bibfield  {title} {\bibinfo {title} {Nonlocal screening, electron-phonon
  coupling, and phonon renormalization in metals},\ }\href
  {https://doi.org/10.1103/PhysRevLett.94.225502} {\bibfield  {journal}
  {\bibinfo  {journal} {Phys. Rev. Lett.}\ }\textbf {\bibinfo {volume} {94}},\
  \bibinfo {pages} {225502} (\bibinfo {year} {2005})}\BibitemShut {NoStop}%
\bibitem [{\citenamefont {Calandra}\ \emph {et~al.}(2007)\citenamefont
  {Calandra}, \citenamefont {Lazzeri},\ and\ \citenamefont
  {Mauri}}]{calandra2007}%
  \BibitemOpen
  \bibfield  {author} {\bibinfo {author} {\bibfnamefont {M.}~\bibnamefont
  {Calandra}}, \bibinfo {author} {\bibfnamefont {M.}~\bibnamefont {Lazzeri}},\
  and\ \bibinfo {author} {\bibfnamefont {F.}~\bibnamefont {Mauri}},\ }\bibfield
   {title} {\bibinfo {title} {Anharmonic and non-adiabatic effects in
  {Mg}{B}$_{2}$: Implications for the isotope effect and interpretation of
  {R}aman spectra},\ }\href
  {https://doi.org/https://doi.org/10.1016/j.physc.2007.01.021} {\bibfield
  {journal} {\bibinfo  {journal} {Physica C: Superconductivity}\ }\textbf
  {\bibinfo {volume} {456}},\ \bibinfo {pages} {38} (\bibinfo {year} {2007})},\
  \bibinfo {note} {recent Advances in ${\mathrm{Mg}}{\mathrm{B}}_{2}$
  Research}\BibitemShut {NoStop}%
\bibitem [{\citenamefont {Mor}\ \emph {et~al.}(2026)\citenamefont {Mor},
  \citenamefont {Boschini}, \citenamefont {Razzoli}, \citenamefont {Zonno},
  \citenamefont {Michiardi}, \citenamefont {Levy}, \citenamefont {Zhigadlo},
  \citenamefont {Canfield}, \citenamefont {Cerullo}, \citenamefont
  {Damascelli}, \citenamefont {Giannetti},\ and\ \citenamefont
  {Conte}}]{mor2025}%
  \BibitemOpen
  \bibfield  {author} {\bibinfo {author} {\bibfnamefont {S.}~\bibnamefont
  {Mor}}, \bibinfo {author} {\bibfnamefont {F.}~\bibnamefont {Boschini}},
  \bibinfo {author} {\bibfnamefont {E.}~\bibnamefont {Razzoli}}, \bibinfo
  {author} {\bibfnamefont {M.}~\bibnamefont {Zonno}}, \bibinfo {author}
  {\bibfnamefont {M.}~\bibnamefont {Michiardi}}, \bibinfo {author}
  {\bibfnamefont {G.}~\bibnamefont {Levy}}, \bibinfo {author} {\bibfnamefont
  {N.~D.}\ \bibnamefont {Zhigadlo}}, \bibinfo {author} {\bibfnamefont {P.~C.}\
  \bibnamefont {Canfield}}, \bibinfo {author} {\bibfnamefont {G.}~\bibnamefont
  {Cerullo}}, \bibinfo {author} {\bibfnamefont {A.}~\bibnamefont {Damascelli}},
  \bibinfo {author} {\bibfnamefont {C.}~\bibnamefont {Giannetti}},\ and\
  \bibinfo {author} {\bibfnamefont {S.~D.}\ \bibnamefont {Conte}},\ }\bibfield
  {title} {\bibinfo {title} {Selective electron-phonon coupling strength from
  nonequilibrium optical spectroscopy: The case of {${\mathrm{MgB}}_{2}$}},\
  }\href {https://doi.org/10.1103/25zv-zv94} {\bibfield  {journal} {\bibinfo
  {journal} {Phys. Rev. B}\ }\textbf {\bibinfo {volume} {113}},\ \bibinfo
  {pages} {064514} (\bibinfo {year} {2026})}\BibitemShut {NoStop}%
\bibitem [{\citenamefont {Adriano}\ \emph {et~al.}(2025)\citenamefont
  {Adriano}, \citenamefont {Xu}, \citenamefont {Huyan}, \citenamefont
  {Pakuszewski}, \citenamefont {Machado}, \citenamefont {Schrunk},
  \citenamefont {Bud'ko}, \citenamefont {Ribeiro}, \citenamefont {Canfield},\
  and\ \citenamefont {Kaminski}}]{Adriano2025}%
  \BibitemOpen
  \bibfield  {author} {\bibinfo {author} {\bibfnamefont {C.}~\bibnamefont
  {Adriano}}, \bibinfo {author} {\bibfnamefont {M.}~\bibnamefont {Xu}},
  \bibinfo {author} {\bibfnamefont {S.}~\bibnamefont {Huyan}}, \bibinfo
  {author} {\bibfnamefont {K.~R.}\ \bibnamefont {Pakuszewski}}, \bibinfo
  {author} {\bibfnamefont {A.~P.}\ \bibnamefont {Machado}}, \bibinfo {author}
  {\bibfnamefont {B.}~\bibnamefont {Schrunk}}, \bibinfo {author} {\bibfnamefont
  {S.~L.}\ \bibnamefont {Bud'ko}}, \bibinfo {author} {\bibfnamefont {R.~A.}\
  \bibnamefont {Ribeiro}}, \bibinfo {author} {\bibfnamefont {P.~C.}\
  \bibnamefont {Canfield}},\ and\ \bibinfo {author} {\bibfnamefont
  {A.}~\bibnamefont {Kaminski}},\ }\bibfield  {title} {\bibinfo {title} {Tuning
  the electronic properties of {${\mathrm{Mg}}{\mathrm{B}}_{2}$} by
  substitution with {Mn and C}},\ }\href
  {https://doi.org/10.1088/1361-648X/adeb27} {\bibfield  {journal} {\bibinfo
  {journal} {Journal of Physics: Condensed Matter}\ }\textbf {\bibinfo {volume}
  {37}},\ \bibinfo {pages} {305502} (\bibinfo {year} {2025})}\BibitemShut
  {NoStop}%
\bibitem [{\citenamefont {Girotto~Erhardt}\ and\ \citenamefont
  {Poncé}(2026)}]{dataset_}%
  \BibitemOpen
  \bibfield  {author} {\bibinfo {author} {\bibfnamefont {N.}~\bibnamefont
  {Girotto~Erhardt}}\ and\ \bibinfo {author} {\bibfnamefont {S.}~\bibnamefont
  {Poncé}},\ }\href
  {https://archive.materialscloud.org/records/rzy5z-c9172?preview=1&token=eyJhbGciOiJIUzUxMiJ9.eyJpZCI6ImUwNjQ4MTM2LTBhOTgtNDQ3Yi1hM2MwLTg5MDZmNjNiM2RkZiIsImRhdGEiOnt9LCJyYW5kb20iOiJjMzBhMGI1YmMxZmM1NGEwNGNlNDg4MWNkNzNkYzVhNyJ9.UqSlpKh7reA4DJ3bmiz0cZE7nREPTnIJQQjyirlFrAXUMMKWZnKT8c6T1PT4GrbtiSkYn4hh3TqugVrUUJausA}
  {\bibinfo {title} {Higher-order nonadiabaticity governs the temperature
  dependence of the phonon spectrum}} (\bibinfo {year} {2026})\BibitemShut
  {NoStop}%
\end{thebibliography}%


\providecommand{\noopsort}[1]{}\providecommand{\singleletter}[1]{#1}%
\begin{thebibliography}{17}%
\makeatletter
\providecommand \@ifxundefined [1]{%
 \@ifx{#1\undefined}
}%
\providecommand \@ifnum [1]{%
 \ifnum #1\expandafter \@firstoftwo
 \else \expandafter \@secondoftwo
 \fi
}%
\providecommand \@ifx [1]{%
 \ifx #1\expandafter \@firstoftwo
 \else \expandafter \@secondoftwo
 \fi
}%
\providecommand \natexlab [1]{#1}%
\providecommand \enquote  [1]{``#1''}%
\providecommand \bibnamefont  [1]{#1}%
\providecommand \bibfnamefont [1]{#1}%
\providecommand \citenamefont [1]{#1}%
\providecommand \href@noop [0]{\@secondoftwo}%
\providecommand \href [0]{\begingroup \@sanitize@url \@href}%
\providecommand \@href[1]{\@@startlink{#1}\@@href}%
\providecommand \@@href[1]{\endgroup#1\@@endlink}%
\providecommand \@sanitize@url [0]{\catcode `\\12\catcode `\$12\catcode
  `\&12\catcode `\#12\catcode `\^12\catcode `\_12\catcode `\%12\relax}%
\providecommand \@@startlink[1]{}%
\providecommand \@@endlink[0]{}%
\providecommand \url  [0]{\begingroup\@sanitize@url \@url }%
\providecommand \@url [1]{\endgroup\@href {#1}{\urlprefix }}%
\providecommand \urlprefix  [0]{URL }%
\providecommand \Eprint [0]{\href }%
\providecommand \doibase [0]{https://doi.org/}%
\providecommand \selectlanguage [0]{\@gobble}%
\providecommand \bibinfo  [0]{\@secondoftwo}%
\providecommand \bibfield  [0]{\@secondoftwo}%
\providecommand \translation [1]{[#1]}%
\providecommand \BibitemOpen [0]{}%
\providecommand \bibitemStop [0]{}%
\providecommand \bibitemNoStop [0]{.\EOS\space}%
\providecommand \EOS [0]{\spacefactor3000\relax}%
\providecommand \BibitemShut  [1]{\csname bibitem#1\endcsname}%
\let\auto@bib@innerbib\@empty
\bibitem [{\citenamefont {Mahan}(2013)}]{mahan}%
  \BibitemOpen
  \bibfield  {author} {\bibinfo {author} {\bibfnamefont {G.~D.}\ \bibnamefont
  {Mahan}},\ }\bibinfo {title} {Physics of solids and liquids},\ in\ \href
  {https://doi.org/https://doi.org/10.1007/978-1-4757-5714-9} {\emph {\bibinfo
  {booktitle} {Many-Particle Physics}}}\ (\bibinfo  {publisher} {Springer New
  York, NY},\ \bibinfo {year} {2013})\ p.\ \bibinfo {pages} {785}\BibitemShut
  {NoStop}%
\bibitem [{\citenamefont {Berges}\ \emph {et~al.}(2023)\citenamefont {Berges},
  \citenamefont {Girotto}, \citenamefont {Wehling}, \citenamefont {Marzari},\
  and\ \citenamefont {Ponc\'e}}]{berges2023}%
  \BibitemOpen
  \bibfield  {author} {\bibinfo {author} {\bibfnamefont {J.}~\bibnamefont
  {Berges}}, \bibinfo {author} {\bibfnamefont {N.}~\bibnamefont {Girotto}},
  \bibinfo {author} {\bibfnamefont {T.}~\bibnamefont {Wehling}}, \bibinfo
  {author} {\bibfnamefont {N.}~\bibnamefont {Marzari}},\ and\ \bibinfo {author}
  {\bibfnamefont {S.}~\bibnamefont {Ponc\'e}},\ }\bibfield  {title} {\bibinfo
  {title} {Phonon self-energy corrections: To screen, or not to screen},\
  }\href {https://doi.org/10.1103/PhysRevX.13.041009} {\bibfield  {journal}
  {\bibinfo  {journal} {Phys. Rev. X}\ }\textbf {\bibinfo {volume} {13}},\
  \bibinfo {pages} {041009} (\bibinfo {year} {2023})}\BibitemShut {NoStop}%
\bibitem [{\citenamefont {Allen}\ and\ \citenamefont
  {Cardona}(1981)}]{allen1981}%
  \BibitemOpen
  \bibfield  {author} {\bibinfo {author} {\bibfnamefont {P.~B.}\ \bibnamefont
  {Allen}}\ and\ \bibinfo {author} {\bibfnamefont {M.}~\bibnamefont
  {Cardona}},\ }\bibfield  {title} {\bibinfo {title} {Theory of the temperature
  dependence of the direct gap of germanium},\ }\href
  {https://doi.org/10.1103/PhysRevB.23.1495} {\bibfield  {journal} {\bibinfo
  {journal} {Phys. Rev. B}\ }\textbf {\bibinfo {volume} {23}},\ \bibinfo
  {pages} {1495} (\bibinfo {year} {1981})}\BibitemShut {NoStop}%
\bibitem [{\citenamefont {Allen}\ and\ \citenamefont
  {Cardona}(1983)}]{allen1983}%
  \BibitemOpen
  \bibfield  {author} {\bibinfo {author} {\bibfnamefont {P.~B.}\ \bibnamefont
  {Allen}}\ and\ \bibinfo {author} {\bibfnamefont {M.}~\bibnamefont
  {Cardona}},\ }\bibfield  {title} {\bibinfo {title} {Temperature dependence of
  the direct gap of {Si and Ge}},\ }\href
  {https://doi.org/10.1103/PhysRevB.27.4760} {\bibfield  {journal} {\bibinfo
  {journal} {Phys. Rev. B}\ }\textbf {\bibinfo {volume} {27}},\ \bibinfo
  {pages} {4760} (\bibinfo {year} {1983})}\BibitemShut {NoStop}%
\bibitem [{\citenamefont {Allen}\ and\ \citenamefont
  {Heine}(1976)}]{Allen_1976}%
  \BibitemOpen
  \bibfield  {author} {\bibinfo {author} {\bibfnamefont {P.~B.}\ \bibnamefont
  {Allen}}\ and\ \bibinfo {author} {\bibfnamefont {V.}~\bibnamefont {Heine}},\
  }\bibfield  {title} {\bibinfo {title} {Theory of the temperature dependence
  of electronic band structures},\ }\href
  {https://doi.org/10.1088/0022-3719/9/12/013} {\bibfield  {journal} {\bibinfo
  {journal} {Journal of Physics C: Solid State Physics}\ }\textbf {\bibinfo
  {volume} {9}},\ \bibinfo {pages} {2305} (\bibinfo {year} {1976})}\BibitemShut
  {NoStop}%
\bibitem [{\citenamefont {Lihm}\ and\ \citenamefont {Park}(2021)}]{lihm2021}%
  \BibitemOpen
  \bibfield  {author} {\bibinfo {author} {\bibfnamefont {J.-M.}\ \bibnamefont
  {Lihm}}\ and\ \bibinfo {author} {\bibfnamefont {C.-H.}\ \bibnamefont
  {Park}},\ }\bibfield  {title} {\bibinfo {title} {Wannier function
  perturbation theory: Localized representation and interpolation of wave
  function perturbation},\ }\href {https://doi.org/10.1103/PhysRevX.11.041053}
  {\bibfield  {journal} {\bibinfo  {journal} {Phys. Rev. X}\ }\textbf {\bibinfo
  {volume} {11}},\ \bibinfo {pages} {041053} (\bibinfo {year}
  {2021})}\BibitemShut {NoStop}%
\bibitem [{\citenamefont {Ponc\'e}\ \emph {et~al.}(2014)\citenamefont
  {Ponc\'e}, \citenamefont {Antonius}, \citenamefont {Gillet}, \citenamefont
  {Boulanger}, \citenamefont {Laflamme~Janssen}, \citenamefont {Marini},
  \citenamefont {C\^ot\'e},\ and\ \citenamefont {Gonze}}]{ponce2014}%
  \BibitemOpen
  \bibfield  {author} {\bibinfo {author} {\bibfnamefont {S.}~\bibnamefont
  {Ponc\'e}}, \bibinfo {author} {\bibfnamefont {G.}~\bibnamefont {Antonius}},
  \bibinfo {author} {\bibfnamefont {Y.}~\bibnamefont {Gillet}}, \bibinfo
  {author} {\bibfnamefont {P.}~\bibnamefont {Boulanger}}, \bibinfo {author}
  {\bibfnamefont {J.}~\bibnamefont {Laflamme~Janssen}}, \bibinfo {author}
  {\bibfnamefont {A.}~\bibnamefont {Marini}}, \bibinfo {author} {\bibfnamefont
  {M.}~\bibnamefont {C\^ot\'e}},\ and\ \bibinfo {author} {\bibfnamefont
  {X.}~\bibnamefont {Gonze}},\ }\bibfield  {title} {\bibinfo {title}
  {Temperature dependence of electronic eigenenergies in the adiabatic harmonic
  approximation},\ }\href {https://doi.org/10.1103/PhysRevB.90.214304}
  {\bibfield  {journal} {\bibinfo  {journal} {Phys. Rev. B}\ }\textbf {\bibinfo
  {volume} {90}},\ \bibinfo {pages} {214304} (\bibinfo {year}
  {2014})}\BibitemShut {NoStop}%
\bibitem [{\citenamefont {Park}(2025)}]{park2025}%
  \BibitemOpen
  \bibfield  {author} {\bibinfo {author} {\bibfnamefont {C.-H.}\ \bibnamefont
  {Park}},\ }\bibfield  {title} {\bibinfo {title} {Nonadiabatic phonon
  self-energy due to electrons with finite linewidths},\ }\href
  {https://doi.org/10.1103/6jq1-cbwq} {\bibfield  {journal} {\bibinfo
  {journal} {Phys. Rev. B}\ }\textbf {\bibinfo {volume} {112}},\ \bibinfo
  {pages} {104314} (\bibinfo {year} {2025})}\BibitemShut {NoStop}%
\bibitem [{\citenamefont {Lihm}\ \emph {et~al.}(2024)\citenamefont {Lihm},
  \citenamefont {Ponc\'e},\ and\ \citenamefont {Park}}]{lihm2024}%
  \BibitemOpen
  \bibfield  {author} {\bibinfo {author} {\bibfnamefont {J.-M.}\ \bibnamefont
  {Lihm}}, \bibinfo {author} {\bibfnamefont {S.}~\bibnamefont {Ponc\'e}},\ and\
  \bibinfo {author} {\bibfnamefont {C.-H.}\ \bibnamefont {Park}},\ }\bibfield
  {title} {\bibinfo {title} {Self-consistent electron lifetimes for
  electron-phonon scattering},\ }\href
  {https://doi.org/10.1103/PhysRevB.110.L121106} {\bibfield  {journal}
  {\bibinfo  {journal} {Phys. Rev. B}\ }\textbf {\bibinfo {volume} {110}},\
  \bibinfo {pages} {L121106} (\bibinfo {year} {2024})}\BibitemShut {NoStop}%
\bibitem [{\citenamefont {Poncé}\ \emph {et~al.}(2025)\citenamefont {Poncé},
  \citenamefont {Lihm},\ and\ \citenamefont {Park}}]{Ponce2025}%
  \BibitemOpen
  \bibfield  {author} {\bibinfo {author} {\bibfnamefont {S.}~\bibnamefont
  {Poncé}}, \bibinfo {author} {\bibfnamefont {J.-M.}\ \bibnamefont {Lihm}},\
  and\ \bibinfo {author} {\bibfnamefont {C.-H.}\ \bibnamefont {Park}},\
  }\bibfield  {title} {\bibinfo {title} {Verification and validation of
  zero-point electron-phonon renormalization of the bandgap, mass enhancement,
  and spectral functions},\ }\href {https://doi.org/10.1038/s41524-025-01587-5}
  {\bibfield  {journal} {\bibinfo  {journal} {npj Computational Materials}\
  }\textbf {\bibinfo {volume} {11}},\ \bibinfo {pages} {117} (\bibinfo {year}
  {2025})}\BibitemShut {NoStop}%
\bibitem [{\citenamefont {Giannozzi}\ \emph {et~al.}(2017)\citenamefont
  {Giannozzi}, \citenamefont {Andreussi}, \citenamefont {Brumme}, \citenamefont
  {Bunau}, \citenamefont {Buongiorno~Nardelli}, \citenamefont {Calandra},
  \citenamefont {Car}, \citenamefont {Cavazzoni}, \citenamefont {Ceresoli},
  \citenamefont {Cococcioni}, \citenamefont {Colonna}, \citenamefont
  {Carnimeo}, \citenamefont {Dal~Corso}, \citenamefont {de~Gironcoli},
  \citenamefont {Delugas}, \citenamefont {DiStasio}, \citenamefont {Ferretti},
  \citenamefont {Floris}, \citenamefont {Fratesi}, \citenamefont {Fugallo},
  \citenamefont {Gebauer}, \citenamefont {Gerstmann}, \citenamefont {Giustino},
  \citenamefont {Gorni}, \citenamefont {Jia}, \citenamefont {Kawamura},
  \citenamefont {Ko}, \citenamefont {Kokalj}, \citenamefont {Küçükbenli},
  \citenamefont {Lazzeri}, \citenamefont {Marsili}, \citenamefont {Marzari},
  \citenamefont {Mauri}, \citenamefont {Nguyen}, \citenamefont {Nguyen},
  \citenamefont {Otero-de-la Roza}, \citenamefont {Paulatto}, \citenamefont
  {Poncé}, \citenamefont {Rocca}, \citenamefont {Sabatini}, \citenamefont
  {Santra}, \citenamefont {Schlipf}, \citenamefont {Seitsonen}, \citenamefont
  {Smogunov}, \citenamefont {Timrov}, \citenamefont {Thonhauser}, \citenamefont
  {Umari}, \citenamefont {Vast}, \citenamefont {Wu},\ and\ \citenamefont
  {Baroni}}]{giannozzi2017qe}%
  \BibitemOpen
  \bibfield  {author} {\bibinfo {author} {\bibfnamefont {P.}~\bibnamefont
  {Giannozzi}}, \bibinfo {author} {\bibfnamefont {O.}~\bibnamefont
  {Andreussi}}, \bibinfo {author} {\bibfnamefont {T.}~\bibnamefont {Brumme}},
  \bibinfo {author} {\bibfnamefont {O.}~\bibnamefont {Bunau}}, \bibinfo
  {author} {\bibfnamefont {M.}~\bibnamefont {Buongiorno~Nardelli}}, \bibinfo
  {author} {\bibfnamefont {M.}~\bibnamefont {Calandra}}, \bibinfo {author}
  {\bibfnamefont {R.}~\bibnamefont {Car}}, \bibinfo {author} {\bibfnamefont
  {C.}~\bibnamefont {Cavazzoni}}, \bibinfo {author} {\bibfnamefont
  {D.}~\bibnamefont {Ceresoli}}, \bibinfo {author} {\bibfnamefont
  {M.}~\bibnamefont {Cococcioni}}, \bibinfo {author} {\bibfnamefont
  {N.}~\bibnamefont {Colonna}}, \bibinfo {author} {\bibfnamefont
  {I.}~\bibnamefont {Carnimeo}}, \bibinfo {author} {\bibfnamefont
  {A.}~\bibnamefont {Dal~Corso}}, \bibinfo {author} {\bibfnamefont
  {S.}~\bibnamefont {de~Gironcoli}}, \bibinfo {author} {\bibfnamefont
  {P.}~\bibnamefont {Delugas}}, \bibinfo {author} {\bibfnamefont {R.~A.}\
  \bibnamefont {DiStasio}}, \bibinfo {author} {\bibfnamefont {A.}~\bibnamefont
  {Ferretti}}, \bibinfo {author} {\bibfnamefont {A.}~\bibnamefont {Floris}},
  \bibinfo {author} {\bibfnamefont {G.}~\bibnamefont {Fratesi}}, \bibinfo
  {author} {\bibfnamefont {G.}~\bibnamefont {Fugallo}}, \bibinfo {author}
  {\bibfnamefont {R.}~\bibnamefont {Gebauer}}, \bibinfo {author} {\bibfnamefont
  {U.}~\bibnamefont {Gerstmann}}, \bibinfo {author} {\bibfnamefont
  {F.}~\bibnamefont {Giustino}}, \bibinfo {author} {\bibfnamefont
  {T.}~\bibnamefont {Gorni}}, \bibinfo {author} {\bibfnamefont
  {J.}~\bibnamefont {Jia}}, \bibinfo {author} {\bibfnamefont {M.}~\bibnamefont
  {Kawamura}}, \bibinfo {author} {\bibfnamefont {H.-Y.}\ \bibnamefont {Ko}},
  \bibinfo {author} {\bibfnamefont {A.}~\bibnamefont {Kokalj}}, \bibinfo
  {author} {\bibfnamefont {E.}~\bibnamefont {Küçükbenli}}, \bibinfo {author}
  {\bibfnamefont {M.}~\bibnamefont {Lazzeri}}, \bibinfo {author} {\bibfnamefont
  {M.}~\bibnamefont {Marsili}}, \bibinfo {author} {\bibfnamefont
  {N.}~\bibnamefont {Marzari}}, \bibinfo {author} {\bibfnamefont
  {F.}~\bibnamefont {Mauri}}, \bibinfo {author} {\bibfnamefont {N.~L.}\
  \bibnamefont {Nguyen}}, \bibinfo {author} {\bibfnamefont {H.-V.}\
  \bibnamefont {Nguyen}}, \bibinfo {author} {\bibfnamefont {A.}~\bibnamefont
  {Otero-de-la Roza}}, \bibinfo {author} {\bibfnamefont {L.}~\bibnamefont
  {Paulatto}}, \bibinfo {author} {\bibfnamefont {S.}~\bibnamefont {Poncé}},
  \bibinfo {author} {\bibfnamefont {D.}~\bibnamefont {Rocca}}, \bibinfo
  {author} {\bibfnamefont {R.}~\bibnamefont {Sabatini}}, \bibinfo {author}
  {\bibfnamefont {B.}~\bibnamefont {Santra}}, \bibinfo {author} {\bibfnamefont
  {M.}~\bibnamefont {Schlipf}}, \bibinfo {author} {\bibfnamefont {A.~P.}\
  \bibnamefont {Seitsonen}}, \bibinfo {author} {\bibfnamefont {A.}~\bibnamefont
  {Smogunov}}, \bibinfo {author} {\bibfnamefont {I.}~\bibnamefont {Timrov}},
  \bibinfo {author} {\bibfnamefont {T.}~\bibnamefont {Thonhauser}}, \bibinfo
  {author} {\bibfnamefont {P.}~\bibnamefont {Umari}}, \bibinfo {author}
  {\bibfnamefont {N.}~\bibnamefont {Vast}}, \bibinfo {author} {\bibfnamefont
  {X.}~\bibnamefont {Wu}},\ and\ \bibinfo {author} {\bibfnamefont
  {S.}~\bibnamefont {Baroni}},\ }\bibfield  {title} {\bibinfo {title} {Advanced
  capabilities for materials modelling with quantum espresso},\ }\href
  {https://doi.org/10.1088/1361-648X/aa8f79} {\bibfield  {journal} {\bibinfo
  {journal} {Journal of Physics: Condensed Matter}\ }\textbf {\bibinfo {volume}
  {29}},\ \bibinfo {pages} {465901} (\bibinfo {year} {2017})}\BibitemShut
  {NoStop}%
\bibitem [{\citenamefont {Hamann}(2013)}]{hamann13}%
  \BibitemOpen
  \bibfield  {author} {\bibinfo {author} {\bibfnamefont {D.~R.}\ \bibnamefont
  {Hamann}},\ }\bibfield  {title} {\bibinfo {title} {Optimized norm-conserving
  vanderbilt pseudopotentials},\ }\href
  {https://doi.org/10.1103/PhysRevB.88.085117} {\bibfield  {journal} {\bibinfo
  {journal} {Phys. Rev. B}\ }\textbf {\bibinfo {volume} {88}},\ \bibinfo
  {pages} {085117} (\bibinfo {year} {2013})}\BibitemShut {NoStop}%
\bibitem [{\citenamefont {{van Setten}}\ \emph {et~al.}(2018)\citenamefont
  {{van Setten}}, \citenamefont {Giantomassi}, \citenamefont {Bousquet},
  \citenamefont {Verstraete}, \citenamefont {Hamann}, \citenamefont {Gonze},\
  and\ \citenamefont {Rignanese}}]{pseudodojo}%
  \BibitemOpen
  \bibfield  {author} {\bibinfo {author} {\bibfnamefont {M.}~\bibnamefont {{van
  Setten}}}, \bibinfo {author} {\bibfnamefont {M.}~\bibnamefont {Giantomassi}},
  \bibinfo {author} {\bibfnamefont {E.}~\bibnamefont {Bousquet}}, \bibinfo
  {author} {\bibfnamefont {M.}~\bibnamefont {Verstraete}}, \bibinfo {author}
  {\bibfnamefont {D.}~\bibnamefont {Hamann}}, \bibinfo {author} {\bibfnamefont
  {X.}~\bibnamefont {Gonze}},\ and\ \bibinfo {author} {\bibfnamefont {G.-M.}\
  \bibnamefont {Rignanese}},\ }\bibfield  {title} {\bibinfo {title} {The
  pseudodojo: Training and grading a 85 element optimized norm-conserving
  pseudopotential table},\ }\href
  {https://doi.org/https://doi.org/10.1016/j.cpc.2018.01.012} {\bibfield
  {journal} {\bibinfo  {journal} {Computer Physics Communications}\ }\textbf
  {\bibinfo {volume} {226}},\ \bibinfo {pages} {39} (\bibinfo {year}
  {2018})}\BibitemShut {NoStop}%
\bibitem [{\citenamefont {Baroni}\ \emph {et~al.}(2001)\citenamefont {Baroni},
  \citenamefont {de~Gironcoli}, \citenamefont {Dal~Corso},\ and\ \citenamefont
  {Giannozzi}}]{Baroni2001}%
  \BibitemOpen
  \bibfield  {author} {\bibinfo {author} {\bibfnamefont {S.}~\bibnamefont
  {Baroni}}, \bibinfo {author} {\bibfnamefont {S.}~\bibnamefont
  {de~Gironcoli}}, \bibinfo {author} {\bibfnamefont {A.}~\bibnamefont
  {Dal~Corso}},\ and\ \bibinfo {author} {\bibfnamefont {P.}~\bibnamefont
  {Giannozzi}},\ }\bibfield  {title} {\bibinfo {title} {Phonons and related
  crystal properties from density-functional perturbation theory},\ }\href
  {https://doi.org/10.1103/RevModPhys.73.515} {\bibfield  {journal} {\bibinfo
  {journal} {Rev. Mod. Phys.}\ }\textbf {\bibinfo {volume} {73}},\ \bibinfo
  {pages} {515} (\bibinfo {year} {2001})}\BibitemShut {NoStop}%
\bibitem [{\citenamefont {Lee}\ \emph {et~al.}(2023)\citenamefont {Lee},
  \citenamefont {Poncé}, \citenamefont {Bushick}, \citenamefont {Hajinazar},
  \citenamefont {Lafuente-Bartolome}, \citenamefont {Leveillee}, \citenamefont
  {Lian}, \citenamefont {Lihm}, \citenamefont {Macheda}, \citenamefont {Mori},
  \citenamefont {Paudyal}, \citenamefont {Sio}, \citenamefont {Tiwari},
  \citenamefont {Zacharias}, \citenamefont {Zhang}, \citenamefont {Bonini},
  \citenamefont {Kioupakis}, \citenamefont {Margine},\ and\ \citenamefont
  {Giustino}}]{lee2023epw}%
  \BibitemOpen
  \bibfield  {author} {\bibinfo {author} {\bibfnamefont {H.}~\bibnamefont
  {Lee}}, \bibinfo {author} {\bibfnamefont {S.}~\bibnamefont {Poncé}},
  \bibinfo {author} {\bibfnamefont {K.}~\bibnamefont {Bushick}}, \bibinfo
  {author} {\bibfnamefont {S.}~\bibnamefont {Hajinazar}}, \bibinfo {author}
  {\bibfnamefont {J.}~\bibnamefont {Lafuente-Bartolome}}, \bibinfo {author}
  {\bibfnamefont {J.}~\bibnamefont {Leveillee}}, \bibinfo {author}
  {\bibfnamefont {C.}~\bibnamefont {Lian}}, \bibinfo {author} {\bibfnamefont
  {J.-M.}\ \bibnamefont {Lihm}}, \bibinfo {author} {\bibfnamefont
  {F.}~\bibnamefont {Macheda}}, \bibinfo {author} {\bibfnamefont
  {H.}~\bibnamefont {Mori}}, \bibinfo {author} {\bibfnamefont {H.}~\bibnamefont
  {Paudyal}}, \bibinfo {author} {\bibfnamefont {W.~H.}\ \bibnamefont {Sio}},
  \bibinfo {author} {\bibfnamefont {S.}~\bibnamefont {Tiwari}}, \bibinfo
  {author} {\bibfnamefont {M.}~\bibnamefont {Zacharias}}, \bibinfo {author}
  {\bibfnamefont {X.}~\bibnamefont {Zhang}}, \bibinfo {author} {\bibfnamefont
  {N.}~\bibnamefont {Bonini}}, \bibinfo {author} {\bibfnamefont
  {E.}~\bibnamefont {Kioupakis}}, \bibinfo {author} {\bibfnamefont {E.~R.}\
  \bibnamefont {Margine}},\ and\ \bibinfo {author} {\bibfnamefont
  {F.}~\bibnamefont {Giustino}},\ }\bibfield  {title} {\bibinfo {title}
  {Electron–phonon physics from first principles using the epw code},\ }\href
  {https://doi.org/10.1038/s41524-023-01107-3} {\bibfield  {journal} {\bibinfo
  {journal} {npj Computational Materials}\ }\textbf {\bibinfo {volume} {9}},\
  \bibinfo {pages} {156} (\bibinfo {year} {2023})}\BibitemShut {NoStop}%
\bibitem [{\citenamefont {Ponc\'e}\ \emph {et~al.}(2016)\citenamefont
  {Ponc\'e}, \citenamefont {Margine}, \citenamefont {Verdi},\ and\
  \citenamefont {Giustino}}]{Ponce2016}%
  \BibitemOpen
  \bibfield  {author} {\bibinfo {author} {\bibfnamefont {S.}~\bibnamefont
  {Ponc\'e}}, \bibinfo {author} {\bibfnamefont {E.}~\bibnamefont {Margine}},
  \bibinfo {author} {\bibfnamefont {C.}~\bibnamefont {Verdi}},\ and\ \bibinfo
  {author} {\bibfnamefont {F.}~\bibnamefont {Giustino}},\ }\bibfield  {title}
  {\bibinfo {title} {{EPW}: Electron--phonon coupling, transport and
  superconducting properties using maximally localized {Wannier} functions},\
  }\href {https://doi.org/10.1016/j.cpc.2016.07.028} {\bibfield  {journal}
  {\bibinfo  {journal} {Comput. Phys. Commun.}\ }\textbf {\bibinfo {volume}
  {209}},\ \bibinfo {pages} {116} (\bibinfo {year} {2016})}\BibitemShut
  {NoStop}%
\bibitem [{\citenamefont {Marzari}\ \emph {et~al.}(2012)\citenamefont
  {Marzari}, \citenamefont {Mostofi}, \citenamefont {Yates}, \citenamefont
  {Souza},\ and\ \citenamefont {Vanderbilt}}]{Marzari2012}%
  \BibitemOpen
  \bibfield  {author} {\bibinfo {author} {\bibfnamefont {N.}~\bibnamefont
  {Marzari}}, \bibinfo {author} {\bibfnamefont {A.~A.}\ \bibnamefont
  {Mostofi}}, \bibinfo {author} {\bibfnamefont {J.~R.}\ \bibnamefont {Yates}},
  \bibinfo {author} {\bibfnamefont {I.}~\bibnamefont {Souza}},\ and\ \bibinfo
  {author} {\bibfnamefont {D.}~\bibnamefont {Vanderbilt}},\ }\bibfield  {title}
  {\bibinfo {title} {Maximally localized {Wannier} functions: Theory and
  applications},\ }\href {https://doi.org/10.1103/RevModPhys.84.1419}
  {\bibfield  {journal} {\bibinfo  {journal} {Rev. Mod. Phys.}\ }\textbf
  {\bibinfo {volume} {84}},\ \bibinfo {pages} {1419} (\bibinfo {year}
  {2012})}\BibitemShut {NoStop}%
\end{thebibliography}%
\onecolumngrid
\section*{End Matter}
\twocolumngrid
Using Feynman diagrams, we show the higher-order NA contributions included in the $\tilde{\text{G}}_1\tilde{\text{G}}_1$ approximation.
The state-of-the-art G$_0$G$_0$ phonon self-energy is computed from the bare electron-hole bubble as
\begin{equation}
    \begin{tikzpicture}
        \pic at (0, 0) {chiPH_bare};
        \pic at (0, 0) {gr};
        \pic at (1.8, 0) {gr};
        \node at (2.1, 0) {$,$};        
    \end{tikzpicture}
\end{equation}
where the thin lines denote the electron and hole propagators and the black dots the statically screened EPC.
In the $\tilde{\text{G}}_1\tilde{\text{G}}_1$ approximation, the electron and hole propagators are dressed by the first-order electron self-energy containing the FM and the DW terms
\begin{equation}
    \begin{tikzpicture}
             \pic at (0.2, 0) {sigma};
        \node at (0.25, 0) {$\Sigma^{\text{G}_0\text{D}_0}$};
            \node at (1.3, 0) {$=$};
         \pic at (2, 0) {chi_bare};
        \pic at (2, 0) {gr};
        \pic at (3.5, 0) {gr};
        \node at (2.7, 1.1) {$\text{D}_0$};
        \node at (2.7, -0.3) {$\text{G}_0$};
        \node at (4., 0) {$+$};
        \pic at (4.5, 0) {wiggle_circle_0};
            \node at (5, 0) {$,$};\nonumber  

    \end{tikzpicture}
\end{equation}
where the thin wiggly line is the bare phonon propagator and the gray rectangle is used to denote the second order EPC present in the DW term.
This results in the $\tilde{\text{G}}_1\tilde{\text{G}}_1$  phonon self-energy 
    \begin{tikzpicture}
        \pic at (0, 0) {chig1};
        \pic at (0, 0) {gr};
        \pic at (2., 0) {gr};
        \node at (2.3, 0) {$=$};
       \pic at (2.7, 0) {chiPH_bare_};
       \pic at (4.5, 0) {gr};
       \pic at (2.7, 0) {gr};
        \node at (4.8, 0) {$+$};
       \pic at (5.2, 0) {chiPH_bare_};
       \pic at (5.2, 0) {gr};
       \pic at (7.0, 0) {gr};
       \pic at (5.7, 0.4) {sigma_small};\nonumber  

    \end{tikzpicture}
    \begin{equation}
            \begin{tikzpicture}
        \node at (0, 0) {};
       \node at (2.5, 0) {$+$};
       \pic at (2.8, 0) {chiPH_bare_};
       \pic at (2.8, 0) {gr};
       \pic at (4.6, 0) {gr};
       \pic at (3.2, 0.35) {sigma_small};
       \pic at (4.16, 0.35) {sigma_small};
        \node at (4.95, 0) {$+$};
       \pic at (5.3, 0) {chiPH_bare_};
       \pic at (5.3, 0) {gr};
       \pic at (7.1, 0) {gr};
       \pic at (5.7, 0.35) {sigma_small};
       \pic at (5.7, -0.35) {sigma_small};
        \node at (7.6, 0) {$+...$  .};\nonumber  

    \end{tikzpicture}
    \end{equation}

In Sec.~S1 of the SI~\cite{supplement} we explicitly show the $\tilde{\text{G}}_2\tilde{\text{G}}_2$ approximation terms.

The scGscG approach gives~\cite{giustino2017}:
\begin{equation}
    \begin{tikzpicture}
             \pic at (0.2, 0) {sigma1};
        \node at (0.23, 0) {$\Sigma^{\text{scG}\text{scD}}$};
            \node at (1.3, 0) {$=$};
        \pic at (2, 0) {G_arrow_thick};
        \pic at (2, 0) {chi_thick};
        \pic at (2, 0) {gr};
        \pic at (3.5, 0) {gr};
        \node at (2.7, 1.1) {$\text{scD}$};
        \node at (2.7, -0.3) {$\text{scG}$};
        \node at (4., 0) {$+$};
        \pic at (4.5, 0) {wiggle_circle};\nonumber  

    \end{tikzpicture}
\end{equation}
where thick lines indicate dressed electron or hole propagators
    \begin{equation}
    \begin{tikzpicture}
        \pic at (0, 0) {G_thick};
        \node at (1.25, 0) {$=$};
        \pic at (1.45, 0) {Gb};
        \node at (2.7, 0) {$+$};
        \pic at (2.95, 0) {Gb};
        \pic at (5.45, 0) {Gb};
        \pic at (3.95, 0) {chi_bare};
            \node at (4.65, -0.4) {G$_0$};   
            \node at (4.65, 1.1) {D$_0$};   
        \pic at (3.95, 0) {gr};
        \pic at (5.45, 0) {gr};
        \node at (6.65, 0) {$+$};
            \pic at (6.85, 0) {Gb};
                    \pic at (7.44, 0) {Gb};
        \pic at (7.5, 0) {wiggle_circle_0}; \nonumber      
\end{tikzpicture}
\end{equation}
\begin{equation}
\begin{tikzpicture}
        \node at (-0.25, 0) {$+$};   
        \pic at (0, 0) {Gb};
        \pic at (3, 0) {Gb};
        \pic at (1, 0) {chi_bare_2};
        \pic at (1, 0) {gr};
        \pic at (3, 0) {gr};
        \pic at (1.8, 0) {gr};
        \pic at (2.2, 0) {gr};
            \node at (1.5, -0.4) {G$_0$};   
            \node at (2.5, -0.4) {G$_0$};   
        \node at (4.25, 0) {$+$};   
        \pic at (4.5, 0) {Gb};
        \pic at (5.5, 0) {rainbow2}; 
        \pic at (5.5, 0) {Gb};
        \pic at (6.5, 0) {Gb};
        \node at (4.25, 0) {$+$}; \nonumber  
\end{tikzpicture}
\end{equation}
\begin{equation}
\begin{tikzpicture}
        \node at (-0.25, 0) {$+$};   
        \pic at (0, 0) {Gb};
        \pic at (3, 0) {Gb};
        \pic at (1, 0) {chi_bare_wbub};
        \pic at (1.5, 0.55) {gr};
                \node at (2, -0.4) {G$_0$};  
        \pic at (3, 0) {gr};
        \pic at (2.4, 0.55) {gr};
        \pic at (1, 0.) {gr};
        \node at (4.5, 0) {$+ ...$};   
\nonumber  
\end{tikzpicture}
\end{equation}
and 
\begin{equation}
    \begin{tikzpicture}
        \pic at (0, 0) {D_thick};
        \node at (1.25, 0) {$=$};
        \pic at (1.5, 0) {Db};
        \node at (2.75, 0) {$+$};
        \pic at (3, 0) {Db};
        \pic at (5.5, 0) {Db};
        \pic at (4, 0) {chiPH_bare_sm};
        \pic at (4, 0) {gr};
        \pic at (5.5, 0) {gr};    
\nonumber
    \end{tikzpicture}
\end{equation}
\begin{equation}
    \begin{tikzpicture}
        \node at (0., 0) {$+$};
        \pic at (0.25, 0) {Db};
        \pic at (2.75, 0) {Db};
        \pic at (1.25, 0) {chiPH_bare_sm};
        \pic at (2.75, 0) {gr};
        \pic at (1.25, 0) {gr};
        \pic at (1.25, 0) {small_wiggle_sm};
        \pic at (1.25, 0) {gr};
        \node at (4., 0) {$+$};
        \pic at (4.25, 0) {Db};
        \pic at (6.75, 0) {Db};
       \pic at (5.25, 0) {chiPH_bare_sm};
       \pic at (5.25, 0) {gr};
       \pic at (6.75, 0) {gr};
       \pic at (5.25, 0) {twosmall_wiggle_sm};\nonumber
    \end{tikzpicture}
\end{equation}
\begin{equation}
    \begin{tikzpicture}
        \node at (0, 0) {$+$};
        \pic at (0.17, 0) {Db};
        \pic at (2.45, 0) {Db};
       \pic at (1.0, 0) {chiPH_bare_sm};
       \pic at (2.45, 0) {gr};
       \pic at (1.0, 0) {gr};
       \pic at (1.0, 0) {rainbow_sm};
        \node at (3.65, 0) {$+$  };
        \pic at (3.85, 0) {Db};
        \pic at (5.95, 0) {Db};
        \pic at (4.65, 0) {chiPH_bare_sm};
        \pic at (6.15, 0) {gr};
        \pic at (4.65, 0) {gr}; 
        \pic at (4.9, 0.25) {wiggle_circle_small0};
        \node at (7.4, 0) {$+...$  .};\nonumber  

    \end{tikzpicture}
\end{equation}
The scGscG phonon self-energy is then 
    \begin{tikzpicture}
        \pic at (0, 0) {chiPH_gn};
        \pic at (0, 0) {gr};
        \pic at (1.8, 0) {gr};
        \node at (2.2, 0) {$=$};
       \pic at (2.5, 0) {chiPH_bare_};
       \pic at (4.3, 0) {gr};
       \pic at (2.5, 0) {gr};
           \node at (4.6, 0) {$+$};
       \pic at (5, 0) {chiPH_bare_};
       \pic at (5, 0) {gr};
       \pic at (6.8, 0) {gr};
       \pic at (5.5, 0.4) {sigma1_small};\nonumber  

    \end{tikzpicture}
    \begin{equation}
            \begin{tikzpicture}
        \node at (0, 0) {};
       \node at (2.5, 0) {$+$};
       \pic at (2.8, 0) {chiPH_bare_};
       \pic at (2.8, 0) {gr};
       \pic at (4.6, 0) {gr};
       \pic at (3.2, 0.35) {sigma1_small};
       \pic at (4.16, 0.35) {sigma1_small};
        \node at (4.95, 0) {$+$};
       \pic at (5.3, 0) {chiPH_bare_};
       \pic at (5.3, 0) {gr};
       \pic at (7.1, 0) {gr};
       \pic at (5.7, 0.35) {sigma1_small};
       \pic at (5.7, -0.35) {sigma1_small};
        \node at (7.6, 0) {$+...$  .};\nonumber  

    \end{tikzpicture}
    \end{equation}

\begin{figure}[t]
    \centering
\includegraphics[width=0.48\textwidth]{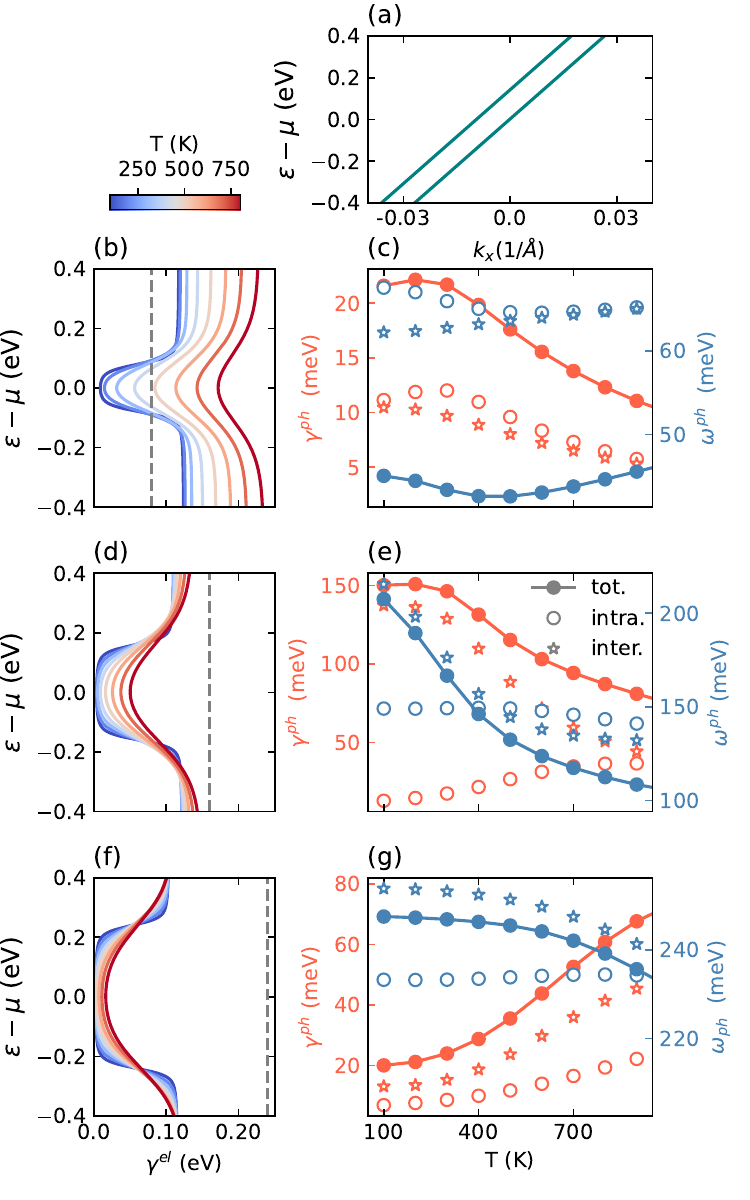}
    \caption{Two parallel linear bands 2D Holstein model results. (a) Model band structure where the spacing between the bands is 160~meV, while the phonon frequency ranges from 80~meV (b-c), to 160~meV (d-e) and 240~meV (f-g). 
    (b,d,f) Electron linewidth for one of the bands. Vertical grey line denotes the phonon frequency (80, 160 and 240~meV).
    (c,e,g) Corresponding phonon linewidth (orange) and frequency renormalization (blue) deriving from electron-mediated anharmonic interband (star symbols), intraband (empty circles) and both (full circles) contributions.}
    \label{fig:models1}
\end{figure}

\begin{figure}[b!]
    \centering
    \includegraphics[width=0.48\textwidth]{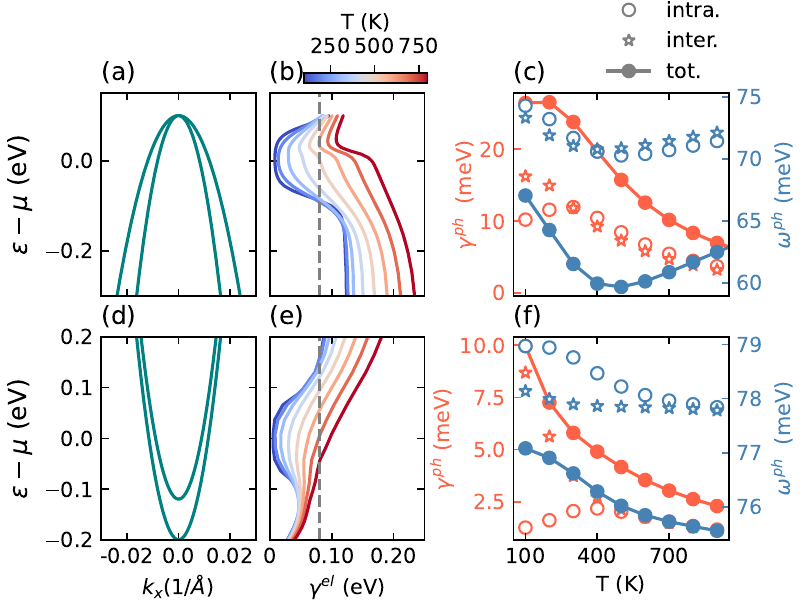}
    \caption{(a-c) Two inverted parabolic bands 2D Holstein model results with doping of $\mu$=-0.1~eV and the phonon frequency fixed to 80~meV.
    (a) Model band structure. (b) Electron linewidth for the upper band. Vertical grey line denotes the phonon frequency. 
    (c) Corresponding phonon linewidth (orange) and frequency renormalization (blue) deriving from higher-order NA interband (star symbols), intraband (empty circles) and both (full circles) contributions. 
    (d-f) Same as (a-c), but applied to a two-parabolic-bands 3D Fröhlich model with doping of $\mu$=0.2~eV.}
    \label{fig:models}
\end{figure}

In Figs.~\ref{fig:models1}  and ~\ref{fig:models}, we apply the formalism to different models. 
The 2D Holstein model is used in conjunction with the linear bands (Fig.~\ref{fig:models1}) and inverted parabolic bands (Fig.~\ref{fig:models}(a-c)).
In Fig.~\ref{fig:models}(d-f), we use a 3D Fröhlich model applied to two parabolic bands.
For every model, we use two bands in order to distinguish inter- and intra- band contributions to the phonon self-energy.
Note that a similar set of higher-order contributions computed in Ref.~\cite{erhardt2025b} refers only to the intraband channel. 
The linewidth for one of the bands shown in panel (b) reaches the scale of half the phonon frequency at 300~K and the scale of the full phonon frequency at 500~K.
For this reason, at 300~K the intraband contribution to the phonon linewidth exhibits a peak, and at 500~K the intraband contribution to the phonon renormalization shows a saturation and a weak phonon hardening with further temperature increase.
For a larger phonon frequency, Fig.~\ref{fig:models1}(d-g), the electron linewidth is smaller because of a reduced Bose-Einstein distribution factor.
For this reason, the intraband linewidth peak is shifted to larger temperatures in panels (e,g), as well as the adiabatic limit for the frequency renormalization, and the overall temperature dependence is less pronounced in the intraband channel. 
The band separation tunes the scale of NA effects in the interband channel and we show three limiting cases by changing the phonon frequency.
In the linear band model in Fig.~\ref{fig:models1}, the band separation is smaller than the phonon frequency in panels (f,g) and larger in other panels. 

In the first-order G$_0$G$_0$ approximation, the largest phonon linewidth and NA effects are expected when  $\omega^{\text{ph}}\approx \Delta$, while the case of $\omega^{\text{ph}}\ll \Delta$ describes the adiabatic limit. 
Note that the first-order interband contributions to the phonon self-energy have a weak temperature dependence and induce a linewidth decrease and frequency hardening with temperature~\cite{Erhardt2025}.
Close to the adiabatic limit, when $\omega^{\text{ph}}\ll \Delta$, the $\tilde{\text{G}}_1\tilde{\text{G}}_1$ interband contributions to the phonon self-energy have qualitatively similar trends to the first-order contributions, but with an enhanced temperature dependence.
In the strongly NA limit, $\omega^{\text{ph}}\approx \Delta$, and a further decrease of $\Delta$, the interband contributions induce phonon softening with temperature, as opposed to the nearly temperature independent first-order NA renormalization~\cite{Erhardt2025}.  
In this limit, the $\tilde{\text{G}}_1\tilde{\text{G}}_1$ interband contributions become qualitatively similar to the already discussed intraband contributions.
Unless $\omega^{\text{ph}}\ll \Delta$, it is worth emphasizing that these contributions are indeed qualitatively and quantitatively significant, and even though there is a finite long wavelength interband contribution in the G$_0$G$_0$ approximation, higher-order corrections in this channel are not negligible.

For the inverted parabolic bands 2D Holstein model, Fig.~\ref{fig:models}(a-c), and the parabolic bands 3D Fröhlich model, Fig.~\ref{fig:models}(d-f),  the intraband contributions to the phonon self-energy have the same temperature trends as in the linear band case in Fig.~\ref{fig:models1}(c).
For larger temperatures, the interband contributions match the intraband ones, as the broadening of the bands becomes large enough for them to overlap. 

As these two models, applied to systems of different dimensions and shapes of the bands, produce qualitatively the same conclusions, this suggests that these are in fact a general feature of the higher-order NA theory. 
We summarize these findings in Fig.~\ref{fig:doodle} of the main text. 
\end{document}


\title{Higher-order nonadiabaticity governs the temperature dependence of the phonon spectrum}

\author{Nina Girotto Erhardt}
\email{nina.girotto@uclouvain.be}
\affiliation{%
European Theoretical Spectroscopy Facility, Institute of Condensed Matter and Nanosciences, Université catholique de Louvain, Chemin des Étoiles 8, B-1348 Louvain-la-Neuve, Belgium. 	
}%

\author{Samuel Ponc\'e}
\email{samuel.ponce@uclouvain.be}
\affiliation{%
European Theoretical Spectroscopy Facility, Institute of Condensed Matter and Nanosciences, Université catholique de Louvain, Chemin des Étoiles 8, B-1348 Louvain-la-Neuve, Belgium. 	
}%
\affiliation{%
WEL Research Institute, avenue Pasteur, 6, 1300 Wavre, Belgique.		
}%

\date{\today}
\maketitle

\section{Formulation of the fully self-consistent problem}
Diagrammatically, the Dyson equations for the interacting electron (G) and phonon (D) propagators can be written as~\cite{mahan},
\begin{equation}
    \begin{tikzpicture}
        \pic at (0, 0) {G_thick};
        \node at (1.25, 0) {$=$};
        \pic at (1.5, 0) {Gb};
        \node at (2.75, 0) {$+$};
        \pic at (3., 0) {Gb};
        \node at (4.1, 0) {$\Bigg[$};
        \pic at (4.4, 0) {G_arrow_thick};
        \pic at (4.4, 0) {chi_thick};
        \pic at (4.4, 0) {gr};
        \pic at (5.4, 0) {grp};
        \pic at (5.7, 0) {gr};
        \node at (6.0, 0) {$+$};
        \pic at (6.4, 0) {wiggle_circle};
        \node at (7.2, 0) {$\Bigg]$};
         \pic at (7.3, 0) {G_thick};
        \node at (8.4, 0) {$,$};
    \end{tikzpicture}
\end{equation}
\begin{equation}
    \begin{tikzpicture}
        \pic at (0, 0) {D_thick};
        \node at (1.25, 0) {$=$};
        \pic at (1.5, 0) {Db};
        \node at (2.75, 0) {$+$};
        \pic at (3, 0) {Db};
        \pic at (6.3, 0) {D_thick};
        \pic at (4, 0) {chiPH_thick};
        \pic at (4, 0) {gb};
        \pic at (6, 0) {grp};
        \pic at (6.3, 0) {gr};
        \node at (7.25, 0) {$,$}; 
    \end{tikzpicture}
\end{equation}
where the interacting electron (phonon) propagators are denoted by a thick straight (wavy) line, bare ones with thin lines, bare EPC with a small black dot, and screened EPC with a larger black dot and the gray rectangle is used as a symbol for the second-order EPC $\mathfrak{g^2}$. 
Bare propagators contain a subscript `0'.

The vertex denoted by an empty circle is currently out of reach for first-principles calculations and is here neglected.
Further, in practice for first-principles calculations, it is advantageous to use the screened-screened approximation for the phonon self-energy~\cite{berges2023}.
This leaves us with a set of Dyson equations for the self-consistent propagators, denoted as scG for electrons and scD for phonons, written as 
\begin{equation}    \label{scG}
    \begin{tikzpicture}
        \pic at (0, 0) {G_thick};
        \node at (0.2, -0.4) {sc};
        \node at (1.25, 0) {$=$};
        \pic at (1.5, 0) {Gb};
        \node at (2.75, 0) {$+$};
        \pic at (3., 0) {Gb};
        \node at (4.1, 0) {$\Bigg[$};
        \pic at (4.4, 0) {G_arrow_thick};
        \node at (4.6, -0.4) {sc};
        \pic at (4.4, 0) {chi_thick};
        \node at (4.6, 0.9) {sc};
        \pic at (4.4, 0) {gr};
        \pic at (5.4, 0) {gr};
        \node at (5.8, 0) {$+$};
        \pic at (6.4, 0) {wiggle_circle};
        \node at (6.2, 1.2) {sc};
        \node at (7.2, 0) {$\Bigg]$};
         \pic at (7.3, 0) {G_thick};
        \node at (7.5, -0.4) {sc};
        \node at (8.4, 0) {$,$};
    \end{tikzpicture}
\end{equation}
\begin{equation}  \label{scD}
    \begin{tikzpicture}
        \pic at (0, 0) {D_thick};
        \node at (0.2, -0.4) {sc};
        \node at (1.25, 0) {$=$};
        \pic at (1.5, 0) {Db};
        \node at (2.75, 0) {$+$};
        \pic at (3, 0) {Db};
        \pic at (6., 0) {D_thick};
        \node at (6.2, -0.4) {sc};
        \node at (4.7, 0.7) {sc};
        \pic at (4, 0) {chiPH_thick};
        \pic at (4, 0) {gr};
        \pic at (6., 0) {gr};
        \node at (4.7, -0.15) {sc};
        \node at (7., 0) {$,$}; 
    \end{tikzpicture}
\end{equation}
First, we note that the exact and the scGscG phonon self-energy in the screened-screened approximation~\cite{berges2023} differ by crossing diagrams and vertex corrections, generated in the following way:
\begin{equation}
\begin{tikzpicture}
        \pic at (0, 0) {chiPH_thick2};
        \pic at (0, 0) {gr};
            \node at (1, 0) {$\Pi^{\text{exact}}$};
        \pic at (2, 0) {grp};
        \pic at (2.25, 0) {gr};
        \node at (2.5, 0) {$-$};
        \pic at (2.8, 0) {chiPH_gn2};
       \pic at (2.8, 0) {gr};
        \pic at (4.8, 0) {gr};%
        \node at (5.1, 0) {$=$};
       \pic at (5.45, 0) {chiPH_bare_};
       \pic at (5.45, 0) {gr};
       \pic at (7.45, 0) {gr};
       \pic at (5.45,0.05) {twosmall_wigglec};\nonumber
    \end{tikzpicture}
\end{equation}
\begin{equation}
\begin{tikzpicture}
       \node at (7., 0) {$+$};
       \pic at (7.35, 0) {chiPH_bare_};
       \pic at (7.35, 0) {gr};
       \pic at (9.35, 0) {gr};
       \pic at (7.35, 0) {vertex};
       \node at (10, 0) {$+...$  .};\nonumber
\end{tikzpicture}
\end{equation}
From the Eqs.~(\ref{scG}-\ref{scD}), the self-consistency of the problem is clear. 
To compute the interacting electron propagator, we need to compute the interacting phonon propagator and vice versa.
To write down the corresponding set of self-consistent equations, we start from the spectral (Lehmann) representation of the interacting electron and phonon propagators
\begin{align}\label{eq:G_spec}
     G^{\text{scGscD}}_{n\mathbf{k}}(i\varepsilon_j)  =& \int_{-\infty}^\infty \mathrm{d}\varepsilon \frac{A^{\text{scGscD}}_{n\mathbf{k}} ( \varepsilon )}{i\varepsilon_j - \varepsilon} \\
      D^{\text{scGscD}}_{\nu \mathbf{q}}(i\omega_j) =& \int_{-\infty}^\infty \text{d}\omega \frac{B^{\text{scGscD}}_{\nu \mathbf{q}}(\omega)}{i\omega_j-\omega},\label{D_spec}   
\end{align}
where $A^{\text{scGscD}}_{n\mathbf{k}} ( \varepsilon )$ is the electron spectral function and $B^{\text{scGscD}}_{\nu \mathbf{q}}(\omega)$ is the phonon spectral function.
The interacting electron and phonon self-energies can be written in terms of their interacting propagators as
\begin{multline}\label{eq:GD}
    \Sigma^{\text{scGscD}- \text{FM}}_{n\mathbf{k}}(i\varepsilon_j) = \text{k}_{\text{B}}\text{T}\sum_{m\nu\mathbf{q}}\sum_{i\omega_j}|g_{mn\nu}(\mathbf{k},\mathbf{q})|^2 \\ \times
      D^{\text{scGscG}}_{\nu \mathbf{q}}(i\omega_j)G^{\text{scGscD}}_{m\mathbf{k}+\mathbf{q}}(i\varepsilon_j+i\omega_j)
\end{multline}
for the Fan-Migdal (FM) contributions for electrons and
\begin{multline}
    \Pi_{\nu \mathbf{q}}^{\text{scGscG}}(i\omega_j) = \text{k}_{\text{B}}\text{T}\sum_{nm\mathbf{k}}\sum_{ip_j}|g_{mn\nu}(\mathbf{k},\mathbf{q})|^2  \\ \times 
     G^{\text{scGscD}}_{n\mathbf{k}}(i\varepsilon_j)G^{\text{scGscD}}_{m\mathbf{k}+\mathbf{q}}(i\varepsilon_j+i\omega_j)
     \label{eq:GG}
\end{multline}
for phonons, where $\text{k}_{\text{B}}$ is the Boltzmann constant and $|g_{mn\nu}(\mathbf{k},\mathbf{q})|^2$ are the statically screened matrix elements.
Injecting Eqs.~(\ref{eq:G_spec}, \ref{D_spec}) into Eqs.~(\ref{eq:GD}, \ref{eq:GG}) gives the following imaginary part:
\begin{multline}
        \Im \Sigma_{n\mathbf{k}}^{\text{scGscD}-\text{FM}}(\varepsilon) = - \pi\sum_{m\nu\mathbf{q}}|{g}_{mn\nu}\left(\mathbf{k},\mathbf{q}\right)|^2 \int_{-\infty}^{\infty} \textrm{d}\omega  \\\times  
    A^{\text{scGscD}}_{m\mathbf{k}+ \mathbf{q}} ( \varepsilon+\omega) B^{\text{scGscG}}_{\nu \mathbf{q}}(\omega) \Big( n(\omega) - f(\varepsilon+\omega) \Big)
        \label{eq:imsigma}
\end{multline}
and
\begin{multline}\label{eq:impi}
\Im\Pi_{\nu \mathbf{q}}^{\text{scGscG}}(\omega)=-\pi\sum_{nm\mathbf{k}}|{g}_{mn\nu}\left(\mathbf{k},\mathbf{q} \right)|^2 \int^{\infty}_{-\infty} \text{d}\varepsilon   \\ \times A^{\text{scGscD}}_{n\mathbf{k}} ( \varepsilon) A^{\text{scGscD}}_{m\mathbf{k}+\mathbf{q}} ( \varepsilon+\omega ) \Big(f(\varepsilon) - f(\varepsilon+\omega)\Big)
\end{multline}
and allows to obtain the real part through the Kramers-Kronig transformation as
\begin{align}
\Re\Sigma^{ \text{scGscD}-\text{FM}}_{n\mathbf{k}}(\varepsilon)&=\frac{1}{\pi} \textit{P} \int^{\infty}_{-\infty} \text{d}\varepsilon' \frac{\Im \Sigma^{ \text{scGscD}-\text{FM}}_{n\mathbf{k}}(\varepsilon')}{\varepsilon'-\varepsilon},
\label{eq:KK} \\
\Re\Pi^{ \text{scGscG}}_{\nu\mathbf{q} }\left(\omega\right)&=\frac{1}{\pi} \textit{P} \int^{\infty}_{-\infty} \text{d}\omega' \frac{\Im \Pi^{ \text{scGscG}}_{\nu\mathbf{q}}\left(\omega'\right)}{\omega'-\omega}.
\label{eq:KK_pi}
\end{align}
The undistorted electronic self-energy is then obtained by adding the static Debye-Waller (DW) self-energy~\cite{allen1981, allen1983, Allen_1976}.
It can be done through Wannier function perturbation theory~\cite{lihm2021}, within the rigid-ion approximation (RIA)~\cite{Allen_1976,ponce2014}
\begin{equation}\label{eq:DW}
\Sigma^{\text{DW}}_{n\mathbf{k}} = \sum_{\nu\mathbf{q} } \frac{\big(n(\omega_{\nu\mathbf{q}})+ \frac{1}{2}\big)}{2\omega_{\nu\mathbf{q}}}\mathfrak{g}^{2,\text{RIA}}_{n\nu}(\mathbf{k},\mathbf{q}).
\end{equation}
where 
\begin{multline}
        \mathfrak{g}^{2,\text{RIA}}_{n\nu}(\mathbf{k},\mathbf{q}) =\sum_{m\neq n, \kappa \alpha \kappa' \beta} g^{*}_{m n \kappa \alpha} (\mathbf{k},\mathbf{\Gamma})g_{m n \kappa' \beta} (\mathbf{k},\mathbf{\Gamma}) \\\times \frac{e^{*}_{\kappa\alpha\nu}(\mathbf{q})e_{\kappa\beta\nu}(\mathbf{q})-e^{*}_{\kappa'\alpha\nu}(\mathbf{q})e_{\kappa'\beta\nu}(\mathbf{q})}{\varepsilon_{n\mathbf{k}}-\varepsilon_{m\mathbf{k}}},
\end{multline}
where $e_{\kappa\alpha\nu}(\mathbf{q})$ are the eigendisplacement vectors of atom $\kappa$ in the Cartesian direction $\alpha$ due to a phonon mode with branch index $\nu$ and wavevector $\bq$, scaled by the atomic mass and $\mathfrak{g}^{2,\text{RIA}}$ is the RIA of the second order EPC vertex. 

Since the imaginary components are computed through the convolution of the two spectral functions, this forms a closed set of equations comprising of the electron self-energy 
\begin{equation}
\!\Sigma_{n\mathbf{k}}^{\text{scGscD}} \!(\varepsilon)  \!\!=  \!\Re\Sigma_{n\mathbf{k}}^{\text{scGscD} \!- \!\text{FM}} \!(\varepsilon)  \!+  \! \Sigma^{\text{DW}}_{n\mathbf{k}} \!+ \!
     i \Im\Sigma_{n\mathbf{k}}^{\text{scGscD} \!- \!\text{FM}} \!(\varepsilon),  
\end{equation}
electron spectral function
\begin{equation}
A^{\text{scGscD}}_{n\mathbf{k}} ( \varepsilon ) = -\frac{1}{\pi} \Im \frac{1}{\varepsilon - \varepsilon_{n\mathbf{k}} - \Sigma^{\text{scGscD}}_{n\mathbf{k}}(\varepsilon)},
\end{equation}
phonon self-energy 
\begin{equation}
\Pi_{\nu\mathbf{q}}^{\text{scGscG}} = \Re\Pi_{\nu\mathbf{q}}^{\text{scGscG}}+ i \Im\Pi_{\nu\mathbf{q}}^{\text{scGscG}},
\end{equation}
and phonon spectral function
\begin{align}
B^{\text{scGscG}}_{\nu\mathbf{q}}(\omega) \!= \!- \frac{2\omega}{\pi}\Im\frac{1}{\omega ^2 \!-\omega_{\nu\mathbf{q}}^2 \!-2\omega_{\nu\mathbf{q}}\Pi^{\text{scGscG}}_{\nu\mathbf{q}}(\omega)}.
    \label{eq:B}
\end{align}
Solving these equations self-consistently with Eqs.~(\ref{eq:imsigma}-\ref{eq:DW}) involves costly iteration loops over energy integrals.
Schematically, the iterations loop can be written as:
\begin{multline}\label{iter}
  \Sigma^{\text{G}_n\text{D}_n} \rightarrow A^{\text{G}_n\text{D}_n} \rightarrow G^{\text{G}_n\text{D}_n}  \rightarrow \Pi^{\text{G}_{n+1}\text{G}_{n+1}} \rightarrow \\ B^{\text{G}_{n+1}\text{G}_{n+1}}  \rightarrow D^{\text{G}_{n+1}\text{G}_{n+1}}\rightarrow \Sigma^{\text{G}_{n+1}\text{D}_{n+1}} ...
\end{multline}
where the first iteration (n=0) is initiated with computing 
\begin{align}\label{g0d0}
\Sigma^{\text{G}_0\text{D}_0}_{n\mathbf{k}} (\varepsilon)=& \sum_{m\nu\mathbf{q}}|g_{mn\nu}(\mathbf{k},\mathbf{q})|^2 
\Bigg(\frac{n(\omega_{\nu\mathbf{q}})+f(\varepsilon_{m \mathbf{k}+\mathbf{q}})}{\varepsilon + i\delta+\omega_{\nu\mathbf{q}}-\varepsilon_{m\mathbf{k}+\mathbf{q}}} \nonumber \\ &+\frac{n(\omega_{\nu\mathbf{q}})+1-f(\varepsilon_{m \mathbf{k}+\mathbf{q}})}{\varepsilon + i\delta-\omega_{\nu\mathbf{q}}-\varepsilon_{m\mathbf{k}+\mathbf{q}}}\Bigg)  + \Sigma_{n\mathbf{k}}^{\text{DW}},
\end{align}
then constructing the interacting electron spectral function
\begin{equation}
    A^{\text{G}_0\text{D}_0}_{n\mathbf{k}} (\varepsilon)= -\frac{1}{\pi}\Im \frac{1}{\varepsilon - \varepsilon_{n\mathbf{k}} -\Sigma^{\text{G}_0\text{D}_0}_{n\mathbf{k}} (\varepsilon)}
\end{equation}
and then the phonon self-energy
\begin{multline}
    \Pi_{\nu \mathbf{q}}^{\text{G$_1$G$_1$}}(i\omega_j) = \text{k}_{\text{B}}\text{T}\sum_{nm\mathbf{k}}\sum_{ip_j}|g_{mn\nu}(\mathbf{k},\mathbf{q})|^2  \\ \times 
     G^{\text{G$_0$D$_0$}}_{n\mathbf{k}}(i\varepsilon_j)G^{\text{G$_0$D$_0$}}_{m\mathbf{k}+\mathbf{q}}(i\varepsilon_j+i\omega_j)
\end{multline}
and spectrum
\begin{equation}
B^{\text{G$_1$G$_1$}}_{\nu\mathbf{q}}(\omega)  \! = \! -\frac{2\omega}{\pi} \!\Im \!\frac{1}{\omega^2  \!- \omega_{\nu\mathbf{q}}^2 \!-2\omega_{\nu\mathbf{q}}\Pi^{\text{G$_1$G$_1$}}_{\nu\mathbf{q}}(\omega) }.
\end{equation}
In this notation, note that the electron propagator dressed by the first-order electron self-energy $G^{\text{G}_0\text{D}_0}$  corresponds to G$_1$.
Had we started the iterations' loop by dressing the phonons first, then the loop would be performed in the opposite order as 
\begin{multline}
\Pi^{\text{G}_{n}\text{G}_{n}} \rightarrow B^{\text{G}_{n}\text{G}_{n}}  \rightarrow D^{\text{G}_{n}\text{G}_{n}}\rightarrow \Sigma^{\text{G}_{n}\text{D}_{n+1}} \rightarrow   \\ A^{\text{G}_n\text{D}_{n+1}}\rightarrow G^{\text{G}_n\text{D}_{n+1}}  \rightarrow \Pi^{\text{G}_{n+1}\text{G}_{n+1}} ...  \text{ .}
\end{multline}
In that case, one would begin by computing the first-order phonon self-energy as 
\begin{equation}
\Pi^{\text{G}_0\text{G}_0}_{\nu\mathbf{q}} \!(\omega)   \!=  \!\sum_{nm\mathbf{k}}  \!|g_{mn\nu}(\mathbf{k},\mathbf{q})|^2 \!\frac{f(\varepsilon_{m \mathbf{k}+\mathbf{q}})-f(\varepsilon_{n \mathbf{k}})}{\varepsilon_{m\mathbf{k}+\mathbf{q}} \! - \!\varepsilon_{n\mathbf{k}} \! - \! \omega \!-  \!i\eta }.
    \label{eq:pi0}
\end{equation}
In this work, we start by dressing the electrons and follow the iteration loop from the schematics in Eq.~\eqref{iter}.
Diagrammatically, we start by computing the one-shot first-order electron self-energy:
\begin{equation}\label{g0d0}
    \begin{tikzpicture}
             \pic at (0.2, 0) {sigma};
        \node at (0.25, 0) {$\Sigma^{\text{G}_0\text{D}_0}$};
            \node at (1.3, 0) {$=$};
         \pic at (2, 0) {chi_bare};
        \pic at (2, 0) {gr};
        \pic at (4, 0) {gr};
        \node at (3, 0.9) {$\text{D}_0$};
        \node at (3, -0.3) {$\text{G}_0$};
        \node at (4.5, 0) {$+$};
        \pic at (5, 0) {wiggle_circle_0};
    \end{tikzpicture}
\end{equation}
and building the interacting electron propagator as
\begin{equation}\label{g1}
    \begin{tikzpicture}
        \pic at (0, 0) {G1};
        \node at (1.25, 0) {$=$};
        \pic at (1.5, 0) {Gb};
        \node at (2.75, 0) {$+$};
        \pic at (3, 0) {Gb};
        \pic at (5.3, 0) {G1};
        \pic at (4.65, 0) {sigma};
        \node at (4.7, 0) {$\Sigma^{\text{G}_0\text{D}_0}$};
        \node at (6.7, 0) {$.$};
    \end{tikzpicture}
\end{equation}
With the interacting propagator lines, we can construct the phonon self-energy
\begin{equation}
    \begin{tikzpicture}
        \pic at (0, 0) {chig1};
        \pic at (0, 0) {gr};
       \pic at (2., 0) {gr};
        \node at (2.5, 0) {$=$};
       \pic at (3., 0) {chiPH_bare};
       \pic at (5, 0) {gr};
       \pic at (3, 0) {gr};
           \node at (5.5, 0) {$+$};
       \pic at (6, 0) {chiPH_bare};
       \pic at (6, 0) {gr};
       \pic at (8, 0) {gr};
       \pic at (6.5, 0.4) {sigma_small};
       \pic at (8, 0) {gr};
       \node at (8.5, 0) {$+$}; \nonumber
    \end{tikzpicture}
\end{equation}
\begin{equation}\label{g1g1}
    \begin{tikzpicture}
       \pic at (3, 0) {chiPH_bare};
       \pic at (3, 0) {gr};
       \pic at (5, 0) {gr};
       \pic at (3.5, -0.4) {sigma_small};
       \node at (5.5, 0) {$+$};
       \pic at (6, 0) {chiPH_bare};
       \pic at (6, 0) {gr};
       \pic at (8, 0) {gr};
           \pic at (7.5, 0.4) {sigma_small};
       \pic at (6.5, 0.4) {sigma_small};
        \node at (8.5, 0) {$+$};
       \pic at (9, 0) {chiPH_bare};
       \pic at (9, 0) {gr};
       \pic at (11, 0) {gr};
        \pic at (9.5, 0.4) {sigma_small};
       \pic at (9.5, -0.4) {sigma_small};
        \node at (11.5, 0) {$+...$};
    \end{tikzpicture}
\end{equation}
and then the phonon propagator
\begin{equation}\label{d1}
    \begin{tikzpicture}
        \pic at (0, 0) {D1};
        \node at (1.25, 0) {$=$};
        \pic at (1.5, 0) {Db};
        \node at (2.75, 0) {$+$};
        \pic at (3, 0) {Db};
        \pic at (6., 0) {D1};
        \pic at (4, 0) {chig1};
        \pic at (4, 0) {gb};
        \pic at (6., 0) {gr};
        \node at (7.25, 0) {$.$};
    \end{tikzpicture}
\end{equation}
The phonon self-energy computed in this work consists of the diagrams in Eq.~\eqref{g1g1}.
For completeness, we show the G$_2$G$_2$ formulation. In that case, the starting point is the electron self-energy built from the interacting electron and phonon propagators from the previous iteration as
\begin{equation}
    \begin{tikzpicture}
         \pic at (0.25, 0) {sigma1};
        \node at (0.3, 0) {$\Sigma^{\text{G}_1\text{D}_1}$};
            \node at (1.3, 0) {$=$};
         \pic at (2, 0) {G_arrow};
        \pic at (2, 0) {chi};
        \pic at (2, 0) {gr};
        \pic at (4, 0) {gr};
        \node at (4.5, 0) {$+$};
            \node at (3, 1) {$\text{D}_1$};
            \node at (3, -0.3) {$\text{G}_1$};
        \pic at (5, 0) {wiggle_circle_1};
        \node at (6, 0) {.};
    \end{tikzpicture}
\end{equation}
This allows to build a new electron propagator 
\begin{equation}
    \begin{tikzpicture}
        \pic at (0, 0) {G2};
        \node at (1.25, 0) {$=$};
        \pic at (1.5, 0) {Gb};
        \node at (2.75, 0) {$+$};
        \pic at (3, 0) {Gb};
        \pic at (5.3, 0) {G2};
             \pic at (4.65, 0) {sigma1};
        \node at (4.7, 0) {$\Sigma^{\text{G}_1\text{D}_1}$};
        \node at (6.7, 0) {$.$};
    \end{tikzpicture}
\end{equation}
To build the phonon self-energy we now have
    \begin{tikzpicture}
        \pic at (0, 0) {chig2};
        \pic at (0, 0) {gr};
        \pic at (2., 0) {gr};
        \node at (2.5, 0) {$=$};
       \pic at (3., 0) {chiPH_bare};
       \pic at (5, 0) {gr};
       \pic at (3, 0) {gr};
           \node at (5.5, 0) {$+$};
       \pic at (6, 0) {chiPH_bare};
       \pic at (6, 0) {gr};
       \pic at (8, 0) {gr};
       \pic at (6.5, 0.4) {sigma1_small};
       \pic at (8, 0) {gr};
       \node at (8.5, 0) {$+$};
    \end{tikzpicture}
\begin{equation}
    \begin{tikzpicture}
      \pic at (3, 0) {chiPH_bare};
       \pic at (3, 0) {gr};
       \pic at (5, 0) {gr};
       \pic at (3.5, -0.4) {sigma1_small};       
       \node at (5.5, 0) {$+$};
       \pic at (6, 0) {chiPH_bare};
       \pic at (6, 0) {gr};
       \pic at (8, 0) {gr};
       \pic at (7.5, 0.4) {sigma1_small};       
       \pic at (6.5, 0.4) {sigma1_small};       
        \node at (8.5, 0) {$+$};
       \pic at (9, 0) {chiPH_bare};
       \pic at (9, 0) {gr};
       \pic at (11, 0) {gr};
       \pic at (9.5, -0.4) {sigma1_small};       
       \pic at (9.5, 0.4) {sigma1_small};       
        \node at (11.5, 0) {$+...$};
    \end{tikzpicture}
\end{equation}
which is written similarly to the G$_1$G$_1$ expression for convenience. 
All the subsequent orders are generated straightforwardly.
Already the G$_2$G$_2$ phonon self-energy is considerably more complex to compute than the G$_1$G$_1$, as it requires integration to compute the G$_1$D$_1$ electron self-energy and then another integration to reach the corresponding phonon self-energy. 
Interestingly, to reach the G$_1$G$_1$ phonon self-energy, one also needs to perform an integral over interacting electron and hole propagators, but as shown in Ref.~\cite{park2025}, this is avoidable within the quasiparticle (QP) approximation.

In this work, we therefore rely on the G$_1$G$_1$ formulation and the QP approximation, denoted $\tilde{\text{G}}_1\tilde{\text{G}}_1$.
\section{Phonon self-energy in the electron quasiparticle approximation}
We start from Eq.~\eqref{eq:GG} and rewrite it as
\begin{align}
\Pi^{\text{scGscG}}_{\nu \mathbf{q}}(i\omega_j)=&\text{k}_{\text{B}}\text{T}\sum_{nm\mathbf{k}}\sum_{ip_j}|{g}_{mn\nu}\left(\mathbf{k},\mathbf{q} \right)|^2  \iint^{\infty}_{-\infty} \text{d}\varepsilon \text{d}\varepsilon' \nonumber\\ &\times  \frac{A^{\text{scGscD}}_{n\mathbf{k}} ( \varepsilon )}{ip_j - \varepsilon} \frac{A^{\text{scGscD}}_{m\mathbf{k}+\mathbf{q}} ( \varepsilon' )}{ip_j +i\omega_j- \varepsilon'}  .\end{align}
Summation over the Matsubara frequencies $p_j$ is performed by solving a complex integral
\begin{equation}
\oint\frac{1}{z-\varepsilon}\frac{1}{z+i\omega_j-\varepsilon}\frac{1}{e^{\frac{z}{\text{k}_{\text{B}}\text{T}}}+1}dz\end{equation}
where $\oint$ denotes a closed loop and the integral is performed along an infinitely large circumference in the complex plane, and $z=\text{Re}^{i\phi}$, R$\rightarrow\infty$. Since the integral vanishes, it is enough to calculate the sum of residua. The retarded phonon self-energy is obtained by letting $i\omega_j \rightarrow \omega + i\eta$ and gives
\begin{align}
    \Pi^{\text{scGscG}}_{\nu \mathbf{q}}(\omega)=&\sum_{nm\mathbf{k}}|{g}_{mn\nu}\left(\mathbf{k},\mathbf{q} \right)|^2 \iint ^{\infty}_{-\infty}\text{d}\varepsilon \text{d}\varepsilon' A^{\text{scGscD}}_{n\mathbf{k}} ( \varepsilon )\nonumber\\ &\times   A^{\text{scGscD}}_{m\mathbf{k}+\mathbf{q}} ( \varepsilon' ) \Big(\frac{f(\varepsilon)-f(\varepsilon')}{\varepsilon+\omega+i\eta-\varepsilon'}  \Big)  .
\end{align}
Since 
\begin{equation}
    \frac{1}{\varepsilon + i\eta} = P \Big(\frac{1}{\varepsilon}\Big) -i\pi \delta(\varepsilon)
\end{equation}
the imaginary part of the phonon self-energy equals 
\begin{align}
\Im\Pi^{\text{scGscG}}_{\nu \mathbf{q}}(\omega) \!=&-\pi \sum_{nm\mathbf{k}}|{g}_{mn\nu}\left(\mathbf{k},\mathbf{q} \right)|^2 \int^{\infty}_{-\infty}  \!\text{d}\varepsilon  A^{\text{scGscD}}_{n\mathbf{k}} ( \varepsilon )\nonumber\\ \times&   A^{\text{scGscD}}_{m\mathbf{k}+\mathbf{q}} ( \varepsilon+\omega )\big( f(\varepsilon) -f(\varepsilon+\omega) \big).\end{align}
Within the QP approximation for electrons, this expression can be written as 
\begin{align}
\!\!\!\Im\Pi^{\text{sc$\tilde{\text{G}}$sc$\tilde{\text{G}}$}}_{\nu\mathbf{q}} \!(\omega) \!=\!&-\frac{1}{\pi}  \!\sum_{m n \mathbf{k}} \! \left|g_{m n \nu}(\mathbf{k}, \mathbf{q})\right|^2 \int^{\infty}_{-\infty}  \!\!\!\text{d} \varepsilon \frac{\gamma_{n \mathbf{k}}}{\left(\varepsilon \!-\varepsilon_{n \mathbf{k}}\right)^2 \!+\gamma^2_{n \mathbf{k}}} \nonumber\\ &\times\!\frac{\gamma_{m \mathbf{k}+\mathbf{q}}}{\left(\varepsilon\!+\!\omega\!-\!\varepsilon_{m \mathbf{k}+\mathbf{q}}\right)^2\!+\!\gamma_{m \mathbf{k}\!+\mathbf{q}}^2}\!\Big(\!f(\varepsilon)\!-\!f(\varepsilon\!+\!\omega)\!\Big)
\end{align}
and explicitly computed as in Ref.~\cite{park2025}
in terms of the digamma functions. 
We introduce the following notation
\begin{align}
    \Psi(\omega + \mathfrak{E}_{n \mathbf{k}}) =& \psi\left[\frac{1}{2}-\frac{1}{2 \pi i\text{k}_{\text{B}}\text{T}}\left(\omega + \mathfrak{E}_{n \mathbf{k}}\right)\right] \nonumber\\ 
    \mathfrak{E}_{n \mathbf{k}} =& E_{n\mathbf{k}}-\mu+i \gamma_{n \mathbf{k}}
\end{align}
with 
\begin{align}
E_{n\mathbf{k}}=&\varepsilon_{n\mathbf{k}} +\Re\Sigma^{\text{scGscD}}_{n\mathbf{k}}(E_{n\mathbf{k}})\\
\gamma_{n\mathbf{k}}=&-\Im\Sigma^{\text{scGscD}}_{n\mathbf{k}}(E_{n\mathbf{k}}),
\end{align} 
while
$\psi(x)$ is the digamma function.

We compute the real part of the $\text{sc$\tilde{\text{G}}$sc$\tilde{\text{G}}$}$ phonon self-energy from the Kramers-Kronig transform 
\begin{align}
\Re \Pi_{\nu\mathbf{q}}^{\text{sc$\tilde{\text{G}}$sc$\tilde{\text{G}}$}}\left(\omega\right)=&-\frac{1}{\pi^2} \sum_{m n \mathbf{k}} \left|g_{m n \nu}(\mathbf{k}, \mathbf{q})\right|^2  
 \iint^{\infty}_{-\infty} \text{d} \varepsilon  \text{d}\omega'\nonumber \\ &\times \frac{1}{\omega'-\omega}   \frac{\gamma_{n \mathbf{k}}}{|\varepsilon-\mathfrak{E}_{n \mathbf{k}}|^2}\frac{\gamma_{m \mathbf{k}+\mathbf{q}}}{|\varepsilon-\mathfrak{E}_{m \mathbf{k}+\mathbf{q}}+\omega'|^2} \nonumber\\ &\times\Big(f(\varepsilon)-f(\varepsilon+\omega')\Big)\nonumber\\
=&-\frac{1}{\pi^2} \sum_{m n\mathbf{k}} \left|g_{m n \nu}(\mathbf{k}, \mathbf{q})\right|^2  \left( \alpha + \beta \right).
\end{align}
With the letters $\alpha$ and $\beta$ we denote the double integrals we need to solve:
\begin{align}\label{alpha}
   \!\! \!\alpha \!\equiv\!&\iint^{\infty}_{-\infty} \!\!\!\text{d}\varepsilon \text{d}\omega'\!\frac{\gamma_{n \mathbf{k}}}{|\varepsilon-\mathfrak{E}_{n \mathbf{k}}|^2}   \frac{f(\varepsilon)}{\omega'\!-\!\omega}   \frac{\gamma_{m \mathbf{k}+\mathbf{q}}}{|\varepsilon\!-\!\mathfrak{E}_{m \mathbf{k}+\mathbf{q}}\!+\omega'|^2}, \\
\label{beta}
\!\!\!\beta \!\equiv\!& \iint^{\infty}_{-\infty}\! \!\!\text{d} \varepsilon \text{d}\omega' \!\frac{\gamma_{n \mathbf{k}}}{|\varepsilon\!-\!\mathfrak{E}_{n \mathbf{k}}|^2}\frac{f(\varepsilon\!+\!\omega')}{\omega\!-\!\omega'} \frac{\gamma_{m \mathbf{k}+\mathbf{q}}}{|\varepsilon\!-\!\mathfrak{E}_{m \mathbf{k}+\mathbf{q}}\!+\!\omega'|^2} .
\end{align}
The frequency integral in Eq.~\eqref{alpha}, concerns only one Lorentzian, and can be obtained as~\cite{lihm2024}
 \begin{equation}
\int^{\infty}_{-\infty}  \frac{\text{d}\omega'}{\omega'-\omega}\frac{\gamma_{m \mathbf{k}+\mathbf{q}}}{|\varepsilon-\mathfrak{E}_{m \mathbf{k}+\mathbf{q}}+\omega'|^2}
= -2\pi\frac{\varepsilon+\omega-E_{m \mathbf{k}+\mathbf{q}}}{|\varepsilon-\mathfrak{E}_{m \mathbf{k}+\mathbf{q}}+\omega|^2}.
\end{equation}
Next, we use the identity~\cite{park2025}
\begin{equation}
    f(\varepsilon) = \sum_{l=-\infty}^{\infty} \left(\frac{1}{2} - \text{k}_{\text{B}}\text{T}\frac{1}{\varepsilon-\mu-i p_l}\right)
\end{equation}
and continue with two integrals over the energy $\varepsilon$:
\begin{align}
    \alpha =&-2\pi\int^{\infty}_{-\infty} \text{d} \varepsilon \sum_{l=-\infty}^{\infty} \left(\frac{1}{2} - \text{k}_{\text{B}}\text{T}\frac{1}{\varepsilon-\mu-i p_l}\right)\nonumber\\&\times\frac{\gamma_{n \mathbf{k}}}{|\varepsilon-\mathfrak{E}_{n \mathbf{k}}|^2} \frac{\varepsilon+\omega-E_{m \mathbf{k}+\mathbf{q}}}{|\varepsilon-\mathfrak{E}_{m \mathbf{k}+\mathbf{q}}+\omega|^2} \nonumber \\=&
    \gamma+2\pi\int^{\infty}_{-\infty} \text{d} \varepsilon \sum_{l=-\infty}^{\infty} \frac{\text{k}_{\text{B}}\text{T}}{\varepsilon-\mu-i p_l}\frac{\gamma_{n \mathbf{k}}}{|\varepsilon-\mathfrak{E}_{n \mathbf{k}}|^2} \nonumber \\& \times \frac{\varepsilon+\omega-E_{m \mathbf{k}+\mathbf{q}}}{|\varepsilon-\mathfrak{E}_{m \mathbf{k}+\mathbf{q}}+\omega|^2}
\end{align}
where with $\gamma$ we denote the contribution:
\begin{multline}
   \!\!\! \gamma \!\equiv  \!-\pi\int^{\infty}_{-\infty} \!\!\text{d} \varepsilon \sum_{l=-\infty}^{\infty} \frac{\gamma_{n \mathbf{k}}}{|\varepsilon-\mathfrak{E}_{n \mathbf{k}}|^2} \frac{\varepsilon+\omega-E_{m \mathbf{k}+\mathbf{q}}}{|\varepsilon-\mathfrak{E}_{m \mathbf{k}+\mathbf{q}}+\omega|^2}.\label{eq:gamma}
\end{multline}
We leave $\gamma$ unsolved and proceed with the second term
\begin{align}
 &\sum_{l=-\infty}^{\infty} \int^{\infty}_{-\infty} \text{d} \varepsilon \frac{1}{\varepsilon-\mu-i p_l} \frac{\gamma_{n \mathbf{k}}}{|\varepsilon-\mathfrak{E}_{n \mathbf{k}}|^2} \frac{\varepsilon+\omega-E_{m \mathbf{k}+\mathbf{q}}}{|\varepsilon-\mathfrak{E}_{m \mathbf{k}+\mathbf{q}}+\omega|^2}\nonumber\\=&\frac{i}{4}
 \sum_{l=-\infty}^{\infty} \int^{\infty}_{-\infty} \text{d} \varepsilon \frac{1}{\varepsilon-\mu-i p_l}\nonumber\\& \times \left[\frac{1}{\varepsilon-E_{n \mathbf{k}}+i \gamma_{n \mathbf{k}}}-\frac{1}{\varepsilon-E_{n \mathbf{k}}-i \gamma_{n \mathbf{k}}}\right]\nonumber\\& \times\!\left[\frac{1}{\varepsilon\!+\!\omega\!-\!E_{m \mathbf{k}+\mathbf{q}}+i \gamma_{m \mathbf{k}+\mathbf{q}}}\!+\!\frac{1}{\varepsilon\!+\!\omega\!-\!E_{m \mathbf{k}+\mathbf{q}}-i \gamma_{m \mathbf{k}+\mathbf{q}}}\right]\nonumber\\  =&\frac{i}{4\text{k}_{\text{B}}\text{T}}(\gamma_1 - \gamma_1^* + \gamma_2  - \gamma_2^*). 
\end{align}
These integrals can be computed as in Ref.~\cite{park2025}, and we obtain 
\begin{align}
    \gamma_1 \equiv&\frac{- \Psi(\mathfrak{E}^*_{n \mathbf{k}}) + \Psi(-\omega + \mathfrak{E}^*_{m \mathbf{k}+\mathbf{q}}) }{\mathfrak{E}^*_{n \mathbf{k}}+\omega-\mathfrak{E}^*_{m \mathbf{k}+\mathbf{q}}} \\
\gamma_2 \equiv&-\frac{ \Psi(\mathfrak{E}^*_{n \mathbf{k}}) - \Psi(\omega - \mathfrak{E}_{m \mathbf{k}+\mathbf{q}}) }{\mathfrak{E}^*_{n \mathbf{k}}+\omega-\mathfrak{E}_{m \mathbf{k}+\mathbf{q}}} .
\end{align}
In the integral $\beta$ from Eq.~\eqref{beta} we  recognize the Kramers-Kronig transform of a Lorentzian multiplied by the Fermi-Dirac function~\cite{lihm2024} and proceed as
\begin{align}
\beta=& 2\pi\int^{\infty}_{-\infty} \text{d} \varepsilon \frac{\gamma_{n \mathbf{k}}}{|\varepsilon-\mathfrak{E}_{n \mathbf{k}}|^2} \Bigg[\frac{\varepsilon+\omega-E_{m \mathbf{k}+\mathbf{q}}}{|\varepsilon-\mathfrak{E}_{m \mathbf{k}+\mathbf{q}}+\omega|^2} \nonumber\\&\times\Big(\frac{1}{2} - \text{k}_{\text{B}}\text{T}\sum_{l=-\infty}^{\infty}\frac{1}{\varepsilon+\omega-\mu-i p_l}\Big)  \nonumber\\&-
 \text{k}_{\text{B}}\text{T} \sum_{l=0}^{\infty}\Big(\frac{1}{ip_l\!+\!\mu\!-\!\omega\!-\!\varepsilon}\frac{1}{ip_l+\mu-\mathfrak{E}^*_{m \mathbf{k}+\mathbf{q}}} \!+ \!\text{(c.c.)}\Big) 
 \Bigg]\nonumber\\
 =&  -\gamma  - 2\pi\text{k}_{\text{B}}\text{T} \delta_1 -2\pi \delta_2
\end{align}
where (c.c.) denotes complex conjugate of the preceding term. $\gamma$ appears with a negative sign and reduces the same contribution from Eq.~\eqref{eq:gamma}, so we can proceed without solving this integral.
By $\delta_1$ we denote the next term 
\begin{align}
\delta_1 \equiv&  \frac{i}{4} \sum_{l=-\infty}^{\infty} \int^{\infty}_{-\infty} \text{d}\varepsilon \frac{1}{\varepsilon+\omega-\mu-i p_l}\left[\frac{1}{\varepsilon-\mathfrak{E}^*_{n \mathbf{k}}}-\frac{1}{\varepsilon-\mathfrak{E}_{n \mathbf{k}}}\right]\nonumber\\&\times \left[\frac{1}{\varepsilon+\omega-\mathfrak{E}^*_{m \mathbf{k}+\mathbf{q}}}+\frac{1}{\varepsilon+\omega-\mathfrak{E}_{m \mathbf{k}+\mathbf{q}}}\right], 
\end{align}
redefine the integration variable 
\begin{align}
\delta_1 =& \frac{i}{4} \sum_{l=-\infty}^{\infty} \int^{\infty}_{-\infty} \text{d} \varepsilon \frac{1}{\varepsilon-\mu-i p_l}\nonumber\\&
\times\left[\frac{1}{\varepsilon-\mathfrak{E}^*_{m \mathbf{k}+\mathbf{q}}}+\frac{1}{\varepsilon-\mathfrak{E}_{m \mathbf{k}+\mathbf{q}}}\right] \nonumber\\&
\times \left[\frac{1}{\varepsilon-\omega-\mathfrak{E}^*_{n \mathbf{k}}}-\frac{1}{\varepsilon-\omega-\mathfrak{E}_{n \mathbf{k}}}\right] \nonumber\\
 =&\frac{i}{4\text{k}_{\text{B}}\text{T}}(\zeta_1 - \zeta_1^* + \zeta_2  - \zeta_2^*)
\end{align}
and compute $\zeta_{1,2}$ as
\begin{align}
    \zeta_1 \equiv&\frac{\Psi(\omega + \mathfrak{E}^*_{n \mathbf{k}})- \Psi(\mathfrak{E}^*_{m \mathbf{k}+\mathbf{q}})}{\mathfrak{E}^*_{m \mathbf{k}+\mathbf{q}}-\omega-\mathfrak{E}^*_{n \mathbf{k}}} \\
    \zeta_2\equiv&\frac{-\Psi(-\omega - \mathfrak{E}_{n \mathbf{k}})+ \Psi(\mathfrak{E}^*_{m \mathbf{k}+\mathbf{q}})}{\mathfrak{E}^*_{m \mathbf{k}+\mathbf{q}}-\omega-\mathfrak{E}_{n \mathbf{k}}}
\end{align}
With the last part of $\beta$ integral we proceed as 
\begin{align}
\delta_2
\equiv& \text{k}_{\text{B}}\text{T} \sum_{n=0}^{\infty}\int^{\infty}_{-\infty} \text{d}\varepsilon
   \frac{\gamma_{n \mathbf{k}}}
        {|\varepsilon-\mathfrak{E}_{n \mathbf{k}}|^2}
   \frac{1}{\mu + i\nu_n - \varepsilon -\omega}
   \nonumber\\
&\times
   \frac{1}{\mu + i\nu_n -\mathfrak{E}^*_{m \mathbf{k}+\mathbf{q}} }
   \nonumber\\
=& \pi\text{k}_{\text{B}}\text{T}\sum_{n=0}^{\infty}
   \frac{1}
        {i\nu_n + \mu -\mathfrak{E}^*_{m \mathbf{k}+\mathbf{q}}}
   \frac{1}
        {\mu + i\nu_n -\mathfrak{E}^*_{n \mathbf{k}}-\omega}
   \nonumber\\
=& \frac{i}{2}
   \frac{
      \psi\left[
         \frac12+\frac{(\mu-\mathfrak{E}^*_{m \mathbf{k}+\mathbf{q}})}{2\pi i\text{k}_{\text{B}}\text{T}}
      \right]-
      \psi\!\left[
         \frac12+\frac{(\mu-\mathfrak{E}^*_{n \mathbf{k}}-\omega)}{2\pi i\text{k}_{\text{B}}\text{T}}
      \right]
   }
   {\mathfrak{E}^*_{m \mathbf{k}+\mathbf{q}}
    -\mathfrak{E}^*_{n \mathbf{k}}
    -\omega}.
\end{align}
This contribution is equal to $-\frac{i}{2}\zeta_1$, and the complex conjugate of $\delta_2$ yields $\frac{i}{2} \zeta_1^*$.
The contributions can be gathered into a final expression as
\begin{align}
\Re\Pi_{\nu\mathbf{q}}^{\text{sc$\tilde{\text{G}}$sc$\tilde{\text{G}}$}}\left(\omega\right)
 =&\frac{1}{\pi}\sum_{m n\mathbf{k}} \left|g_{m n \nu}(\mathbf{k}, \mathbf{q})\right|^2 \nonumber\\ &\times\bigg(\Im[\gamma_1] + \Im[\gamma_2] -\Im[\zeta_1] -\Im[\zeta_2] + 2\Im[\zeta_1]  \bigg)
 \end{align}
leading to the final result
 \begin{multline}
\Re\Pi_{\nu\mathbf{q}}^{\text{sc$\tilde{\text{G}}$sc$\tilde{\text{G}}$}}\left(\omega\right)
=\frac{1}{2 \pi} \sum_{m n \mathbf{k}}\left|g_{m n \nu}(\mathbf{k}, \mathbf{q})\right|^2 \\ \times\Im\Bigg\{\!
\frac{\Psi(\mathfrak{E}^*_{m \mathbf{k}+\mathbf{q}}-\omega)-\Psi(\mathfrak{E}^*_{n \mathbf{k}})}{\mathfrak{E}^*_{n \mathbf{k}}+\omega-\mathfrak{E}^*_{m \mathbf{k}+\mathbf{q}}} 
-\frac{\Psi(\mathfrak{E}^*_{n \mathbf{k}})-\Psi(\omega-\mathfrak{E}_{m \mathbf{k}+\mathbf{q}})}{\mathfrak{E}^*_{n \mathbf{k}}+\omega-\mathfrak{E}_{m \mathbf{k}+\mathbf{q}}} \\
 -\frac{\Psi(\mathfrak{E}^*_{n \mathbf{k}}+\omega)-\Psi(\mathfrak{E}^*_{m \mathbf{k}+\mathbf{q}})}{\mathfrak{E}^*_{n \mathbf{k}}+\omega-\mathfrak{E}^*_{m \mathbf{k}+\mathbf{q}}} 
+\frac{\Psi(\mathfrak{E}^*_{m \mathbf{k}+\mathbf{q}})-\Psi(\!-\omega\!-\mathfrak{E}_{n \mathbf{k}})}{\mathfrak{E}_{n \mathbf{k}}+\omega-\mathfrak{E}^*_{m \mathbf{k}+\mathbf{q}}}\!\Bigg\}.
\end{multline}
In this work we use the one-shot first-order approximation, and therefore we use the derived expression with  $\text{sc$\tilde{\text{G}}$sc$\tilde{\text{G}}$} \rightarrow \tilde{\text{G}}_1\tilde{\text{G}}_1$. 

We now verify that the expression takes the first-order form in the limit of noninteracting electrons. If $\Sigma_{n\mathbf{k}} \longrightarrow 0$

\begin{align}
 \Re\Pi_{\nu\mathbf{q}}^{\text{G}_0\text{G}_0}(\omega)=&\frac{1}{2 \pi} \sum_{m n\mathbf{k}} \left|g_{m n \nu}(\mathbf{k}, \mathbf{q})\right|^2\nonumber   \\&\times
 \Im\Bigg\{\frac{\psi\left[\frac{1}{2}-\frac{\left(\varepsilon_{m \mathbf{k}+\mathbf{q}}-\omega\right)}{2 \pi i\text{k}_{\text{B}}\text{T}}\right]-\psi\left[\frac{1}{2}-\frac{\varepsilon_{n \mathbf{k}}}{2 \pi i\text{k}_{\text{B}}\text{T}}\right]}{\varepsilon_{n \mathbf{k}}+\omega-\varepsilon_{m \mathbf{k}+\mathbf{q}}} \nonumber \\&
 -\frac{\psi\left[\frac{1}{2}-\frac{\varepsilon_{n \mathbf{k}}}{2 \pi i\text{k}_{\text{B}}\text{T}}
 \right]-\psi\left[\frac{1}{2}+\frac{\left(\varepsilon_{m \mathbf{k}+\mathbf{q}}-\omega\right)}{2 \pi i\text{k}_{\text{B}}\text{T}}\right]}{\varepsilon_{n \mathbf{k}}+\omega-\varepsilon_{m \mathbf{k}+\mathbf{q}}}\nonumber  \\&
 -\frac{\psi\left[\frac{1}{2}-\frac{\left(\varepsilon_{n \mathbf{k}}+\omega\right)}{2 \pi i\text{k}_{\text{B}}\text{T}}\right]-\psi\left[\frac{1}{2}-\frac{\varepsilon_{m \mathbf{k}+\mathbf{q}}}{2 \pi i\text{k}_{\text{B}}\text{T}}\right]}{\varepsilon_{n \mathbf{k}}+\omega-\varepsilon_{m \mathbf{k}+\mathbf{q}}}\nonumber \\&
 +\frac{\psi\left[\frac{1}{2}-\frac{\varepsilon_{m \mathbf{k}+\mathbf{q}}}{2 \pi i\text{k}_{\text{B}}\text{T}}\right]-\psi\left[\frac{1}{2}+\frac{\left(\varepsilon_{n \mathbf{k}}+\omega\right)}{2 \pi i\text{k}_{\text{B}}\text{T}}\right]}{\varepsilon_{n \mathbf{k}}+\omega-\varepsilon_{m \mathbf{k}+\mathbf{q}}}\Bigg\}
\end{align}
where the energies $\varepsilon_{n \mathbf{k}}$ are measured with respect to the chemical potential $\mu$.
Using the identity~\cite{park2025}
\begin{equation}
    f(\varepsilon) = \frac{1}{2} +\frac{1}{\pi} \Im \Psi\Bigg(\frac{1}{2}+\frac{\varepsilon}{2 \pi i\text{k}_{\text{B}}\text{T}}\Bigg),
\end{equation}
we recover the first-order dynamical phonon self-energy
\begin{align}
 \Re\Pi^{\text{G}_0\text{G}_0}_{\nu\mathbf{q}}(\omega)=&\frac{1}{2} \sum_{m n\mathbf{k}} \left|g_{m n \nu}(\mathbf{k}, \mathbf{q})\right|^2  \nonumber\\ \nonumber &\times
\left\{\frac{f(\varepsilon_{n \mathbf{k}}) - \frac{1}{2}-f(\varepsilon_{m \mathbf{k}+\mathbf{q}}-\omega) + \frac{1}{2}}{\varepsilon_{n \mathbf{k}}+\omega-\varepsilon_{m \mathbf{k}+\mathbf{q}}}\right. \\\nonumber
 &+\frac{f(\varepsilon_{n \mathbf{k}}) - \frac{1}{2}+f(\varepsilon_{m \mathbf{k}+\mathbf{q}}-\omega) + \frac{1}{2}}{\varepsilon_{n \mathbf{k}}+\omega-\varepsilon_{m \mathbf{k}+\mathbf{q}}} \\\nonumber
 &+\frac{f(\varepsilon_{n \mathbf{k}}+\omega) - \frac{1}{2}-f(\varepsilon_{m \mathbf{k}+\mathbf{q}}) + \frac{1}{2}}{\varepsilon_{n \mathbf{k}}+\omega-\varepsilon_{m \mathbf{k}+\mathbf{q}}} \\\nonumber
 &+\frac{-f(\varepsilon_{n \mathbf{k}}+\omega) + \frac{1}{2}-f(\varepsilon_{m \mathbf{k}+\mathbf{q}}) + \frac{1}{2}}{\varepsilon_{n \mathbf{k}}+\omega-\varepsilon_{m \mathbf{k}+\mathbf{q}}}\\ 
 =&\sum_{m n\mathbf{k}} \left|g_{m n \nu}(\mathbf{k}, \mathbf{q})\right|^2 \frac{f(\varepsilon_{n \mathbf{k}}) -f(\varepsilon_{m \mathbf{k}+\mathbf{q}}) }{\varepsilon_{n \mathbf{k}}+\omega-\varepsilon_{m \mathbf{k}+\mathbf{q}}} .
\end{align}
\section{Static limit}
\begin{figure*}
    \centering
    \includegraphics[width=0.9\textwidth]{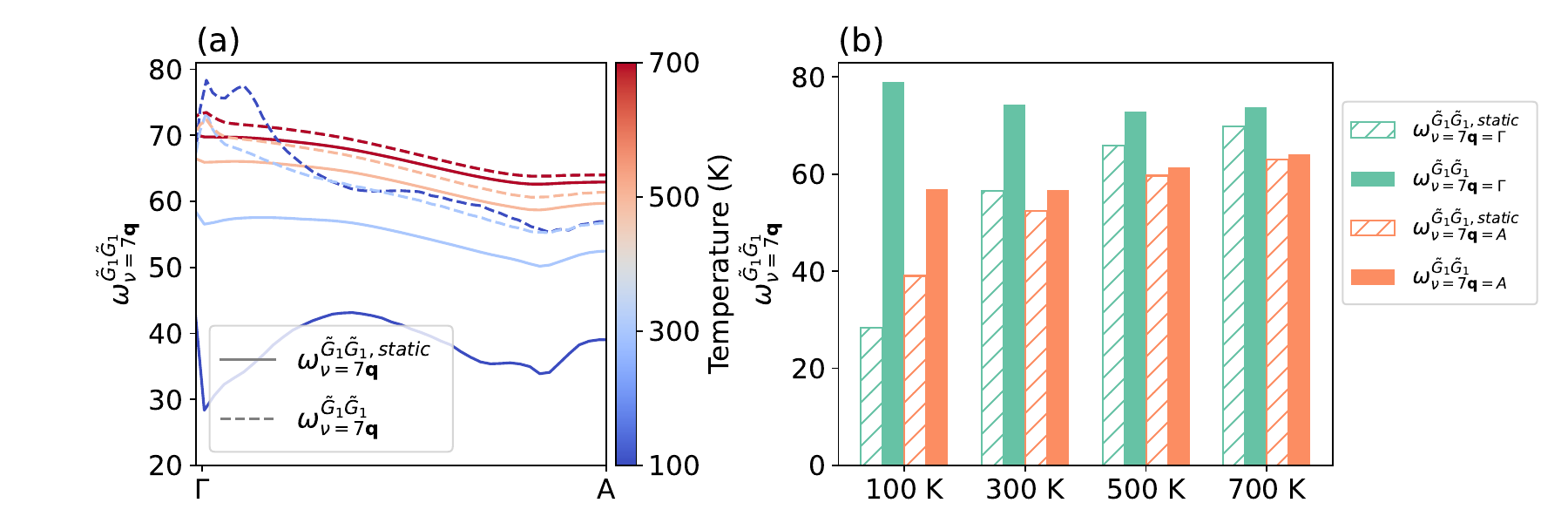}
    \caption{Static and dynamical E$_{2g}$ phonon renormalization (a) along the $\Gamma$ - A path and (b) at  $\mathbf{q}$=$\Gamma$ and $\mathbf{q}$=A points for various temperatures. }
    \label{fig:static_vs_dyn}
\end{figure*}
While the static first-order phonon correction underestimates the E$_{2g}$ phonon frequency, the first-order dynamical phonon correction overestimates it.
Further, both are largely temperature independent in MgB$_2$ if we neglect the temperature dependence of the EPC matrix elements. 
The large dynamical correction is diminished when the higher-orders are accounted for, and even more so as the temperature is elevated.
However, we find that at a certain temperature this trend saturates and even reverses, and that the higher-order dynamical phonon renormalization can also lead to phonon hardening. 
To understand this, we will examine the static limit of the first- and higher- order phonon self-energies. 
At large temperatures, where  $\omega\ll$ T, phonon dispersion approaches its static limit. 
Within the G$_0$G$_0$ approach, the high T limit changes the Fermi-Dirac distribution of electrons into a classical Maxwell-Boltzmann distribution and the phonon self-energy assumes the form
\begin{align}
  \Pi^{\rm \text{G}_0 \text{G}_0}_{\nu\mathbf{q}}(\omega) =& \sum_{nm\mathbf{k}} |g_{mn\nu}(\mathbf{k},\mathbf{q})|^2\frac{f(\varepsilon_{m \mathbf{k}+\mathbf{q}})-f(\varepsilon_{n \mathbf{k}})}{\varepsilon_{m\mathbf{k}+\mathbf{q}} - \varepsilon_{n\mathbf{k}}-\omega - i\eta } \nonumber\\
   = & \sum_{nm\mathbf{k}} |g_{mn\nu}(\mathbf{k},\mathbf{q})|^2\frac{-\frac{\varepsilon_{m\mathbf{k}+\mathbf{q}}}{k_BT} +\frac{\varepsilon_{n\mathbf{k}}}{k_BT}}{\varepsilon_{m\mathbf{k}+\mathbf{q}}- \varepsilon_{n\mathbf{k}}-\omega -i\eta } \nonumber\\
    =& \Pi^{\rm \text{G}_0 \text{G}_0}_{\nu\mathbf{q}}(0)  -\sum_{nm\mathbf{k}} \frac{|g_{mn\nu}(\mathbf{k},\mathbf{q})|^2}{k_BT} \nonumber\\& \times \frac{\omega}{\varepsilon_{m\mathbf{k}+\mathbf{q}}- \varepsilon_{n\mathbf{k}}- \omega- i\eta }.
\end{align}
This expression reveals that in the high T limit, the dynamical part is dominated by the static part because the dynamical part scales as $\frac{\omega}{T}$. 
The static limit of the $\text{sc$\tilde{\text{G}}$sc$\tilde{\text{G}}$}$ approach can be computed as
\begin{align}
\Re\Pi^{\text{sc$\tilde{\text{G}}$sc$\tilde{\text{G}}$}}_{\nu\mathbf{q}}(0)
 =&\frac{1}{\pi} \sum_{m n \mathbf{k}}\left|g_{m n \nu}(\mathbf{k}, \mathbf{q})\right|^2 \nonumber \\ &\times \Im\Bigg\{
\frac{\Psi(\mathfrak{E}_{m \mathbf{k}+\mathbf{q}})-\Psi(\mathfrak{E}_{n \mathbf{k}})}{\mathfrak{E}_{n \mathbf{k}}-\mathfrak{E}_{m \mathbf{k}+\mathbf{q}}}\Bigg\}.\label{eq:static}
\end{align}
In Fig.~\ref{fig:static_vs_dyn}, we show the higher-order dynamic and static corrections to the DFPT phonon dispersion for the strongly-coupled E$_{2g}$ phonon branch in MgB$_2$ for different temperatures.
The dynamical higher-order correction that we show is performed by computing the  $\tilde{\text{G}}_1\tilde{\text{G}}_1$ phonon self-energy and removing the static contribution from DFPT as
\begin{align}
\omega^{\tilde{\text{G}}_1\tilde{\text{G}}_1,2}_{\nu\bq}\!= \!\omega^2_{\nu\bq } \!+ \! 2  \omega_{\nu\bq } \Re\Big(\!\Pi^{\tilde{\text{G}}_1\tilde{\text{G}}_1}_{\nu\bq}(\omega^{\tilde{\text{G}}_1\tilde{\text{G}}_1}_{\nu\bq}\!,\!\text{T})\! -\!\Pi^{\text{G}_0\text{G}_0}_{\nu\bq}(0,\sigma) \!\Big)
\end{align}
where $\omega_{\nu\bq}$ is the DFPT dispersion, the temperature dependence is explicitly denoted, and $\sigma$ is the smearing used in the DFPT calculation (in our case 1578 K).
The dynamical first-order correction can be computed by replacing $\tilde{\text{G}}_1\tilde{\text{G}}_1$ by $\text{G}_0\text{G}_0$.
The higher-order static correction is obtained in the same way as 
\begin{align}
\!\omega^{\tilde{\text{G}}_1\tilde{\text{G}}_1,2}_{\nu\bq } \!=\! \omega^2_{\nu\bq }\! +\!2  \omega_{\nu\bq } \Big(\!\Re\Pi^{\tilde{\text{G}}_1\tilde{\text{G}}_1}_{\nu\bq }(0,\!\text{T})\! -\!\Pi^{\text{G}_0\text{G}_0}_{\nu\bq}(0,\sigma) \!\Big).
\end{align}
Comparing the higher-order dynamically and statically corrected dispersions, it is clear how the difference between the two reduces with T and the system becomes more adiabatic in the sense of phonon renormalization. 
Since the full $\tilde{\text{G}}_1\tilde{\text{G}}_1$ dynamical correction at all the investigated temperatures satisfies $|\Pi^{\tilde{\text{G}}_1\tilde{\text{G}}_1}_{\nu=7\bq }(\omega^{\tilde{\text{G}}_1\tilde{\text{G}}_1}_{\nu=7\bq },T ) |< |\Pi^{\text{G}_0\text{G}_0}_{\nu=7\bq}(0,\sigma)|$ and hardens the adiabatic DFPT phonons, the nonomonotonous T trend corresponds to the competition between the dynamical and statical contributions to the $\tilde{\text{G}}_1\tilde{\text{G}}_1$ self-energy.
At approximately 300~K, away from $\mathbf{q}= \Gamma$ the higher-order static correction and the first-order static correction become equal, and the DFPT phonon dispersion remains largely uncorrected in the static limit.
Further increase in temperature decreases the absolute value of the static higher-order contribution, subsequently making the renormalized phonon frequency larger because 
$|\Pi^{\tilde{\text{G}}_1\tilde{\text{G}}_1}_{\nu=7\bq}(0,\text{T} > 300 \text{ K} ) |< |\Pi^{\text{G}_0\text{G}_0}_{\nu=7\bq}(0,\sigma)|$.
The dynamical part of the higher-order self-energy increases in absolute value with T, but eventually a stronger temperature decrease of the phonon-hardening correction coming from static contributions prevails (see Fig.~\ref{fig:static_vs_dyn}).

\section{Phonon spectral function}

\begin{figure}[b]
    \centering
    \includegraphics[width=0.5\textwidth]{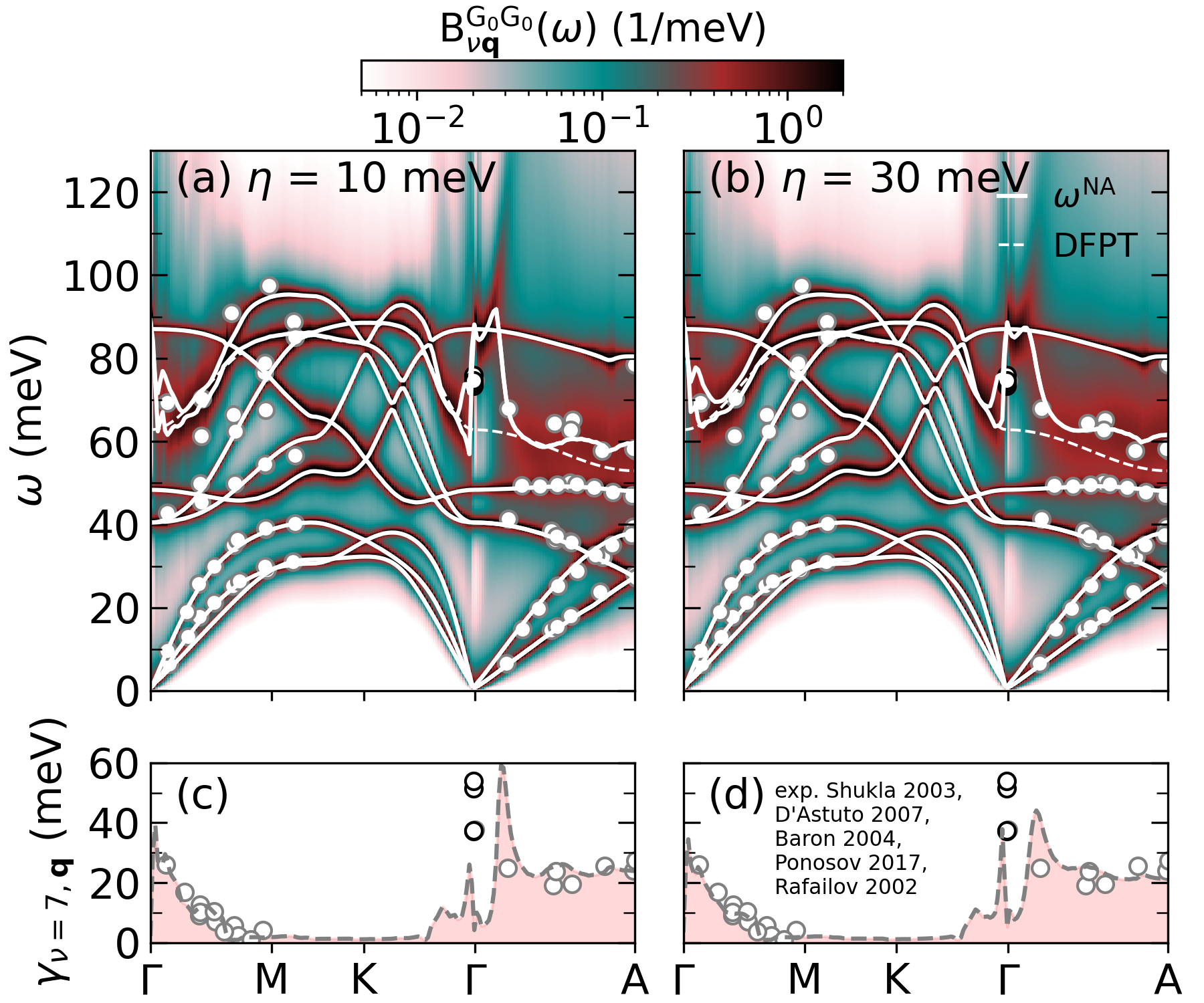}
    \caption{Smearing parameter dependence of the first-order (a,b) phonon spectral function and (c,d) phonon linewidth of the E$_{2g}$ mode in MgB$_2$.}
    \label{fig:supp_fig1}
\end{figure}

\begin{figure*}[ht]
    \centering
    \includegraphics[width=\textwidth]{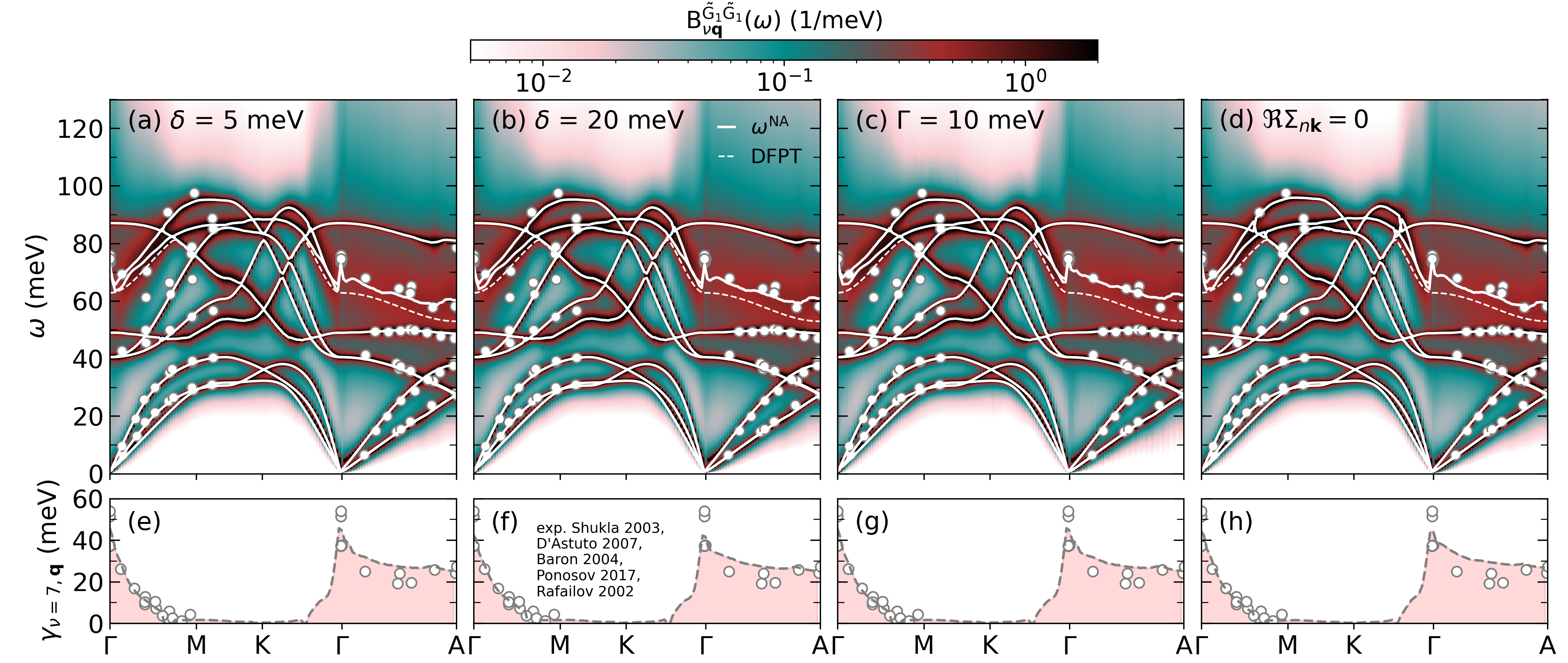}
    \caption{Various modifications of the higher-order result. (a) Reference calculation with $\delta$ = 5 meV. (b) Calculation with the electron self-energy computed with $\delta$ = 20~meV. (c) Calculation with an extra energy-independent broadening of 10 meV, coming from electron-impurity scattering. (d) Result as in (a) but without accounting for the real part of electron self-energy. }
    \label{fig:supp_fig2}
\end{figure*}
The first-order expression for the phonon self-energy, Eq.~\eqref{eq:pi0}, has a smearing parameter in the denominator, denoted as $\eta$.
In Fig.~\ref{fig:supp_fig1} we show how the phonon spectrum depends on this smearing parameter. 
Phonon splitting and nonanalytic behaviour of the frequency and linewidth as a function of the wavevector is greatly reduced by increasing the smearing parameter. However, in order to achieve experimental agreement at the zone centre, one would need to use even larger values than 30~meV, which is excessive since $\eta$ is a convergence parameter.

For the higher-order calculation, the smearing parameter appears only when calculating the electron self-energy since we approximate it with the one-shot FM and DW terms.
With $\delta$, we denote the smearing parameter in the Fan-Migdal expression for the electron self-energy in Eq.~\eqref{g0d0}.
In Fig.~\ref{fig:supp_fig2} we verify how the higher-order spectrum depends on the choice of this parameter. The first two panels show the spectrum obtained with $\delta$ = 5~meV and with $\delta$ = 20~meV, with the effect of a larger smearing $\delta$ mainly being an increase of the electron linewidth at the Fermi level. The two spectral functions are qualitatively similar, but there are slight quantitative differences. The white line in Fig.~\ref{fig:supp_fig2} follows the spectral function maxima and remains mainly unchanged by changing the smearing. The linewidth of the strongly coupled mode is in a slightly better experimental agreement with $\delta$ = 20 meV for $\mathbf{q}=\Gamma$.  Stronger differences with respect to the choice of smearing might arise at lower temperatures because electron broadening is more sensitive at lower temperatures. With a larger smearing $\delta$ the electron broadening increases around the Fermi level and would reach the scale of the phonon frequency at a lower temperature. One can then expect the Raman linewidth of the E$_{2g}$ mode shown in the main text to peak at a smaller temperature. 

Electron broadening can also come from impurity scattering, so in Fig.~\ref{fig:supp_fig2} (c) we add a constant broadening of 10 meV. This extra broadening hardens the phonon frequencies and brings them in closer experimental agreement. The linewidth is slightly reduced, again in closer agreement to the experiments.
Finally, by comparing Fig.~\ref{fig:supp_fig2} (a) and (d) one can see that the real part of the electron self-energy has a small effect on the phonon spectrum of MgB$_2$, but can be important in general~\cite{Ponce2025}.

\section{Computational details}
The DFT ground state calculations were performed with the plane-wave based software \textsc{Quantum ESPRESSO}\,\cite{giannozzi2017qe}, where we used optimized norm-conserving pseudopotentials\,\cite{hamann13} from the \textsc{Pseudo Dojo} library v0.5\,\cite{pseudodojo} with standard accuracy and the PBE exchange-correlation functional.
The plane wave energy cutoff is set to $80$\,Ry.
The convergence criterion for energy is 10$^{-12}$\,eV and the atomic positions are relaxed with a force threshold of 10$^{-4}$\,eV/\AA. 
We obtain the relaxed unit cell determined by the lattice parameters of $a=3.083\,\text{\AA}$ and $c=3.521\,\text{\AA}$. The ground state charge density is obtained by sampling the Brillouin zone with a $12 \times12\times 12$ $\mathbf{k}$-point grid with a Fermi-Dirac smearing function at 1578\,K. The density functional perturbation theory (DFPT)\,\cite{Baroni2001} calculation was performed on a 3$\times$3$\times$3 coarse $\mathbf{q}$-point grid. The Wannier function perturbation theory~\cite{lihm2021} calculation includes 9 bands. 
For the EPC calculations we use the EPW code\,\cite{lee2023epw, Ponce2016}. We obtain  5 maximally localized Wannier functions~\cite{Marzari2012} from the initial projections of B-$pz$. The fine $\mathbf{k}$- and $\mathbf{q}$-point meshes are $84\times84\times84$ and $60\times60\times60$, respectively. For electron self-energy calculations we use a smearing of 5 meV, and for the first-order phonon self-energy we use the smearing value of 10 meV. The calculation of the electron self-energy in EPW is performed by distinguishing the active and rest space. In the former the self-energy is dynamical, while in the latter it is static. In our calculations, the active space spans from -10 eV to 8.8 eV and the Fermi level is located at 7.45 eV.   
\bibliography{ref}